\documentclass[12pt]{article}

\usepackage[]{caption} 

\usepackage{geometry}
\usepackage{array}
\usepackage{longtable}
\usepackage{afterpage}
\usepackage{pdflscape}
\usepackage{arydshln}
\usepackage{multirow}
\usepackage{float}
\usepackage{url}
\usepackage{amsmath}
\usepackage{amsthm}
\usepackage{eufrak}
\usepackage{bbm}
\usepackage{graphicx}
\usepackage{esint}
\usepackage{setspace}
\usepackage{titlesec}
\usepackage{color}
\usepackage[acronym,nonumberlist]{glossaries}
\usepackage{natbib}

\titleformat{\section}{\large\bfseries}{\thesection}{1em}{}
\titleformat{\subsection}{\bfseries}{\thesubsection}{1em}{}
\titleformat{\subsubsection}{\itshape}{\thesubsubsection}{1em}{}

\newcommand{\Keywords}[1]{\par\noindent
	{\small{\em Keywords\/}: #1}}

\begin{document}

\title{\Large 
	Analytic Moments for GARCH Processes}
\author{ Carol Alexander,\footnote{School of Business, Management and Economics, University of Sussex, United Kingdom. c.alexander@sussex.ac.uk}\hspace{0.2cm} Emese Lazar\footnote{ICMA Centre, Henley Business School, University of Reading. e.lazar@rdg.ac.uk},\hspace{0.2cm}Silvia Stanescu\footnote{Kent Business School, University of Kent. stanescu.silvia@gmail.com}}

\maketitle

\doublespacing

\thispagestyle{empty}

\begin{abstract}\normalsize \singlespacing
	\noindent 	For a GJR-GARCH specification with a generic innovation distribution we derive analytic expressions for the first four conditional moments of the forward and aggregated returns and variances. Moment for the most commonly used GARCH models are stated as special cases. We also the limits of these moments as the time horizon increases, establishing regularity conditions for the moments of aggregated returns to converge to normal moments. Our empirical study yields excellent approximate predictive distributions from these analytic moments, thus precluding the need for time-consuming simulations.\\

	\Keywords{Approximate predictive distributions, conditional and unconditional moments, GARCH, kurtosis, skewness, simulation}
		
	\vspace{1cm}
	
	\noindent\textbf{JEL Code:} C53
	
\end{abstract}

\clearpage
\glsresetall
\setcounter{page}{1}

\newpage
\doublespacing
\setcounter{page}{1}

\section{Introduction}
Forward-looking physical return distributions attract a vast academic research interest because they have a great variety of financial applications to market risk assessment and portfolio optimization techniques.
Since Mandelbrot (1963)  and Fama (1965) it is recognized that time series of asset returns are not well described by normal, independent processes. Typically, their conditional distributions are non-normal and they exhibit volatility clustering, so they are not independent. Hence, we require forecasts of the entire  distribution, not only of the first two moments of returns. 

The family of generalized autoregressive conditional heteroskedasticity (GARCH) models is highly successful in capturing (at least partially) the salient empirical features of both conditional and unconditional returns distributions. 
Following the pioneering work of Engle (1982), Bollerslev (1986) and Taylor (1986) numerous alternative specifications for GARCH processes have been proposed. In many financial markets, especially equities and commodities, the GARCH conditional variance equation captures the asymmetric response of volatility to innovations with different signs. Well-known asymmetric GARCH models include the EGARCH model of Nelson (1991), the AGARCH model of Engle (1990) and Engle and Ng (1993), the NGARCH model also proposed by Engle and Ng (1993), and the model of Glosten, Jagannathan and Runkle (1993), henceforth denoted GJR. Additionally, GARCH models with non-normal innovation distributions have been developed by Bollerslev (1987), Nelson (1991), Haas, Mittnik, Paolella (2004) and many others.

The performance of various GARCH models has been emprically assessed by numerous authors, following Andersen and Bollerslev (1998), Marcucci (2005) and many others.\footnote{Also, Andersen, Bollerslev and Diebold (2009) give a broad overview of volatility modelling procedures, focusing on the GARCH methodology and Bauwens, Laurent and Rombouts (2006) review some important contributions to the multivariate ARCH literature.} Virtually all this literature refers to the accuracy of forward or aggregated returns distributions when a point GARCH variance forecast is used. However, only the one-step-ahead GARCH variance forecast is deterministic: due to the uncertainty about future returns, the forward returns variances, and variances of aggregated future returns, are stochastic. So a point GARCH variance forecast represents only an expected value of the GARCH variance, under its distribution. Until now, the only papers to examine this conditional distribution are Ishida and Engle (2002), who derive the conditional variance of the forward conditional variance for a symmetric GARCH(1,1) model with symmetric innovations, and Christoffersen et al. (2010) who derive the second conditional moment of the two-step-ahead forward variance for eight GARCH processes with affine vs. non-affine and conditionally-normal vs. conditionally GED alternatives.

By contrast, there is considerable research on the unconditional moments of returns generated by GARCH processes.\footnote{See Engle (1982), Nemec (1985), Milhoj (1985), Bollerslev (1986), He and Terasvirta (1999a, 1999b), Karanasos (1999, 2001), He, Terasvirta and Malmsten (2002), Demos (2002), Ling and McAleer (2002a, 2002b), Karanasos and Kim (2003), Bai, Russell and Tiao (2003) and Karanasos, Psaradakis and Sola (2004).} However, since returns are not identically distributed, the conditional moments and their dynamics are most important for many financial applications. Knowledge of the dynamics of the conditional mean and variance is sufficient only when conditional distributions are normal: more generally, the dynamics of higher order conditional moments are needed. Hence,  most recent research focuses on the first four conditional moments of forward and aggregated returns for some specific GARCH processes.

Duan et al.(1999) derive expressions for the first four conditional moments of the aggregated returns generated by the normal NGARCH model under the risk-neutral probability measure, and Duan et al.(2006) extend these results to the risk-neutral moments of aggregated returns under normal GJR and normal EGARCH processes.
Wong and So (2003) derive an expression for the variance of aggregated QGARCH  returns and, under the additional assumption that the innovation is symmetric, expressions for the third and fourth order conditional moments of aggregate returns.\footnote{Their QGARCH model is the same as an AGARCH($p$,$q$). Of some relation to this research is the paper by Christoffersen et al.(2008), who propose a new two-component GARCH volatility model for European options valuation, for which the authors also derive the conditional moment generating function of the (log) price distribution.} Breuer and Jandacka (2010) derive the limit of the variance and kurtosis of forward and aggregated returns for a generic symmetric GARCH(1,1) process, for which the forward and aggregated skewness are both zero. 

 We extend this previous research in a unified framework for which the results cited above may be derived as special cases. Assuming a GJR specification and assuming a generic conditional distribution that can accommodate both skewness and kurtosis in the innovations, we derive formulae for the first four conditional moments of forward and aggregated returns and the first four conditional moments of forward and aggregated variances. We also derive the limits of these moments as the returns horizon increases.

A simulation experiment derives approximate predictive distributions from our analytic  moments, showing that they are very close  to the distributions that are generated directly, by Monte Carlo simulation. Thus our results may be use to generate accurate forward and aggregated GARCH distributions without time-consuming simulations. Finally, an empirical study estimates a variety of GARCH models for the S\&P 500 index, Euro--US dollar exchange rate and the 3-month Treasury bill rate and again demonstrates a very good fit between the approximate and simulated predictive returns distributions.

The formulae for the moments are presented in Section \ref{sec:Mom} and those for the limits in Section \ref{sec:Lim}. The proofs are lengthy and are presented in a separate on-line technical appendix, which also details the results for important special cases of the generic model. Section \ref{sec:App} derives approximate predictive distributions for the forward and aggregated returns, using a simulation experiment and then an extensive empirical study to examine the goodness-of-fit between these distributions and those that are generated via simulation; Section \ref{sec:Conc} concludes.

\section{Moments of Generic GJR Returns and Variances}\label{sec:Mom}
Here we present analytic expressions for the first four conditional moments of both forward one-period and future aggregated (also called cumulative) returns and variances for the GJR model with a generic innovation distribution having zero mean, unit variance and finite higher moments. 
We assume that the one-period log return ${r_t} = \log \left(\frac{P_{t+1}}{P_t}\right)$ on a financial asset with market price $P_t$ follows a stationary processes with no significant autocorrelation (if returns exhibit autocorrelation, we can de-autocorrelate the data before estimating the GARCH parameters). The mathematical specification of the generic GJR model is:
\vspace{-0.5cm}
\begin{equation}
{r_t} = \mu  + {\varepsilon _t}, \quad
{\varepsilon _t} = {z_t}h_t^{1/2}, \quad {z_t}\mathop  \sim \limits_{i.i.d.} D\left( {0,1} \right),\quad
{h_t} = \omega  + \alpha {\kern 1pt} {\kern 1pt} \varepsilon _{t - 1}^2 + \lambda {\kern 1pt} {\kern 1pt} \varepsilon _{t - 1}^2I_{t - 1}^ -  + \beta {\kern 1pt} {\kern 1pt} {h_{t - 1}},
\label{gjr_spec}
\end{equation}
where $I_{t}^ -$ is an indicator function which equals 1 if $\varepsilon _{t}$ < 0 and zero otherwise. We also assume that $z_t$ and $h_t$ are independent and  $D(0,1)$ is a generic conditional distribution with zero mean, unit variance, constant skewness ${\tau _z}$ and kurtosis ${\kappa _z}$, and with constant higher moments $\mu _z^{\left( i \right)} = E\left( {z_t^i} \right)$ for any $i > 4,\;i \in N$.\footnote{To be more precise, we have: $$\mu _z^{\left( i \right)} = E\left( {z_t^i} \right) = {E_t}\left( {z_t^i} \right) = {E_t}\left( {z_t - {E_t}\left( {z_t} \right)} \right)^i = {E_t}\left( {z_t - {E_t}\left( {z_t} \right)} \right)^i{E_t}{\left( {z_t^2} \right)^{ - i/2}}$$ since un-centred, centred and standardized moments are all equal for a zero mean, unit variance distribution. Also, since the $z$ process is i.i.d., conditional and unconditional moments of $z$ are also identical. Actually, for the fourth conditional moment of returns to exist we need the first four moments of $z$ to be finite, while we require up to the eighth moment of $z$ to be finite in order to have a finite fourth conditional moment of future variances.}

The aggregated return over $n$ consecutive time periods is denoted ${R_{tn}} = \sum\limits_{s = 1}^n {{r_{t + s}}}$, and for the conditional un-centred and centred moments of the forward and aggregated returns and variances the following notation is used:
\begin{center}
$\tilde \mu _{x,s}^{\left( i \right)} = {E_t}\left( {x_{t + s}^i} \right),\quad \mu _{x,s}^{\left( i \right)} = {E_t}\left( {{{\left( {{x_{t + s}} - \tilde \mu _{x,s}^{\left( 1 \right)}} \right)}^i}} \right)$ \\
$\tilde M_{x,n}^{\left( i \right)} = {E_t}\left[ {{{\left( {\sum\limits_{s = 1}^n {{x_{t + s}}} } \right)}^i}} \right],\quad M_{x,n}^{\left( i \right)} = {E_t}\left( {{{\left( {\sum\limits_{s = 1}^n {\left( {{x_{t + s}} - \tilde \mu _{x,s}^{\left( 1 \right)}} \right)} } \right)}^i}} \right)$
 \vspace{0.5cm}
\end{center}
for $x=r$ and $h$ in turn, $s=1,2,$...$,n$ and $i = 1, 2, 3, 4$. Thus, the skewness and kurtosis of the forward (return or variance) distributions are:
\begin{center}
${\tau _{x,s}} = \mu _{x,s}^{\left( 3 \right)}{\left( {\mu _{x,s}^{\left( 2 \right)}} \right)^{ - {3/2}}}\quad{\rm{ and}}\quad {\kappa _{x,s}} = \mu _{x,s}^{\left( 4 \right)}{\left( {\mu _{x,s}^{\left( 2 \right)}} \right)^{ - 2}}$
\end{center}
and the skewness and kurtosis of the aggregated return or variance distributions are:
\begin{center}
${{\rm T}_{x,n}} = M_{x,n}^{\left( 3 \right)}{\left( {M_{x,n}^{\left( 2 \right)}} \right)^{ - {3/2}}}\quad{\rm{ and }}\quad{{\rm K}_{x,n}} = M_{x,n}^{\left( 4 \right)}{\left( {M_{x,n}^{\left( 2 \right)}} \right)^{ - 2}}.$
\vspace{0.25cm}
\end{center}
We start with the un-centred moments, namely:
\begin {center}
${E_t}\left( {x_{t + s}^i} \right)\quad {\rm{and}}\quad {E_t}\left[ {{{\left( {\sum\limits_{s = 1}^n {{x_{t + s}}} } \right)}^i}} \right]$
\end{center}
again, for $x=r$ and $h$ in turn, $s=1,2,$...$,n$ and $i = 1, 2, 3, 4$. Subsequently, we obtain the centred and standardized moments of the GJR process with a generic innovation distribution.
The derivations rely on the observation that, although ${E_t}\left( {{h_{t + 1}}} \right) = {h_{t + 1}}$ (i.e. ${V_t}\left( {{h_{t + 1}}} \right) = 0$) both $\left\{ {{h_{t + s}}\left| {{\Omega _t}:s \in N\backslash \left\{ {0,1} \right\}} \right.} \right\}$ and $\left\{ {\sum\limits_{s = 1}^n {{h_{t + s}}} \left| {{\Omega _t}:n \in N\backslash \left\{ {0,1} \right\}} \right.} \right\}$ are random. Moreover, both $\left\{ {{r_{t + s}}\left| {{\Omega _t}:s \in N\backslash \left\{ 0 \right\}} \right.} \right\}$ and \\$\left\{ {{R_{tn}}\left| {{\Omega _t}:n \in N\backslash \left\{ 0 \right\}} \right.} \right\}$ are random and have distributions that can also be approximated using moments that we derive.

The following results and notation will be used: \\
 $\varphi  = \alpha  + \lambda {F_0} + \beta $, with ${F_0}$  being the distribution function for $D(0,1)$  evaluated at zero; \\
 $\bar h = \omega {\left( {1 - \varphi } \right)^{ - 1}}$; if $\varphi  \in \left( {0,1} \right)$, then $\bar h$  is the steady-state variance towards which the conditional variance mean reverts;
 \begin{equation}\label{mu_h_2}
\tilde \mu _{h,s}^{\left( 2 \right)} = {c_1} + \left( {h_{t + 1}^2 - {c_3}} \right){\gamma ^{s - 1}} + {c_2}{\varphi ^{s - 1}},
\end{equation}
where
$\;\gamma  = {\varphi ^2} + \left( {{\kappa _z} - 1} \right){\left( {\alpha  + \lambda {F_0}} \right)^2} + {\kappa _z}{\lambda ^2}{F_0}\left( {1 - {F_0}} \right)$,\,\,
${c_1} = \left( {{\omega ^2} + 2\omega \varphi \bar h} \right){\left( {1 - \gamma } \right)^{ - 1}}$, \\${c_2} = 2\omega \varphi \left( {{h_{t + 1}} - \bar h} \right){\left( {\varphi  - \gamma } \right)^{ - 1}}$ and ${c_3} = {c_1} + {c_2}$;
 \begin{equation}\label{eq:2.6}
{E_t}\left( {{\varepsilon _{t + s}}\varepsilon _{t + s + u}^2} \right) = {\varphi ^{u - 1}}\left( {\alpha {\tau _{z}} + \lambda \int\limits_{x =  - \infty }^0 {{x^3}f\left( x \right)dx} } \right){E_t}\left( {h_{t + s}^{3/2}} \right),
\end{equation}
where $f$ is the density function of $D(0,1)$ and ${E_t}\left( {h_{t + s}^{3/2}} \right)\simeq \frac{5}{8}{\left( {\tilde \mu _{h,s}^{\left( 1 \right)}} \right)^{3/2}} + \frac{3}{8}\tilde \mu _{h,s}^{\left( 2 \right)}{\left( {\tilde \mu _{h,s}^{\left( 1 \right)}} \right)^{ - 1/2}}$, with $\tilde \mu _{h,s}^{\left( 2 \right)}$ given in (\ref{mu_h_2}) and $\tilde \mu _{h,s}^{\left( 1 \right)}$ given in Theorem 1 below;
\[{E_t}\left( {{\varepsilon _{t + s}}\varepsilon _{t + s + u}^3} \right) = {\tau _z}\theta _{su}^{\left( {3/2} \right)},\]
where
\[\theta _{su}^{\left( {3/2} \right)} = \frac{3}{4}{\left( {\tilde \mu _{h,s + u}^{\left( 1 \right)}} \right)^{1/2}}\left( \begin{array}{l}
 {c_9}{\varphi ^{u - 1}}{E_t}\left( {h_{t + s}^{3/2}} \right) + \frac{1}{2}{\left( {\tilde \mu _{h,s + u}^{\left( 1 \right)}} \right)^{ - 1}} \\
 \left( {{c_{10}}{\gamma ^{u - 1}}{E_t}\left( {h_{t + s}^{5/2}} \right) + 2\omega {c_9}\left( {\varphi {{\left( {\varphi  - \gamma } \right)}^{ - 1}}\left( {{\varphi ^{u - 1}} - {\gamma ^{u - 1}}} \right) + {\gamma ^{u - 1}}} \right){E_t}\left( {h_{t + s}^{3/2}} \right)} \right) \\
 \end{array} \right),\]
with
\[{c_9} = \left( {\alpha {\tau _ z} + \lambda \int\limits_{x =  - \infty }^0 {{x^3}f\left( x \right)dx} } \right),\]
\[{c_{10}} = \alpha \left( {\alpha \mu _ z^{\left( 5 \right)} + 2\beta {\tau _z}} \right) + \lambda \left( {2\alpha  + \lambda} \right)\int\limits_{x =  - \infty }^0 {{x^5}f\left( x \right)dx} + 2\beta\lambda\int\limits_{x =  - \infty }^0 {{x^3}f\left( x \right)dx},\] and ${E_t}\left( {h_{t + s}^{5/2}} \right)$ is given approximately using a second order Taylor expansion for ${h_{t + s}^{5/2}}$ around ${E_t}\left( {h_{t + s}} \right)$: ${E_t}\left( {h_{t + s}^{5/2}} \right) \simeq \frac{1}{8}{\left( {\tilde \mu _{h,s}^{\left( 1 \right)}} \right)^{1/2}}\left( {15\tilde \mu _{h,s}^{\left( 2 \right)} - 7{{\left( {\tilde \mu _{h,s}^{\left( 1 \right)}} \right)}^2}} \right)$;
 \[{E_t}\left( {\varepsilon _{t + s}^2\varepsilon _{t + s + u}^2} \right) = \bar h\left( {1 - {\varphi ^u}} \right)\tilde \mu _{h,s}^{\left( 1 \right)} + {\varphi ^{u - 1}}{\kappa _z}\left( {\alpha  + \lambda {F_0} + \kappa _z^{ - 1}\beta } \right)\tilde \mu _{h,s}^{\left( 2 \right)};\]
 \[{E_t}\left( {{\varepsilon _{t + s}}{\varepsilon _{t + s + u}}\varepsilon _{t + s + u + v}^2} \right) = {c_9}{\varphi ^{v - 1}}\theta _{su}^{\left( {3/2} \right)}.\]


\noindent \textbf{Theorem 1: Moments of Forward and Aggregated Returns}\\
The conditional moments of forward one-period returns of model (\ref{gjr_spec}) are:
\begin{center}
$\tilde \mu _{r,s}^{\left( 1 \right)} = \mu , \quad
 \mu _{r,s}^{\left( 2 \right)} = \tilde \mu _{h,s}^{(1)} = \bar h + {\varphi ^{s - 1}}\left( {{h_{t + 1}} - \bar h} \right),$ \\
${\tau _{r,s}} = {\tau _{z}}{E_t}\left( {h_{t + s}^{3/2}} \right){\left( {\tilde \mu _{h,s}^{\left( 1 \right)}} \right)^{ - 3/2}} \simeq \tau _{z}\left( {\frac{5}{8} + \frac{3}{8}\tilde \mu _{h,s}^{\left( 2 \right)}{{\left( {\tilde \mu _{h,s}^{\left( 1 \right)}} \right)}^{ - 2}}} \right),$ \\
 ${\kappa _{r,s}} = {\kappa _{z}}\tilde \mu _{h,s}^{\left( 2 \right)}{\left( {\tilde \mu _{h,s}^{\left( 1 \right)}} \right)^{ - 2}}.$
\vspace{0.25cm}
\end{center}

The conditional moments of the aggregated returns of model (\ref{gjr_spec}) are:
\begin{center}
$\tilde M_{r,n}^{\left( 1 \right)} = n\mu,  \quad
M_{r,n}^{\left( 2 \right)} = n\bar h + {\left( {1 - \varphi } \right)^{ - 1}}\left( {1 - {\varphi ^n}} \right)\left( {{h_{t + 1}} - \bar h} \right),$ \\
$ {{\rm T}_{r,n}} \simeq \left( {{\tau _{z}}\sum\limits_{s = 1}^n {\left( {\frac{5}{8}{{\left( {\tilde \mu _{h,s}^{\left( 1 \right)}} \right)}^{3/2}} + \frac{3}{8}\tilde \mu _{h,s}^{\left( 2 \right)}{{\left( {\tilde \mu _{h,s}^{\left( 1 \right)}} \right)}^{ - 1/2}}} \right) + }\,\, 3\sum\limits_{s = 1}^n {\sum\limits_{u = 1}^{n - s} {{E_t}\left( {{\varepsilon _{t + s}}\varepsilon _{t + s + u}^2} \right)} } } \right){\left( {M_{r,n}^{\left( 2 \right)}} \right)^{ - 3/2}},$ \\
${{\rm K}_{r,n}} = \left( \begin{array}{l}
 {\kappa _{z}}\sum\limits_{s = 1}^n {\tilde \mu _{h,s}^{\left( 2 \right)} + \sum\limits_{s = 1}^n {\sum\limits_{u = 1}^{n - s} {\left( {4{E_t}\left( {{\varepsilon _{t + s}}\varepsilon _{t + s + u}^3} \right) + 6{E_t}\left( {\varepsilon _{t + s}^2\varepsilon _{t + s + u}^2} \right)} \right)} } }  \\
  + 12\sum\limits_{s = 1}^n {\sum\limits_{u = 1}^{n - s} {\sum\limits_{v = 1}^{n - s - u} {{E_t}\left( {{\varepsilon _{t + s}}{\varepsilon _{t + s + u}}\varepsilon _{t + s + u + v}^2} \right)} } }  \\
 \end{array} \right){\left( {M_{r,n}^{\left( 2 \right)}} \right)^{ - 2}}.$ \\
\end{center}

\vspace{0.5cm}
\noindent The first conditional moments $\tilde \mu _{r,s}^{\left( 1 \right)}$ and  $\tilde M_{r,n}^{\left( 1 \right)}$ simply state that with a constant conditional mean equation the time $t$ conditional expectation of the $s$-step-ahead one-period return is equal to the constant conditional mean, whereas the expected return aggregated over $n$ periods scales with time. The second moment of the forward return $\mu _{r,s}^{\left( 2 \right)}$ shows that the conditional expectation of the $s$-step-ahead variance is equal to the steady state variance $\bar h$, plus an exponentially decreasing correction term to account for the distance between the one-step-ahead variance $h_{t+1}$ and the steady state variance $\bar h$. Because we assume that the returns are not autocorrelated, the variance of aggregated returns over $n$ time periods $M_{r,n}^{\left( 2 \right)}$ is simply equal to the sum of the $s$-step-ahead variances for $s$ = 1, 2, ..., $n$.

The expressions for the forward and aggregated skewness are obtained using a second order Taylor series expansion for ${E_t}\left( {h_{t + s}^{3/2}}\right)$, as detailed in the Technical Appendix T.A.1. It is easily observed that if the innovation is symmetric ($ {\tau _{z}}=0$) then the forward returns distribution is also symmetric.
By contrast, considering the expression for ${E_t}\left( {{\varepsilon _{t + s}}\varepsilon _{t + s + u}^2} \right)$ in (\ref{eq:2.6}), cumulative returns have an independent source of skewness in addition to that of the innovations ${\tau _{z}}$, due to the asymmetric response parameter $\lambda$ in the conditional variance equation. As a result, aggregated returns can exhibit skewness even if the innovation is symmetric.

For $s=1$, the forward kurtosis equals the kurtosis of the innovation process. But for $s>1$, the forward excess kurtosis can be non-zero even when the innovation has zero excess kurtosis, due to the uncertainty in forward variance. The conditional variance of the conditional variance varies with $s$ and, as it must be positive, the forward kurtosis will be greater than the kurtosis of the innovation, whenever $s>1$, and will itself be time-varying. The net effect of uncertainty in variance is a greater weight in the tails of forward one-period returns. Also, the time-varying conditional variance of the conditional variance introduces dynamics in the higher moments of the forward returns.

The moments of variances require the following results (proved in the Technical Appendix T.A.2):
\begin{center}
$\tilde \mu _{h,s}^{\left( 3 \right)} = \sum\limits_{i = 0}^{s - 2} {c_4^i} \left( {{\omega ^3} + 3{\omega ^2}\varphi \tilde \mu _{h,s - i - 1}^{(1)} + 3\omega \gamma \tilde \mu _{h,s - i - 1}^{(2)}} \right) + c_4^{s - 1}h_{t + 1}^3,$
\end{center}
with
\vspace{-0.5cm}
\begin{eqnarray}\label{eqn:c4}
{c_4} &=& \mu _z^{\left( 6 \right)}\left( {{\alpha ^3} + 3\alpha \lambda \left( {\alpha  + \lambda } \right){F_0} + {\lambda ^3}{F_0}} \right) + 3\beta \gamma  - {\beta ^2}\left( {2\beta  + 3\left( {\alpha  + \lambda {F_0}} \right)} \right)\\ \nonumber
\tilde \mu _{h,s}^{(4)} &=& \sum\limits_{j = 0}^{s - 2} {c_7^j} \left( {{\omega ^4} + 4{\omega ^3}\varphi \tilde \mu _{h,s - j - 1}^{(1)} + {c_5}\tilde \mu _{h,s - j - 1}^{(2)} + {c_6}\tilde \mu _{h,s - j - 1}^{(3)}} \right) + c_7^{s - 1}h_{t + 1}^4, \nonumber
\end{eqnarray}
with ${c_5} = 6{\omega ^2}\gamma, \, {c_6} = 4\omega {c_4}$, and
\vspace{-0.5cm}
\begin{eqnarray*}
{c_7} &=& \mu _z^{\left( 8 \right)}\left( {{\alpha ^4} + {F_0}\left( {{\lambda ^4} + 4\left( {{\alpha ^3}\lambda  + \alpha {\lambda ^3}} \right) + 6{\alpha ^2}{\lambda ^2}} \right)} \right) + {\beta ^4} \\
&+& 4\left[ {\mu _z^{\left( 6 \right)}\beta \left( {{\alpha ^3} + {F_0}\left( {{\lambda ^3} + 3\left( {{\alpha ^2}\lambda  + \alpha {\lambda ^2}} \right)} \right)} \right) + {\beta ^3}\left( {\alpha  + \lambda {F_0}} \right)} \right]+ 6{\kappa _z}{\beta ^2}\left( {{\alpha ^2} + {\lambda ^2}{F_0} + 2\alpha \lambda {F_0}} \right).
\end{eqnarray*}
Expressions for $\tilde \mu _{h,suv}^{\left( {i,j,k} \right)}$, with $i$, $j$, $k$ $\in \left\{ {0,1,2} \right\}$ are also required but since most are rather lengthy they are only stated in the Technical Appendix T.A.2, with the proof of the following result:

\noindent \textbf{Theorem 2: Moments of Forward and Aggregated Variances}\\
The conditional moments of forward one-period variances of model (\ref{gjr_spec}) are:
\begin{center}
$\tilde \mu _{h,s}^{\left( 1 \right)} = \bar h + {\varphi ^{s - 1}}\left( {{h_{t + 1}} - \bar h} \right), \quad
\mu _{h,s}^{\left( 2 \right)} = \tilde \mu _{h,s}^{\left( 2 \right)} - {\left( {\tilde \mu _{h,s}^{\left( 1 \right)}} \right)^2},$ \\
${\tau _{h,s}} = \left[ {\tilde \mu _{h,s}^{\left( 3 \right)} - 3\tilde \mu _{h,s}^{\left( 2 \right)}\tilde \mu _{h,s}^{\left( 1 \right)} + 2{{\left( {\tilde \mu _{h,s}^{\left( 1 \right)}} \right)}^3}} \right]{\left( {\tilde \mu _{h,s}^{\left( 2 \right)} - \tilde \mu _{h,s}^{\left( 1 \right)}} \right)^{-{3/2}}},$ \\
${\kappa _{h,s}} = \left( {\tilde \mu _{h,s}^{\left( 4 \right)} - 4\tilde \mu _{h,s}^{\left( 1 \right)}\tilde \mu _{h,s}^{\left( 3 \right)} + 6{{\left( {\tilde \mu _{h,s}^{\left( 1 \right)}} \right)}^2}\tilde \mu _{h,s}^{\left( 2 \right)} - 3{{\left( {\tilde \mu _{h,s}^{\left( 1 \right)}} \right)}^4}} \right){\left( {\tilde \mu _{h,s}^{\left( 2 \right)} - {{\left( {\tilde \mu _{h,s}^{\left( 1 \right)}} \right)}^2}} \right)^{ - 2}}.$
\end{center}
The conditional moments of the aggregated future variances of model (\ref{gjr_spec}) are:
\begin{center}
$\tilde M_{h,n}^{\left( 1 \right)} = n\bar h + \left( {{h_{t + 1}} - \bar h} \right){\left( {1 - \varphi } \right)^{ - 1}}\left( {1 - {\varphi ^n}} \right),$
\end{center}
\begin{center}
$M_{h,n}^{\left( 2 \right)} = \sum\limits_{s = 1}^n {\left( {\tilde \mu _{h,s}^{\left( 2 \right)} - {{\left( {\tilde \mu _{h,s}^{\left( 1 \right)}} \right)}^2}} \right)}  + 2\sum\limits_{s = 1}^n {\sum\limits_{u = 1}^{n - s} {\left( {\tilde \mu _{h,su}^{\left( {1,1} \right)} - \tilde \mu _{h,s}^{\left( 1 \right)}\tilde \mu _{h,s + u}^{\left( 1 \right)}} \right)} } ,$
\end{center}
\begin{center}
${{\rm T}_{h,n}} = M_{h,n}^{\left( 3 \right)}{\left( {M_{h,n}^{\left( 2 \right)}} \right)^{-{3/2}}},$
\end{center}
\begin{center}
$M_{h,n}^{\left( 3 \right)} = \sum\limits_{s = 1}^n {\left( \tilde \mu _{h,s}^{\left( 3 \right)} - 3\tilde \mu _{h,s}^{\left( 2 \right)}\tilde \mu _{h,s}^{\left( 1 \right)} + 2{{\left( {\tilde \mu _{h,s}^{\left( 1 \right)}} \right)}^3}\right)}  + 3\sum\limits_{s = 1}^n {\sum\limits_{u = 1}^{n - s} {{A_{h,s,u}}} }  + 6\sum\limits_{s = 1}^n {\sum\limits_{u = 1}^{n - s} {\sum\limits_{v = 1}^{n - s - u} {{B_{h,s,u,v}}} } } ,$ \\
${A_{h,s,u}} = \tilde \mu _{h,su}^{\left( {2,1} \right)} + \tilde \mu _{h,su}^{\left( {1,2} \right)} + 2\left( {\tilde \mu _{h,s}^{\left( 1 \right)} + \tilde \mu _{h,s + u}^{\left( 1 \right)}} \right)\left( {\tilde \mu _{h,s}^{\left( 1 \right)}\tilde \mu _{h,s + u}^{\left( 1 \right)} - \tilde \mu _{h,su}^{\left( {1,1} \right)}} \right) - \tilde \mu _{h,s}^{\left( 1 \right)}\tilde \mu _{h,s + u}^{\left( 2 \right)} - \tilde \mu _{h,s + u}^{\left( 1 \right)}\tilde \mu _{h,s}^{\left( 2 \right)},$ \\
${B_{h,s,u,v}} = \tilde \mu _{h,suv}^{\left( {1,1,1} \right)} - \tilde \mu _{h,s}^{\left( 1 \right)}\tilde \mu _{h,\left( {s + u} \right)v}^{\left( {1,1} \right)} - \tilde \mu _{h,\left( {s + u} \right)}^{\left( 1 \right)}\tilde \mu _{h,s\left( {u + v} \right)}^{\left( {1,1} \right)} - \tilde \mu _{h,\left( {s + u + v} \right)}^{\left( 1 \right)}\tilde \mu _{h,su}^{\left( {1,1} \right)} + 2\tilde \mu _{h,s}^{\left( 1 \right)}\tilde \mu _{h,\left( {s + u} \right)}^{\left( 1 \right)}\tilde \mu _{h,\left( {s + u + v} \right)}^{\left( 1 \right)},$ \\
\end{center}
\begin{center}
${{\rm K}_{h,n}} = M_{h,n}^{\left( 4 \right)}{\left( {M_{h,n}^{\left( 2 \right)}} \right)^{ - 2}}\,$ \\
$M_{h,n}^{\left( 4 \right)} = \sum\limits_{s = 1}^n {\mu _{h,s}^{\left( 4 \right)} + } \sum\limits_{s = 1}^n {\sum\limits_{u = 1}^{n - s} {{C_{h,s,u}}} }  + 12\sum\limits_{s = 1}^n {\sum\limits_{u = 1}^{n - s} {\sum\limits_{v = 1}^{n - s - u} {{D_{h,s,u,v}}} } }  + 24\sum\limits_{s = 1}^n {\sum\limits_{u = 1}^{n - s} {\sum\limits_{v = 1}^{n - s - u} {\sum\limits_{w = 1}^{n - s - u - v} {{E_{h,s,u,v,w}}} } } }, $
\end{center}
\[\begin{array}{l}
 {C_{h,s,u}} = 4\left( \begin{array}{l}
 \tilde \mu _{h,su}^{\left( {3,1} \right)} + \tilde \mu _{h,su}^{\left( {1,3} \right)} - 3\left( {\tilde \mu _{h,s}^{\left( 1 \right)}\tilde \mu _{h,su}^{\left( {2,1} \right)} + \tilde \mu _{h,s + u}^{\left( 1 \right)}\tilde \mu _{h,su}^{\left( {1,2} \right)}} \right) - \left( {\tilde \mu _{h,s + u}^{\left( 1 \right)}\tilde \mu _{h,s}^{\left( 3 \right)} + \tilde \mu _{h,s}^{\left( 1 \right)}\tilde \mu _{h,s + u}^{\left( 3 \right)}} \right) \\
  + 3\tilde \mu _{h,s}^{\left( 1 \right)}\left( {\tilde \mu _{h,s}^{\left( 1 \right)}\tilde \mu _{h,su}^{\left( {1,1} \right)} + \tilde \mu _{h,s + u}^{\left( 1 \right)}\tilde \mu _{h,s}^{\left( 2 \right)} - {{\left( {\tilde \mu _{h,s}^{\left( 1 \right)}} \right)}^2}\tilde \mu _{h,s + u}^{\left( 1 \right)}} \right) \\
  + 3\tilde \mu _{h,s + u}^{\left( 1 \right)}\left( {\tilde \mu _{h,s + u}^{\left( 1 \right)}\tilde \mu _{h,su}^{\left( {1,1} \right)} + \tilde \mu _{h,s}^{\left( 1 \right)}\tilde \mu _{h,s + u}^{\left( 2 \right)} - \tilde \mu _{h,s}^{\left( 1 \right)}{{\left( {\tilde \mu _{h,s + u}^{\left( 1 \right)}} \right)}^2}} \right) \\
 \end{array} \right) \\
 \quad \quad\quad + 6\left( \begin{array}{l}
 \tilde \mu _{h,su}^{\left( {2,2} \right)} - 2\left( {\tilde \mu _{h,s}^{\left( 1 \right)}\tilde \mu _{h,su}^{\left( {1,2} \right)} + \tilde \mu _{h,s + u}^{\left( 1 \right)}\tilde \mu _{h,su}^{\left( {2,1} \right)}} \right) + {\left( {\tilde \mu _{h,s}^{\left( 1 \right)}} \right)^2}\tilde \mu _{h,s + u}^{\left( 2 \right)} \\
  + {\left( {\tilde \mu _{h,s + u}^{\left( 1 \right)}} \right)^2}\tilde \mu _{h,s}^{\left( 2 \right)} + 4\tilde \mu _{h,s}^{\left( 1 \right)}\tilde \mu _{h,s + u}^{\left( 1 \right)}\tilde \mu _{h,su}^{\left( {1,1} \right)} - 3{\left( {\tilde \mu _{h,s}^{\left( 1 \right)}\tilde \mu _{h,s + u}^{\left( 1 \right)}} \right)^2} \\
 \end{array} \right), \\
 \end{array}\]
\[\begin{array}{l}
 {D_{h,s,u,v}} = \tilde \mu _{h,suv}^{\left( {2,1,1} \right)} - 2\tilde \mu _{h,s}^{\left( 1 \right)}\tilde \mu _{h,suv}^{\left( {1,1,1} \right)} - \tilde \mu _{h,s + u}^{\left( 1 \right)}\tilde \mu _{h,s\left( {u + v} \right)}^{\left( {2,1} \right)} - \tilde \mu _{h,s + u + v}^{\left( 1 \right)}\tilde \mu _{h,su}^{\left( {2,1} \right)} + {\left( {\tilde \mu _{h,s}^{\left( 1 \right)}} \right)^2}\tilde \mu _{h,\left( {s + u} \right)v}^{\left( {1,1} \right)} \\
 \quad \quad \;\; + 2\tilde \mu _{h,s}^{\left( 1 \right)}\tilde \mu _{h,s + u}^{\left( 1 \right)}\tilde \mu _{h,s\left( {u + v} \right)}^{\left( {1,1} \right)} + 2\tilde \mu _{h,s}^{\left( 1 \right)}\tilde \mu _{h,s + u + v}^{\left( 1 \right)}\tilde \mu _{h,su}^{\left( {1,1} \right)} + \tilde \mu _{h,s + u}^{\left( 1 \right)}\tilde \mu _{h,s + u + v}^{\left( 1 \right)}\tilde \mu _{h,s}^{\left( 2 \right)} \\
 \quad \quad \;\; - 3{\left( {\tilde \mu _{h,s}^{\left( 1 \right)}} \right)^2}\tilde \mu _{h,s + u}^{\left( 1 \right)}\tilde \mu _{h,s + u + v}^{\left( 1 \right)} + \tilde \mu _{h,suv}^{\left( {1,2,1} \right)} - 2\tilde \mu _{h,s + u}^{\left( 1 \right)}\tilde \mu _{h,suv}^{\left( {1,1,1} \right)} - \tilde \mu _{h,s}^{\left( 1 \right)}\tilde \mu _{h,\left( {s + u} \right)v}^{\left( {2,1} \right)} - \tilde \mu _{h,s + u + v}^{\left( 1 \right)}\tilde \mu _{h,su}^{\left( {1,2} \right)} \\
 \quad \quad \;\; + {\left( {\tilde \mu _{h,s + u}^{\left( 1 \right)}} \right)^2}\tilde \mu _{h,s\left( {u + v} \right)}^{\left( {1,1} \right)} + 2\tilde \mu _{h,s}^{\left( 1 \right)}\tilde \mu _{h,s + u}^{\left( 1 \right)}\tilde \mu _{h,\left( {s + u} \right)v}^{\left( {1,1} \right)} + 2\tilde \mu _{h,s + u}^{\left( 1 \right)}\tilde \mu _{h,s + u + v}^{\left( 1 \right)}\tilde \mu _{h,su}^{\left( {1,1} \right)} \\
 \quad \quad \;\; + \tilde \mu _{h,s}^{\left( 1 \right)}\tilde \mu _{h,s + u + v}^{\left( 1 \right)}\tilde \mu _{h,s + u}^{\left( 2 \right)} - 3\tilde \mu _{h,s}^{\left( 1 \right)}{\left( {\tilde \mu _{h,s + u}^{\left( 1 \right)}} \right)^2}\tilde \mu _{h,s + u + v}^{\left( 1 \right)} + \tilde \mu _{h,suv}^{\left( {1,1,2} \right)} - 2\tilde \mu _{h,s + u + v}^{\left( 1 \right)}\tilde \mu _{h,suv}^{\left( {1,1,1} \right)} \\
 \quad \quad \;\; - \tilde \mu _{h,s}^{\left( 1 \right)}\tilde \mu _{h,\left( {s + u} \right)v}^{\left( {1,2} \right)} - \tilde \mu _{h,s + u}^{\left( 1 \right)}\tilde \mu _{h,s\left( {u + v} \right)}^{\left( {1,2} \right)} + {\left( {\tilde \mu _{h,s + u + v}^{\left( 1 \right)}} \right)^2}\tilde \mu _{h,su}^{\left( {1,1} \right)} + 2\tilde \mu _{h,s}^{\left( 1 \right)}\tilde \mu _{h,s + u + v}^{\left( 1 \right)}\tilde \mu _{h,\left( {s + u} \right)v}^{\left( {1,1} \right)} \\
 \quad \quad \;\; + 2\tilde \mu _{h,s + u}^{\left( 1 \right)}\tilde \mu _{h,s + u + v}^{\left( 1 \right)}\tilde \mu _{h,s\left( {u + v} \right)}^{\left( {1,1} \right)} + \tilde \mu _{h,s}^{\left( 1 \right)}\tilde \mu _{h,s + u}^{\left( 1 \right)}\tilde \mu _{h,s + u + v}^{\left( 2 \right)} - 3\tilde \mu _{h,s}^{\left( 1 \right)}\tilde \mu _{h,s + u}^{\left( 1 \right)}{\left( {\tilde \mu _{h,s + u + v}^{\left( 1 \right)}} \right)^2}, \\
 \end{array}\]
\[\begin{array}{l}
 {E_{h,s,u,v,w}} = \tilde \mu _{h,suvw}^{\left( {1,1,1,1} \right)} - \tilde \mu _{h,s + u + v}^{\left( 1 \right)}\tilde \mu _{h,su\left( {v + w} \right)}^{\left( {1,1,1} \right)} - \tilde \mu _{h,s + u + v + w}^{\left( 1 \right)}\tilde \mu _{h,suv}^{\left( {1,1,1} \right)} + \tilde \mu _{h,s + u + v}^{\left( 1 \right)}\tilde \mu _{h,s + u + v + w}^{\left( 1 \right)}\tilde \mu _{h,su}^{\left( {1,1} \right)} - \tilde \mu _{h,s}^{\left( 1 \right)}\tilde \mu _{h,\left( {s + u} \right)vw}^{\left( {1,1,1} \right)} \\
 \quad \quad \;\;\; + \tilde \mu _{h,s}^{\left( 1 \right)}\tilde \mu _{h,s + u + v}^{\left( 1 \right)}\tilde \mu _{h,\left( {s + u} \right)\left( {v + w} \right)}^{\left( {1,1} \right)} + \tilde \mu _{h,s}^{\left( 1 \right)}\tilde \mu _{h,s + u + v + w}^{\left( 1 \right)}\tilde \mu _{h,\left( {s + u} \right)v}^{\left( {1,1} \right)} - \tilde \mu _{h,s + u}^{\left( 1 \right)}\tilde \mu _{h,\left( {s + u} \right)vw}^{\left( {1,1,1} \right)} + \tilde \mu _{h,s + u}^{\left( 1 \right)}\tilde \mu _{h,s + u + v}^{\left( 1 \right)}\tilde \mu _{h,s\left( {u + w + v} \right)}^{\left( {1,1} \right)} \\
 \quad \quad \;\;\; + \tilde \mu _{h,s + u}^{\left( 1 \right)}\tilde \mu _{h,s + u + v + w}^{\left( 1 \right)}\tilde \mu _{h,s\left( {u + v} \right)}^{\left( {1,1} \right)} + \tilde \mu _{h,s}^{\left( 1 \right)}\tilde \mu _{h,s + u}^{\left( 1 \right)}\tilde \mu _{h,\left( {s + u + v} \right)w}^{\left( {1,1} \right)} - 3\tilde \mu _{h,s}^{\left( 1 \right)}\tilde \mu _{h,s + u}^{\left( 1 \right)}\tilde \mu _{h,s + u + v}^{\left( 1 \right)}\tilde \mu _{h,s + u + v + w}^{\left( 1 \right)}. \\
 \end{array}\]

\vspace{0.5cm}
\noindent It has been argued, for instance in Ishida and Engle (2002), that the conditional variance of the conditional variance grows faster than linearly with the current variance. Our formula for $\mu _{h,s}^{\left( 2 \right)}$ shows that, in the context of model (\ref{gjr_spec}), it grows quadratically, as from this it easily follows that the conditional variance of the forward conditional variance is a quadratic function of the current variance ${h_{t + 1}}\,.$ Hence, the uncertainty around the point variance forecast increases much more than linearly when variance levels are high, much reducing the reliability of the point forecast. This highlights the importance of our analytic formulae for the higher moments of GARCH variances.\footnote{Intuitively, we expect distributions of forward variances to be positively skewed, since jumps in variance are usually positive rather than negative. In an empirical implementation of the moments formulae derived in this section we find that the skewness of forward variance is indeed positive, for all horizons and all three samples considered; we also find that the excess kurtosis of variance is always positive. These empirical results are excluded from this paper for reasons of space, but can be obtained from the authors on request.}


\section{Limits of the Moments}\label{sec:Lim}
The convergence behaviour of the conditional moments as the time horizon increases
is of interest both theoretically and empirically. The  following result summarizes the limits of the moments of returns (the limits of the forward and aggregated mean are trivial and immediate and are thus excluded). \\

\noindent\textbf{Theorem 3: Limiting Behaviour of Moments of Forward and Aggregated Returns}

\noindent Suppose $\varphi \in (0,1)$ and $\varphi\neq \gamma$. The limiting behaviour of the conditional moments of the forward one-period and aggregated returns of model (\ref{gjr_spec}) when we increase the time horizon is:\footnote{${\mathop{\rm sgn}} \left( x \right) = \left\{ \begin{array}{l}
  - 1\quad {\rm{if}}\;x < 0, \\
 0\quad \; \; \; {\rm{if}}\;x = 0, \\
 1\quad \; \; \; {\rm{if}}\;x > 0 \\
 \end{array} \right.$ and we use the convention that $\mathop{\rm sgn}\left( 0 \right)\infty = 0$.}
\begin{equation}\label{lim_var}
\mathop {\lim }\limits_{s \to \infty } \mu _{r,s}^{\left( 2 \right)} = \bar h,
\end{equation}
\begin{equation}\label{lim_skew}
\mathop {\lim }\limits_{s \to \infty } {\tau _{r,s}} = \left\{ \begin{array}{l}
 {\tau _z}\left( {\frac{5}{8} + \frac{3}{8}\left( {{\omega ^2} + 2\omega \varphi \bar h} \right){{\left( {1 - \gamma } \right)}^{ - 1}}{{\left( {\bar h} \right)}^{ - 2}}} \right)\quad {\rm{if}}\;\gamma  \in \left( {0,1} \right), \\
 {\mathop{\rm sgn}} \left( {{\tau _z}} \right)\infty \qquad \qquad \qquad \qquad \qquad \qquad \quad \; \;\, {\rm{if}}\;\gamma  \in \left[ {1,\infty } \right), \\
 \end{array} \right.
\end{equation}
\begin{equation}\label{lim_kurt}
\mathop {\lim }\limits_{s \to \infty } {\kappa _{r,s}} = \left\{ \begin{array}{l}
 {\kappa _z}\left( {{\omega ^2} + 2\omega \varphi \bar h} \right){\left( {1 - \gamma } \right)^{ - 1}}{\left( {\bar h} \right)^{ - 2}}\qquad \qquad {\rm{if}}\;\gamma  \in \left( {0,1} \right), \\
 \infty \qquad \qquad \qquad \qquad \qquad \qquad \qquad \qquad \; {\rm{if}}\;\gamma  \in \left[ {1,\infty } \right), \\
 \end{array} \right.
\end{equation}
\begin{equation}\label{lim_agg_var}
\mathop {\lim }\limits_{n \to \infty } \frac{{M_{r,n}^{\left( 2 \right)}}}{n} = \bar h,
\end{equation}
\begin{equation}\label{lim_agg_skew}
\mathop {\lim }\limits_{n \to \infty } {{\rm T}_{r,n}} = \left\{ \begin{array}{l}
 0\qquad \qquad \qquad \qquad \qquad \qquad \qquad \qquad \qquad \quad \; \; {\rm{if}}\;\gamma  \in \left( {0,1} \right), \\
 {\mathop{\rm sgn}} \left( {{\tau _z}\left( {\alpha  + \frac{{\gamma  - \varphi }}{3}} \right) + \lambda \int\limits_{x =  - \infty }^0 {{x^3}f\left( x \right)dx} } \right)\infty \quad \quad {\rm{if}}\;\gamma  \in \left[ {1,\infty } \right), \\
 \end{array} \right.
\end{equation}
\begin{equation}\label{lim_agg_kurt}
\mathop {\lim }\limits_{n \to \infty } {{\rm K}_{r,n}} = \left\{ \begin{array}{l}
 3\qquad \qquad \qquad \qquad \qquad \qquad \qquad \qquad \qquad \qquad \qquad \quad \qquad \qquad \qquad \; \; \; \,\, \;{\rm{if}}\;\gamma  \in \left( {0,1} \right), \\
 3 + \frac{{{\kappa _{z}}}}{2}\left( {1 - {\varphi ^2}} \right)\left( {1 + 6\left( {\alpha  + \lambda {F_0} + \kappa _z^{ - 1}\beta } \right){{\left( {1 - \varphi } \right)}^{ - 1}}} \right)+ {\rm{ sgn}}\left( {\left| \lambda  \right| + \left| {{\tau _z}} \right|} \right)\infty \quad {\rm{if}}\;\gamma  = 1, \\
 \infty \qquad \qquad \qquad \qquad \qquad \qquad \qquad \qquad \qquad \qquad \qquad \quad \qquad \qquad \qquad \; \;\; \, \, {\rm{if}}\;\gamma  \in \left( {1,\infty } \right). \\
 \end{array} \right.
\end{equation}

\vspace{0.5cm}
\noindent Hence, under suitable parameter conditions, the conditional moments of forward one-period returns converge to finite limits that are the unconditional counterparts of the respective conditional moments, and these parameter conditions are the necessary and sufficient conditions for the existence of the corresponding unconditional moments.
Indeed, $\varphi \in \left( {0,1} \right)$ is a necessary and sufficient condition for the existence of the unconditional variance, as can be shown using Theorem 2.2 (and Example 2.1) in Ling and McAleer(2002b), 
and for  $\varphi \in \left( {0,1}   \right)$ a steady-state level of variance exists, i.e. $\exists \; h_0$ such that $E\left( {{h_{t }}} \right) = h_0$, for any $t \in N$. It is easy to show that, when it exists, the unconditional variance $h_0$ is given by: $h_0= \frac{\omega }{{1 - \varphi }} = \bar h$.\footnote{Applying the expectation operator on both sides of the equation for the GJR conditional variance and using that the indicator $I_t^ - $ and the even powers of the contemporaneous innovations $\varepsilon _t^{2k}$, where $k \in N$, are independent, we get: $E\left( {{h_t}} \right) = \omega  + \alpha {\kern 1pt} {\kern 1pt} E\left( {\varepsilon _{t - 1}^2} \right) + \lambda {\kern 1pt} E\left( {\varepsilon _{t - 1}^2} \right){F_0} + \beta {\kern 1pt} {\kern 1pt} E\left( {{h_t}} \right)$. Using that ${\kern 1pt} {E_{t - 2}}\left( {\varepsilon _{t - 1}^2} \right) = {h_{t - 1}}$ and the tower law of expectations, we can write: $E\left( {{h_t}} \right) = \omega  + \left( {\alpha {\kern 1pt} {\kern 1pt}  + \lambda {\kern 1pt} {F_0} + \beta } \right){\kern 1pt} {\kern 1pt} E\left( {{h_t}} \right)$, which yields $E\left( {{h_t}} \right) = \frac{\omega }{{1 - \varphi }} = \bar h$.} Thus we obtain that for $\varphi  \in \left( {0,1} \right)$, $\mathop {\lim }\limits_{s \to \infty } \mu _{r,s}^{\left( 2 \right)} = \mathop {\lim }\limits_{s \to \infty } {E_t}\left( {{h_{t + s}}} \right) = \bar h = E\left( {{h_{t + s}}} \right)$.

Again, using Theorem 2.2 from Ling and McAleer(2002b), a necessary and sufficient condition for the existence of the fourth unconditional moment is $\gamma \in \left( {0,1} \right)$. Again, this is the condition required in Theorem 3 for the fourth conditional moment to converge to a finite limit. It is easy to show that, when it exists, the fourth unconditional moment is given by:
\vspace{-0.5cm}
$$E\left( {\varepsilon _t^4} \right) = {\kappa _z}E\left( {h_t^2} \right)= {\kappa _z}\frac{{{\omega ^2} + 2\omega \varphi \bar h}}{{1 - \gamma }}={\kappa _{z}}{c_1}
\vspace{-0.5cm}
$$ and, as a result, the unconditional kurtosis is given by
the same expression as in (\ref{lim_kurt}) above, for $\gamma \in \left( {0,1} \right)$. A special case of this result is the unconditional kurtosis for a GARCH(1,1) process with symmetric innovations, derived by Ishida and Engle (2002).

The unconditional skewness is given by:
\vspace{-0.5cm}
 $$\frac{{E\left( {\varepsilon _t^3} \right)}}{{{{\left[ {E\left( {\varepsilon _t^2} \right)} \right]}^{3/2}}}} = \frac{{E\left( {z_t^3h_t^{3/2}} \right)}}{{{{\left[ {E\left( {{h_t}} \right)} \right]}^{3/2}}}} = {\tau _z}\frac{{E\left( {h_t^{3/2}} \right)\,}}{{{{\left[ {E\left( {{h_t}} \right)} \right]}^{3/2}}}}.
\vspace{-0.5cm}
$$
Since $E\left( {h_t^{3/2}} \right)$ cannot be computed analytically in this framework, we use a second order Taylor series expansion to approximate it. Thus, $E\left( {h_t^{3/2}} \right) \simeq \frac{5}{8}{\left( {E\left( {{h_t}} \right)} \right)^{3/2}} + \frac{3}{8}E\left( {h_t^2} \right){\left( {E\left( {{h_t}} \right)} \right)^{ - 1/2}}$ and a resulting approximation of the unconditional skewness is
the same as the expression given in (\ref{lim_skew}) for $\gamma \in \left( {0,1} \right)$.

The classical central limit theorem does not apply here, because the variables are not independent. Nevertheless, the conditional moments of the aggregated returns converge to the corresponding moments of a normal distribution, provided that certain parameter conditions are met. Outside of the regularity conditions, the conditional skewness of aggregated returns diverges to $\pm \infty$ and the conditional kurtosis of aggregated returns diverges to $+ \infty$. This is similar to a result of Diebold (1988) who shows that the unconditional distribution of the aggregated returns for a conditionally normal AR-ARCH $(m,p)$ process also converges to a normal distribution, under suitable parameter conditions. 


Interestingly, identical convergence conditions apply for the moments of both forward and aggregated returns. Whenever the moments of forward returns converge to the unconditional moments, the aggregated moments converge to the corresponding moments of a normal distribution. 
Moreover, for a special case of the generic framework, namely for the normal GARCH(1,1) model with $\gamma=1$, the limit of the kurtosis of forward returns is infinite whilst the kurtosis of aggregated returns converges to a constant value different from $3$. In fact, this additional convergence case for $\gamma=1$ is not specific to the normal GARCH(1,1): it applies to any GARCH(1,1) model with symmetric innovations. This result, for the symmetric special case, is in agreement with Breuer and Jandacka (2010) even though our proof is different from theirs. \\

\noindent\textbf{Theorem 4: Limiting Behaviour of Moments for Forward and Aggregated Variances}

\noindent Suppose  $\varphi \in (0,1)$ and $\gamma \neq \varphi$ (as above); additionally $c_4 \neq \gamma$ and $c_4 \neq \varphi$. Then we have:\\
a)	The limit of the conditional variance of the forward conditional variance of model (\ref{gjr_spec}) is:
\vspace{-0.25cm}
\begin{equation}
\mathop {\lim }\limits_{s \to \infty } \mu _{h,s}^{\left( 2 \right)} = \left\{ \begin{array}{l}
 \left( {\left( {{\omega ^2} + 2\omega \varphi \bar h} \right){{\left( {1 - \gamma } \right)}^{ - 1}} - {{\bar h}^2}} \right)\quad {\rm{if}}\;\gamma  \in \left( {0,1} \right), \\
 \infty \quad \quad \quad \;\quad \quad \quad \;\qquad \qquad \qquad \;\,{\rm{if}}\;\gamma  \in \left[ {1,\infty } \right). \\
 \end{array} \right.
\label{eq:3.14}
\vspace{-0.25cm}
\end{equation}
b)	The limit of  the conditional variance of the aggregated conditional variance (per unit of time) of model (\ref{gjr_spec}) is:
\vspace{-0.25cm}
\begin{equation}
\mathop {\lim }\limits_{n \to \infty } \frac{{M_{h,n}^{\left( 2 \right)}}}{n} = \left\{ \begin{array}{l}
 \left( {\left( {{\omega ^2} + 2\omega \varphi \bar h} \right){{\left( {1 - \gamma } \right)}^{ - 1}} - {{\bar h}^2}} \right)\left( {1 + 2\varphi {{\left( {1 - \varphi } \right)}^{ - 1}}} \right)\quad {\rm{if}}\,\gamma  \in \left( {0,1} \right), \\
 \infty \quad \quad \quad \quad \quad \quad \quad \quad \quad \quad \quad \quad \qquad \qquad \qquad \qquad \;\; \;{\rm{if}}\,\gamma  \in \left[ {1,\infty } \right). \\
 \end{array} \right.
\label{eq:3.15}
\vspace{-0.25cm}
\end{equation}
c)	The limit of the conditional skewness of the forward conditional variance of model (\ref{gjr_spec}) is:
\begin{equation}
\mathop {\lim }\limits_{s \to \infty } {\tau _{h,s}} = \left\{ \begin{array}{l}
 {M_1}\quad {\rm{if}}\;\gamma  \in \left( {0,1} \right)\;{\rm{and}}\;{c_4} \in \left( {0,1} \right), \\
 0\quad \;{\rm{if}}\;\gamma  \in \left[ {1,\infty } \right)\;{\rm{and}}\;{c_4} \in \left( {0,{\gamma ^{3/2}}} \right), \\
 {M_2}\quad {\rm{if}}\;\gamma  \in \left( {1,\infty } \right)\;{\rm{and}}\;{c_4} = {\gamma ^{3/2}}, \\
 \infty \quad \;{\rm{if}}\;\left\{ {\gamma  \in \left( {0,1} \right),\;{c_4} \in \left[ {1,\infty } \right)} \right\}\;{\rm{or}}\;\left\{ {\gamma  = {\rm{1}},\;{c_4} \in \left( {1,\infty } \right)} \right\}, \\
\qquad {\rm{or}}\;\left\{ {\gamma  \in \left( {1,\infty } \right),\;{c_4} \in \left( {{\gamma ^{3/2}},\infty } \right)} \right\}, \\
 \end{array} \right.
\label{eq:3.16}
\end{equation}
where
\[M_1 = \frac{{\omega \left( {{\omega ^2} + 3\omega \varphi \bar h + 3\gamma {c_1}} \right){{\left( {1 - {c_4}} \right)}^{ - 1}} - 3\bar h{c_1} + 2{{\bar h}^3}}}{{{{\left( {{c_1} - {{\bar h}^2}} \right)}^{3/2}}}}\]
\[M_2 = \frac{{h_{t + 1}^3 - \omega \left( {{\omega ^2} + 3\omega \varphi \bar h + 3\gamma {c_1}} \right){{\left( {1 - {c_4}} \right)}^{ - 1}} - \left( {3{\omega ^2}\varphi \left( {{h_{t + 1}} - \bar h} \right) + 3\omega \gamma \left( { - {c_1} + h_{t + 1}^2} \right)} \right){{\left( {\varphi  - {c_4}} \right)}^{ - 1}}}}{{{{\left( { - {c_3} + h_{t + 1}^2} \right)}^{3/2}}}}.\]
%
d)	For $ \gamma \in \left (0, 1 \right )$
the conditional skewness of the aggregated conditional variance of model (\ref{gjr_spec}) has limit:\footnote{Since proofs become increasingly lengthy we only state the limit of the conditional skewness of the aggregated conditional variance in the case that $\gamma \in \left (0, 1 \right )$, and for $\gamma  \geq 1$ we present the principles of the derivation in the Technical Appendix T.A.4.}
\begin{equation}
\mathop {\lim }\limits_{n \to \infty } {{\rm T}_{h,n}} = \left\{ \begin{array}{l}
 {0}  \qquad \qquad {\rm{if}}\;{c_4} \in \left( {0,1} \right), \\
 \infty \quad \quad \quad \,\;\,{\rm{if}}\;{c_4} \in \left( {1,\infty } \right), \\
 {\mathop{\rm sgn}} (N)\infty \quad {\rm{if}}\,{c_4} = 1, \\
 \end{array} \right.
\label{eq:3.19}
\end{equation}
where
\[\begin{array}{l}
N =  \frac{{\omega \left( {{\omega ^2} + 3\omega \varphi \bar h + 3\gamma {c_1}} \right)}}{2} + 3\frac{{\bar h}}{2}\left( {{c_1} + \varphi \left( {{\omega ^2} + 3\omega \varphi \bar h + 3\gamma {c_1}} \right)+ {\omega ^2}{{\left( {1 - \gamma } \right)}^{ - 1}}} + 2\omega \varphi {\bar h}\right) \\
 \quad + 3 {\gamma {{\left( {1 - \gamma } \right)}^{ - 1}}\frac{{\left( {{\omega ^2} + 3\omega \varphi \bar h + 3\gamma {c_1}} \right)}}{2}}\left(\omega +2\varphi\bar h\right)  \\
  \quad + 3{\left( {1 - \varphi } \right)^{ - 1}}\bar h\left[ {{c_1}\left( {1 + \varphi } \right) - 2\varphi {{\bar h}^2} + \bar h\left( {\left( {{h_{t + 1}} - \bar h} \right) - \omega } \right)} - \varphi \left( {2{c_1} - {{\bar h}^2}} \right)\right]. \\
 \end{array}\]

\vspace{0.5cm}
\noindent The returns process has no autocorrelation so the variance of aggregated returns is just the sum of the forward one-period variances. However, the variance process is autocorrelated. As a result the variances of the two processes have different limiting behaviour. The limit of the variance of aggregated returns per unit of time is equal to the limit of forward variance (i.e. $\mathop {\lim }\limits_{n \to \infty } \frac{{{\rm M}_{r,n}^{\left( 2 \right)}}}{n} = \mathop {\lim }\limits_{s \to \infty } \mu _{r,s}^{\left( 2 \right)}$), but the same does not hold for the variance of variance. Indeed, $\mathop {\lim }\limits_{n \to \infty } \frac{{{\rm M}_{h,n}^{\left( 2 \right)}}}{n} > \mathop {\lim }\limits_{s \to \infty } \mu _{h,s}^{\left( 2 \right)}$.

Of course, the returns and variance processes are not independent, and certain aspects of this dependence are reflected in a similar behaviour in their limiting distributions. In particular, recall that the forward and aggregated returns had identical regularity conditions and that whenever a moment of forward returns converges to a finite limit, the corresponding moment of aggregated returns converges to the normal value. Theorem 4 yields a similar result for forward and aggregated variances: the skewness of the aggregated variance converges to zero when the skewness of forward variance converges to a finite value. But for the variance, the skewness regularity condition is slightly stronger: it is required that not only $\gamma \in \left({0,1}\right)$ but also ${c_4} \in  \left({0,1}\right)$ and $c_4 \ne \gamma$.

\section{An Application: Approximate Predictive Distributions}\label{sec:App}
Density forecasting is a prime application of our moment formulae. Whilst one must know all the moments of a random variable to determine its distribution,\footnote{To be precise, a distribution is uniquely determined by its moments only if the Carleman condition holds, i.e. only if $\sum\limits_{n = 1}^\infty  {\alpha _{2n}^{ - 0.5n}} \to \infty $, where $\left\{ {{\alpha _k}} \right\}$ is the moment sequence (see Serfling, 1980, p.46). In the following we assume that this condition is met.} it can be approximated based on the first few moments alone. Based on their relative merits and drawbacks, and given the frequency of their use in similar applications as well as the feasibility of obtaining an approximate distribution function in closed form, we have selected two approximation methods: the Edgeworth expansion and the Johnson SU distribution.

\subsection{Distribution Approximations Methods}\label{sec:distapprox}
The Edgeworth expansion can approximate a density of interest around a base density, usually the standard normal density.\footnote{For the general theory and expansion see Edgeworth (1905), Wallace (1958) and Bhattacharya and Ghosh (1978).} It belongs to the class of Gram-Charlier expansions, being a rearrangement of a Gram-Charlier A series.\footnote{See Chebyshev (1860), Chebyshev (1890), Gram (1883), Charlier (1905) and Charlier (1906).}  However, Gram-Charlier A series and Edgeworth series are only equivalent asymptotically, when an infinite number of terms enter the expansions. In empirical applications using finite order approximations they can differ significantly, and the Edgeworth version is preferred since it is an asymptotic expansion.\footnote{An asymptotic expansion is defined in Wallace (1958) as one where the  error of approximation approaches zero as one of its parameters, e.g. the sample size for approximations of the sampling distribution of a random sample of size $T$, approaches infinity. Furthermore, Walace (1958) calls the Edgeworth expansion a 'formal' asymptotic expansion.} Nevertheless, the Edgeworth expansion may have monotonicity and convergence problems, i.e. the distribution function is not guaranteed to be monotonic and the error of approximation does not necessarily improve when we increase the order of the expansion.\footnote{See Jasche (2002) and Wallace (1958).}

The first four terms of the Edgeworth expansion are:
\begin{center}
${f_x}\left( x \right) \simeq f_x^E\left( x \right) = \varphi \left( x \right) - \frac{{{\tau _x}}}{6}{\varphi ^{\left( 3 \right)}}\left( x \right) + \frac{{\left( {{\kappa _x} - 3} \right)}}{{24}}{\varphi ^{\left( 4 \right)}}\left( x \right) + \frac{{\tau _x^2}}{{72}}{\varphi ^{\left( 6 \right)}}\left( x \right),$
\end{center}
where $f_x^E\left( x \right)$ is the second-order Edgeworth approximation of $f_x$, so moments (cumulants) of order higher than four (kurtosis) are ignored, $\varphi$ is the standard normal density and ${\varphi ^{\left( k \right)}}$ is its $k^{th}$ derivative, and ${\tau _x}$ and ${\kappa _x}$ denote the skewness and kurtosis of ${f_x}$. For our purposes ${f_x}$ will be the density of the normalised forward returns.

A random variable $x$ is said to follow a Johnson SU distribution (Johnson, 1949) if:
\begin{center}
$z = \gamma  + \delta {\sinh ^{ - 1}}\left( {\frac{{x - \xi }}{\lambda }} \right),$
\end{center}
where $z$ is a standard normal variable. The four parameters $\gamma$, $\delta$, $\xi$ and $\lambda$ may be estimated using the moment-matching algorithm described in Tuenter (2001). Although flexible, the main disadvantage of this approach is that a Johnson SU distribution is not guaranteed to exist for any set of mean, variance, skewness and (positive) excess kurtosis.

\subsection{Evaluation Methods}\label{sec:eval}
To assess how well these approximate distributions serve their purpose we should investigate whether they provide an adequate representation of the conditional distributions of forward returns. But these distributions are not observable, even ex-post, so we shall use simulated distributions as proxies. The null hypothesis is $H_0$: ${F_m} = {F_s}$, where $F_m$ is the cumulative distribution function for the approximate distribution constructed using the first four moments and a specific approximation method, and $F_s$ is the  distribution function for the simulated forward returns based on the GARCH process.  The simulated distribution $F_s$ is given by the step-function of the sample. Thus, ${F_s}\left( {{x_i}} \right) = {T^{ - 1}}i$, where  \\$i$ = number of returns less than or equal to ${x_i}$; ${x_i}$ with $i  \in \left \{1, ..., T\right\}$ is an increasingly ordered sample, and $T$ is the number of simulations.

Standard hypothesis tests where the null is the equality of two distributions are the Kolmogorov-Smirnov (KS) and the Anderson Darling (AD) tests. The KS test, proposed by Kolmogorov (1933) and Smirnov (1939), is, in effect, a simple hypothesis test which is based on the maximum difference between an empirical and a hypothetical cumulative distribution. The test statistic is given by $\mbox{KS} = \sqrt T D$, where 
$D$ is the maximum distance between the two distributions, i.e.
$D = \mathop {\max }\limits_{x} \left| {{F_m}\left( {{x}} \right) - {F_s}\left( {{x}} \right)} \right|\,.$
For practical implementations, we use the following version of the statistic for the increasingly ordered sample:\footnote{See Anderson and Darling (1952) and Pearson and Hartley (1972).}
\begin{center}
$\mbox{KS} = \sqrt T \mathop {\max }\limits_{1 \le i \le T} \left\{ {\max \left[ {{F_m}\left( {{x_i}} \right) - \frac{{i - 1}}{T},\frac{i}{T} - {F_m}\left( {{x_i}} \right)\,} \right]} \right\}.$
\end{center}
When comparing alternative models, the one with the lowest KS value is deemed the most accurate for predicting the distribution in question.

The framework proposed by Anderson and Darling (1952) is more flexible, allowing for different weighting of the observations. They propose two distance measures, which are actually generalisations of the KS and Cramer von Mises (CVM) statistics.\footnote{The KS and Cramer-von Mises tests are obtained when ${\psi {{\left( t \right)} = 1}}$ in (\ref{AD_1}) and (\ref{AD_2}) respectively.}  The respective test statistics are given by:
\begin{equation}\label{AD_1}
\mbox{AD}_1 = {T^{1/2}}\mathop {\max }\limits_{x} \left| {{F_m}\left( {{x}} \right) - {F_s}\left( {{x}} \right)} \right|\left( {\psi {{\left( {{F_m}\left( {{x}} \right)} \right)}^{1/2}}} \right),
\end{equation}
\begin{equation}\label{AD_2}
\mbox{AD}_2 = T\int\limits_{x} {\left[ {{{\left( {{F_m}\left( {{x}} \right) - {F_s}\left( {{x}} \right)} \right)}^2}\psi \left( {{F_m}\left( {{x}} \right)} \right)} \right]},
\end{equation}
where $\psi$ is a weighting function. Following convention we refer to the Anderson-Darling (AD) test as (\ref{AD_2}) with a weighting function $\psi \left( x \right) = {\left( {x\left( {1 - x} \right)} \right)^{ - 1}}$.\footnote{For practical implementations and for an increasingly ordered sample, we use the following versions of the CVM and AD test statistics (see Anderson and Darling, 1952 and Pearson and Hartley, 1972):
\begin{center}
$\mbox{CVM} = {\sum\limits_{i = 1}^T {\left[ {{F_m}\left( {{x_i}} \right) - \frac{{2i - 1}}{{2T}}} \right]} ^2} + \frac{1}{{12T}}, \quad \mbox{AD} =  - \sum\limits_{i = 1}^T {\frac{{2i - 1}}{T}\left[ {\ln \left( {{F_m}\left( {{x_i}} \right)} \right) + \ln \left( {1 - {F_m}\left( {{x_{T + 1 - i}}} \right)} \right)} \right]}  - T.$
\end{center}}

Conducting these tests in our setting requires the simulation of critical values. The statistics only have standard distributions if the distribution under the null hypothesis is entirely pre-specified, but in our case the ${F_m}$ distribution relies on estimated parameter values so the theoretical critical values are no longer applicable.

\subsection{Data and Methodology}
The performance of our proposed distribution forecasts is
tested using  daily observations on an equity index (S\&P 500), a foreign exchange rate (Euro/dollar) and an interest rate (3-month Treasury bill). These series represent three major market risk types and within each class they represent the most important risk factors in terms of volumes of exposures. The three data sets used in this application were obtained from Datastream and each comprise 20 years of daily data from 1st January 1990 to 31st December 2009.\footnote{The Euro was only introduced in 1999, so the ECU/dollar exchange rate is used before this date.} Figure 1 plots the daily log returns for the equity and exchange rate data and the daily changes in the interest rate.\footnote{First differences in fixed maturity interest rates are the equivalent of log returns on corresponding bonds.} 

\begin{figure}[ht]
\centering{\includegraphics[width=\textwidth]{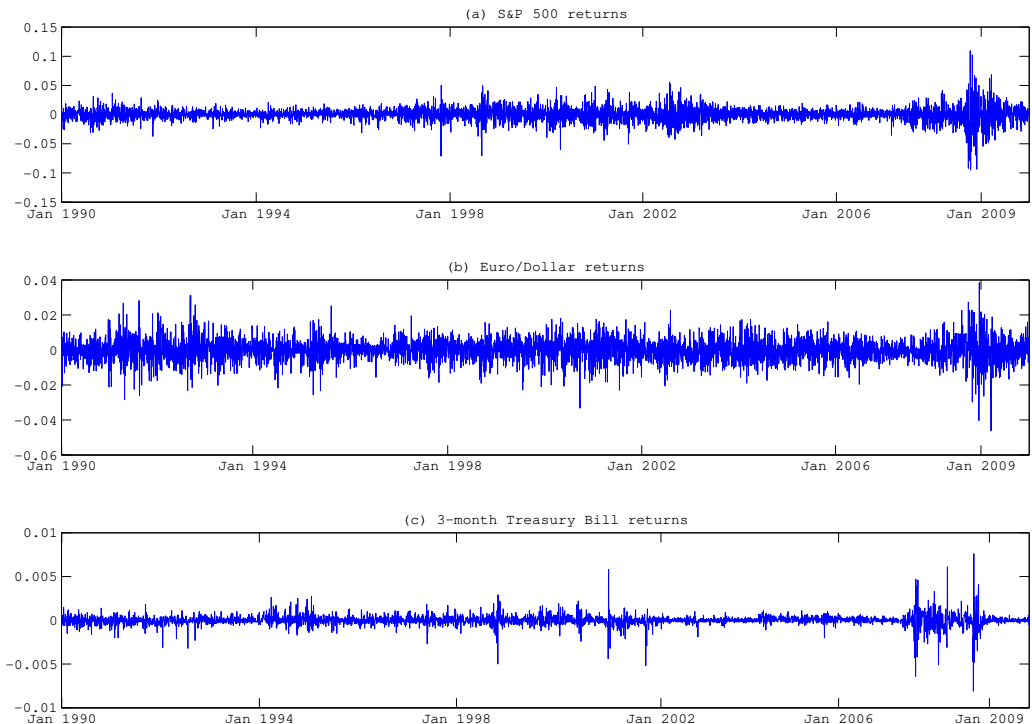}}
	\caption{\footnotesize \textbf{Returns} The equity and exchange rate daily (log) returns are computed as the first differences of the logarithm of the S\&P 500 index values and Euro/dollar exchange rates, respectively. The interest rate returns are computed as first differences in interest rate values. 
}
\end{figure}

Table 1 presents the sample statistics of the empirical unconditional daily returns distribution over the entire sample. In accordance with stylized facts the mean of every series is not statistically different from zero and the unconditional volatility is highest for the equity and lowest for the interest rates. Skewness is negative and low (in absolute value) but significant for all three series, so extreme negative returns are more likely than extreme positive returns of the same magnitude, while excess kurtosis is always positive and highly significant, so the unconditional distributions of the series have a greater probability mass in the tails than the normal distribution with the same variance.

\begin{table}[htbp]
  \centering
\scalebox{1}{
    \begin{tabular}{cccc}
          & S\&P 500 & EUR/USD & 3M Bill \\
    \hline
    Mean  & 0.00022 & -2E-05 & -2E-05 \\
    \hline
    Maximum & 0.1   & 0.0384 & 0.0076 \\
    \hline
    Minimum & -0.0947 & -0.0462 & -0.0081 \\
    \hline
    Volatility & 0.1862 & 0.1   & 0.0101 \\
    \hline
    Skewness & -0.1981*** & -0.0992** & -0.5007*** \\
    \hline
    Excess kurtosis & 9.1643*** & 2.7136*** & 28.2342*** \\
    \end{tabular}}
  \caption{\footnotesize \textbf{Summary statistics.}The summary statistics are of the equity and exchange rate daily log returns, and of the daily changes in interest rates from 1990 to 2009. Asterisks denote significance at 5\% (*), 1\% (**) and 0.1\%(***). The standard error of the sample mean is equal to the sample standard deviation, divided by sample size, while the standard errors are approximately  ${\left( {6/T} \right)^{1/2}}$ and ${\left( {24/T} \right)^{1/2}}$ for the sample skewness and excess kurtosis, respectively, where T is the sample size. We used 252 risk days per year to annualize the standard deviation into volatility.}
\end{table}

Four different GARCH models, namely the baseline GARCH(1,1) and the asymmetric GJR, each with normal and Student $t$ error distributions, are estimated for each of the three time series.
Their parameters are repeatedly estimated, approximately 2500 times, on a sample consisting of ten years of daily data which is rolled daily for an additional ten years. The resulting time series of model parameters are subsequently used to compute corresponding time series for the higher moments of forward and aggregated returns, based on the analytic formulae derived in Section \ref{sec:Mom}.
This way we may construct time series of conditional moments for the forward and aggregated returns and variances, for any time horizon $h$, from 3rd January 2000 to 31st December 2009.

For the symmetric models -- the normal and Student $t$ GARCH(1,1) -- the skewness of the returns, both forward and aggregated, is zero by construction. However, the asymmetric specifications -- the normal and Student $t$ GJR -- lead to non-zero skewness forecasts for the aggregated returns.  The skewness of the (forward and aggregated) variances is non-zero even for the symmetric models. All four models yield non-zero, positive excess kurtosis for all time series. 

Now we apply the distribution approximation methods reviewed in Section \ref{sec:distapprox} to derive distributions for the $h$-day forward and aggregated returns. We combine four GARCH specifications (the normal and Student $t$ GARCH(1,1) and GJR models) with two approximation methods (the Johnson SU distribution and the Edgeworth expansion) and thus have eight alternative approximate distributions to evaluate and compare with the simulated distributions, each based on 10,000 simulations. We test the accuracy of the approximations using the KS distance, the CVM and AD test statistics described in Section \ref{sec:eval}. To capture any differences between market regimes, the tests are performed for 150 days from a low volatility period (i.e. January to August 2006), 150 days from a high volatility period (i.e. August 2008 to March 2009) and the last 150 observations from 2009. In our results these periods are labelled 'low vol', 'high vol' and 'current' respectively. Also the label 'total' refers to the results obtained for all three out-of-sample periods, considered together.
%
Finally, the time horizon we consider here is $h$ = 5 days.

\subsection{Empirical Results}
Tables 2 and 3 summarize the results of the distribution tests for each of the eight approximate distributions considered.\footnote{For the interest rate sample, fitting the Johnson SU distribution using the moments of the aggregated returns estimated for the Student $t$ GJR was problematic and hence, for this sample, we do not report results for the Johnson SU Student $t$ GJR in Table 3.} The AD test gave results very similar to the CVM test, hence we only report the results for the KS and CVM tests. We report the mean values and the associated standard deviations of the test statistics and also the percentage of times when the computed test statistic was higher than the asymptotic 5\% critical value. Since we perform the tests at the 5\% significance level we expect a 5\% rejection rate.

The 5\% critical values are 0.0136 for the KS distance and 0.461 for the CVM  statistic.\footnote{These are asymptotic results for a test where the distribution being tested for is continuous, fully known and generic (no particular family of distributions assumed). Stephens (1970) derives modified statistics for the finite sample case; however, with a sample size of 10,000 these modifications are not actually needed and the asymptotic results would apply, if the hypothetical distribution were fully specified. However, in our case this distribution is based on estimated results and we would need to simulate the correct critical values if we were to properly carry out the tests. Still, we report the percentage of times the test statistics are greater than the asymptotic critical values, so that we can infer, approximately, if the test results are at least in the vicinity of these asymptotic critical values. We also note that the results have to be interpreted with care since it is likely that the appropriate (simulated) critical values for this testing exercise are lower than the asymptotic critical values reported above (see Massey, 1951).} Although the asymptotic critical values do not apply exactly in our case, the model that produces the lowest values of the test statistic is still the best among the alternatives.\footnote{What we mean by "best among alternatives" here means "closest to the (respective) simulated distribution". However, one has to interpret the results with care since the simulated distribution is obviously not the same for all alternative approximate distributions.} We now discuss the results in greater detail for the equity, exchange rate and interest rate distributions in turn. \\

\vspace{-0.5cm}
\noindent\underline{(a) S\&P 500 Index} \\
The Johnson SU approximation appears to be a more suitable approximation than the Edgeworth expansion for the forward returns, especially when applied in combination with the moments produced by the Student $t$ GARCH(1,1) and GJR models. For the normal GARCH(1,1) and GJR models the average values of both the KS and CVM test statistics are still greater, but only marginally, for the Edgeworth approximations, compared with their Johnson SU counterparts. 
Indeed, if the Edgeworth expansion is used, then the models with normal innovations fit the simulated distributions significantly better than their Student $t$ counterparts. This is not the case for the Johnson SU distribution, where all four GARCH specifications provide very close fits to the simulated distributions (although again the normal models, and especially the normal GARCH(1,1), do give slightly better fits than the Student $t$ models). Overall, the Johnson SU normal GARCH(1,1) model produces the lowest average values of both the KS distance (0.0088) and CVM test statistic (0.1719). For the aggregated returns, the results improve even further, with the Johnson SU methodology still proving superior overall to the Edgeworth expansion (when combined with the Student $t$ models), but to a lesser extent than in the case of the forward returns. \\


\vspace{-0.5cm}
\noindent\underline{(b) Euro/Dollar Exchange Rate}\\
From the results in Table 2 there is not much to choose between the two alternative approximation methods when the innovations are normal. Indeed, the results for both distribution tests are virtually identical (and good) for the normal GARCH(1,1) and GJR, using either the Johnson SU or Edgeworth expansion. For the Johnson SU approach, the fit is closest when the innovations are normal but the fit is almost as good based on the Student $t$ models. However, when the Edgeworth expansion is employed, the results obtained with the Student $t$ models are significantly worse than when the innovations are normal. For the distributions of aggregated Euro/dollar returns, all models produce similar and good results. \\

\vspace{-0.5cm}
\noindent\underline{(c) 3-month Treasury Bill Rate} \\
As in (a), the Johnson SU should be preferred to the Edgeworth expansion. The Johnson SU approximation yields the lowest average KS distance as well as the lowest value for the CVM test statistic for the normal GARCH(1,1) model, the  normal GJR model being second best. The fits deteriorate when innovations have a  Student $t$ distribution. \\


\vspace{-0.5cm}
\noindent To summarize, the predictive distributions of forward and aggregated returns on major risk factors may be well approximated using the analytic expressions for the first four conditional moments that we have derived in this paper. The best distribution approximation method overall is the Johnson SU and all four GARCH models that we have tested have predictive forward and aggregated returns distributions that can be well approximated, the easiest being that generated by the normal GARCH(1,1).

\begin{landscape}
\begin{table}[htbp]
  \centering
 \small
\scalebox{0.75}{

\begin{tabular}{cccccccccccccccccc}
\hline
 &  & \multicolumn{4}{c}{Normal GARCH(1,1)} & \multicolumn{4}{c}{Student $t$ GARCH(1,1)} & \multicolumn{4}{c}{Normal GJR} & \multicolumn{4}{c}{Student $t$ GJR}\\
\hline
 &  & total & low vol & high vol & current & total & low vol & high vol & current & total & low vol & high vol & current & total & low vol & high vol & current\\
\hline
S\&P 500 & \multicolumn{17}{c}{Johnson SU}\\
\hline
 & KS-average & 0.0088 & 0.0088 & 0.0086 & 0.0089 & 0.0092 & 0.0091 & 0.0091 & 0.0093 & 0.0088 & 0.0088 & 0.0087 & 0.0089 & 0.009 & 0.009 & 0.0089 & 0.0091\\
 & KS-stdev & 0.0027 & 0.0027 & 0.0026 & 0.0027 & 0.0027 & 0.0028 & 0.0026 & 0.0029 & 0.0027 & 0.0027 & 0.0025 & 0.0027 & 0.0027 & 0.0028 & 0.0025 & 0.0028\\
 & KS-rejections@5\% & 6.00\% & 6.00\% & 4.67\% & 7.33\% & 7.33\% & 8.67\% & 6.00\% & 7.33\% & 6.00\% & 6.67\% & 5.33\% & 6.00\% & 6.89\% & 8.00\% & 4.67\% & 8.00\%\\
 & CVM-average & 0.1719 & 0.1765 & 0.1577 & 0.1814 & 0.1897 & 0.1935 & 0.1809 & 0.1948 & 0.1737 & 0.1803 & 0.1593 & 0.1814 & 0.1824 & 0.1896 & 0.1688 & 0.1889\\
 & CVM-stdev & 0.1496 & 0.1628 & 0.1298 & 0.1543 & 0.1544 & 0.1664 & 0.1346 & 0.1611 & 0.1504 & 0.1635 & 0.1284 & 0.157 & 0.1528 & 0.1652 & 0.1296 & 0.1611\\
 & CVM-rejections@5\% & 6.44\% & 8.00\% & 4.67\% & 6.67\% & 6.00\% & 8.00\% & 4.00\% & 6.00\% & 6.89\% & 8.67\% & 4.67\% & 7.33\% & 6.00\% & 8.00\% & 4.00\% & 6.00\%\\
\hline
 & \multicolumn{17}{c}{Edgeworth}\\
\hline
 & KS-average & 0.0088 & 0.0088 & 0.0086 & 0.009 & 0.0171 & 0.01629 & 0.01676 & 0.0182 & 0.0089 & 0.0089 & 0.0088 & 0.009 & 0.0142 & 0.0142 & 0.0139 & 0.0144\\
 & KS-stdev & 0.0027 & 0.0027 & 0.0026 & 0.0027 & 0.0031 & 0.0029 & 0.0031 & 0.0031 & 0.0027 & 0.0027 & 0.0025 & 0.0028 & 0.003 & 0.0028 & 0.003 & 0.003\\
 & KS-rejections@5\% & 5.78\% & 6.00\% & 4.00\% & 7.33\% & 86.89\% & 80.67\% & 85.33\% & 94.67\% & 6.67\% & 8.00\% & 5.33\% & 6.67\% & 54.00\% & 54.67\% & 50.67\% & 56.67\%\\
 & CVM-average & 0.1721 & 0.1768 & 0.1579 & 0.1815 & 0.8535 & 0.7698 & 0.8121 & 0.9785 & 0.1764 & 0.1846 & 0.1614 & 0.1832 & 0.5387 & 0.5516 & 0.5059 & 0.5587\\
 & CVM-stdev & 0.1497 & 0.1629 & 0.1296 & 0.1546 & 0.3256 & 0.2887 & 0.2912 & 0.3561 & 0.1513 & 0.1647 & 0.1284 & 0.1582 & 0.2471 & 0.244 & 0.2278 & 0.2663\\
 & CVM-rejections@5\% & 6.22\% & 8.00\% & 4.67\% & 6.00\% & 91.33\% & 86.67\% & 91.33\% & 96.00\% & 6.67\% & 8.67\% & 4.67\% & 6.67\% & 56.89\% & 59.33\% & 50.00\% & 61.33\%\\
\hline
Euro/dollar & \multicolumn{17}{c}{Johnson SU}\\
\hline
 & KS-average & 0.0088 & 0.0088 & 0.0086 & 0.009 & 0.009 & 0.009 & 0.0088 & 0.0091 & 0.0088 & 0.0088 & 0.0086 & 0.009 & 0.009 & 0.009 & 0.0088 & 0.0091\\
 & KS-stdev & 0.0027 & 0.0027 & 0.0026 & 0.0028 & 0.0027 & 0.0028 & 0.0025 & 0.0028 & 0.0027 & 0.0027 & 0.0026 & 0.0027 & 0.0027 & 0.0027 & 0.0025 & 0.0028\\
 & KS-rejections@5\% & 5.11\% & 4.67\% & 4.67\% & 6.00\% & 6.44\% & 8.00\% & 4.00\% & 7.33\% & 5.11\% & 4.67\% & 4.67\% & 6.00\% & 6.22\% & 8.00\% & 4.00\% & 6.67\%\\
 & CVM-average & 0.1713 & 0.1754 & 0.1571 & 0.1814 & 0.1802 & 0.1883 & 0.1636 & 0.1888 & 0.1713 & 0.1754 & 0.1571 & 0.1814 & 0.1803 & 0.1886 & 0.1646 & 0.1878\\
 & CVM-stdev & 0.1495 & 0.1618 & 0.1302 & 0.1546 & 0.1519 & 0.1647 & 0.129 & 0.1591 & 0.1496 & 0.1618 & 0.1303 & 0.1547 & 0.1519 & 0.1648 & 0.1297 & 0.1587\\
 & CVM-rejections@5\% & 6.44\% & 7.33\% & 4.67\% & 7.33\% & 6.89\% & 9.33\% & 4.67\% & 6.67\% & 6.44\% & 7.33\% & 4.67\% & 7.33\% & 5.55\% & 7.33\% & 3.33\% & 6.00\%\\
\hline
 & \multicolumn{17}{c}{Edgeworth}\\
\hline
 & KS-average & 0.0088 & 0.0088 & 0.0086 & 0.009 & 0.0136 & 0.0151 & 0.0122 & 0.0134 & 0.0088 & 0.0088 & 0.0086 & 0.009 & 0.0135 & 0.015 & 0.0122 & 0.0133\\
 & KS-stdev & 0.0027 & 0.0027 & 0.0026 & 0.0028 & 0.0032 & 0.0029 & 0.0029 & 0.003 & 0.0027 & 0.0027 & 0.0026 & 0.0027 & 0.0032 & 0.003 & 0.0029 & 0.0029\\
 & KS-rejections@5\% & 5.11\% & 4.67\% & 4.67\% & 6.00\% & 45.78\% & 67.33\% & 29.33\% & 40.67\% & 5.11\% & 4.67\% & 4.67\% & 6.00\% & 45.33\% & 66.67\% & 29.33\% & 40.00\%\\
 & CVM-average & 0.1713 & 0.1754 & 0.1571 & 0.1814 & 0.4941 & 0.6435 & 0.3761 & 0.4628 & 0.1713 & 0.1754 & 0.1571 & 0.1814 & 0.4886 & 0.644 & 0.3735 & 0.4484\\
 & CVM-stdev & 0.1495 & 0.1618 & 0.1302 & 0.1547 & 0.2628 & 0.2735 & 0.1876 & 0.2462 & 0.1496 & 0.1618 & 0.1303 & 0.1547 & 0.2621 & 0.2754 & 0.1878 & 0.2379\\
 & CVM-rejections@5\% & 6.44\% & 7.33\% & 4.67\% & 7.33\% & 52.89\% & 78.67\% & 32.00\% & 48.00\% & 6.44\% & 7.33\% & 4.67\% & 7.33\% & 44.89\% & 74.67\% & 24.00\% & 36.00\%\\
\hline
3M Bill & \multicolumn{17}{c}{Johnson SU}\\
\hline
 & KS-average & 0.0106 & 0.0103 & 0.0104 & 0.0111 & 0.164 & 0.1315 & 0.2498 & 0.0989 & 0.0124 & 0.0113 & 0.0127 & 0.0132 & 0.0234 & 0.0212 & 0.0263 & 0.0227\\
 & KS-stdev & 0.0031 & 0.0031 & 0.0029 & 0.0032 & 0.0898 & 0.0238 & 0.0996 & 0.0214 & 0.0033 & 0.0032 & 0.003 & 0.0034 & 0.0047 & 0.0041 & 0.0046 & 0.0034\\
 & KS-rejections@5\% & 17.02\% & 16.67\% & 12.00\% & 23.58\% & 100\% & 100\% & 100\% & 1 & 34.04\% & 26.00\% & 34.67\% & 43.09\% & 99.53\% & 98.67\% & 100\% & 100\%\\
 & CVM-average & 0.2546 & 0.2478 & 0.236 & 0.2856 & 155.3 & 85.868 & 312.3 & 48.57 & 0.3651 & 0.3038 & 0.3786 & 0.4234 & 1.782 & 1.4011 & 2.3533 & 1.5497\\
 & CVM-stdev & 0.1822 & 0.1865 & 0.1548 & 0.204 & 173.7 & 33.399 & 211.8 & 21.649 & 0.2154 & 0.2006 & 0.1921 & 0.2409 & 0.8393 & 0.6283 & 0.9624 & 0.435\\
 & CVM-rejections@5\% & 12.77\% & 13.33\% & 8.00\% & 17.89\% & 100\% & 100\% & 100\% & 100\% & 26.24\% & 19.33\% & 27.33\% & 33.33\% & 99.29\% & 98.00\% & 100\% & 100\%\\
\hline
 & \multicolumn{17}{c}{Edgeworth}\\
\hline
 & KS-average & 0.0267 & 0.0187 & 0.0325 & 0.0294 & 0.2662 & 0.2097 & 0.3491 & 0.234 & 0.0321 & 0.0217 & 0.0395 & 0.0356 & 0.1997 & 0.1496 & 0.254 & 0.1944\\
 & KS-stdev & 0.0074 & 0.0033 & 0.0054 & 0.004 & 0.0812 & 0.0247 & 0.0824 & 0.0181 & 0.0094 & 0.0035 & 0.0068 & 0.0044 & 0.0537 & 0.0179 & 0.0471 & 0.0122\\
 & KS-rejections@5\% & 98.35\% & 95.33\% & 100\% & 100\% & 100\% & 100\% & 100\% & 100\% & 99.76\% & 99.33\% & 100\% & 100\% & 100\% & 100\% & 100\% & 100\%\\
 & CVM-average & 2.6567 & 1.0831 & 3.9245 & 3.0298 & 286.1 & 181.8 & 438.1 & 227.9 & 4.0869 & 1.5392 & 6.1278 & 4.7048 & 188.6 & 99.35 & 295.4 & 167.1\\
 & CVM-stdev & 1.5653 & 0.4047 & 1.3952 & 0.8581 & 152.7 & 40.51 & 163.1 & 31 & 2.5031 & 0.5157 & 2.2873 & 1.2126 & 106.3 & 25.08 & 105.4 & 21.35\\
 & CVM-rejections@5\% & 99.29\% & 98.00\% & 100\% & 100\% & 100\% & 100\% & 100\% & 100\% & 100\% & 100\% & 100\% & 100\% & 100\% & 100\% & 100\% & 100\%\\
\hline
\end{tabular}}

\caption{\small \textbf{Distribution tests for the approximate distributions of 5-day forward returns} \\
We report the average KS distance (KS-average) and CVM test statistic (CVM-average), with associated standard deviations (KS-stdev and CVM-stdev, respectively) and the percentage of cases where the test statistics are greater than the asymptotic 5\% CVs(KS-rejections@5\% and CVM-rejections@5\%, respectively) for the 5-day forward returns for the S\&P 500, Euro/dollar and 3-month Treasury Bills, respectively. Labels 'low vol', 'high vol' and 'current' refer to the sub-periods: January to August 2006, August 2008 to March 2009 and last 150 observations from 2009, respectively. The label 'total' refers to all three sub-periods, considered together.}
\end{table}
\end{landscape}

\begin{landscape}
\begin{table}[htbp]
  \centering
 \small
\scalebox{0.75}{

\begin{tabular}{cccccccccccccccccc}
\hline
 &  & \multicolumn{4}{c}{Normal GARCH(1,1)} & \multicolumn{4}{c}{Student $t$ GARCH(1,1)} & \multicolumn{4}{c}{Normal GJR} & \multicolumn{4}{c}{Student $t$ GJR}\\
\hline
 &  & total & low vol & high vol & current & total & low vol & high vol & current & total & low vol & high vol & current & total & low vol & high vol & current\\
\hline
S\&P 500 & \multicolumn{17}{c}{Johnson SU}\\
\hline
 & KS-average & 0.0113 & 0.0085 & 0.0086 & 0.0169 & 0.0088 & 0.0086 & 0.009 & 0.0088 & 0.009 & 0.0088 & 0.0091 & 0.0092 & 0.0097 & 0.0094 & 0.0098 & 0.0099\\
 & KS-stdev & 0.0144 & 0.0025 & 0.0024 & 0.0238 & 0.0026 & 0.0025 & 0.0026 & 0.0027 & 0.0026 & 0.0024 & 0.0027 & 0.0027 & 0.0028 & 0.0027 & 0.0028 & 0.0029\\
 & KS-rejections@5\% & 8.67\% & 4.67\% & 4.67\% & 16.67\% & 5.78\% & 6.00\% & 5.33\% & 6.00\% & 5.78\% & 3.33\% & 6.00\% & 8.00\% & 9.78\% & 8.00\% & 9.33\% & 12.00\%\\
 & CVM-average & 1.1472 & 0.1554 & 0.1608 & 3.1253 & 0.1686 & 0.1592 & 0.174 & 0.1726 & 0.1796 & 0.1704 & 0.1818 & 0.1867 & 0.2127 & 0.1957 & 0.2206 & 0.2218\\
 & CVM-stdev & 5.8194 & 0.1351 & 0.1396 & 9.8033 & 0.1408 & 0.1397 & 0.1382 & 0.145 & 0.1498 & 0.1398 & 0.1587 & 0.1509 & 0.1716 & 0.1546 & 0.1811 & 0.1778\\
 & CVM-rejections@5\% & 9.33\% & 4.00\% & 5.33\% & 18.67\% & 4.67\% & 4.67\% & 4.00\% & 5.33\% & 5.78\% & 7.33\% & 4.67\% & 5.33\% & 7.56\% & 6.67\% & 6.67\% & 9.33\%\\
\hline
 & \multicolumn{17}{c}{Edgeworth}\\
\hline
 & KS-average & 0.0114 & 0.0086 & 0.0087 & 0.017 & 0.0107 & 0.0102 & 0.0111 & 0.0108 & 0.0095 & 0.0095 & 0.0095 & 0.0096 & 0.0113 & 0.0112 & 0.0113 & 0.0113\\
 & KS-stdev & 0.0142 & 0.0025 & 0.0025 & 0.0235 & 0.0028 & 0.0028 & 0.0028 & 0.0029 & 0.0027 & 0.0026 & 0.0028 & 0.0028 & 0.003 & 0.0028 & 0.0031 & 0.0031\\
 & KS-rejections@5\% & 9.56\% & 6.00\% & 4.67\% & 18.00\% & 15.33\% & 11.33\% & 17.33\% & 17.33\% & 8.44\% & 8.00\% & 6.67\% & 10.67\% & 18.67\% & 14.00\% & 19.33\% & 22.67\%\\
 & CVM-average & 1.1322 & 0.1595 & 0.1643 & 3.0728 & 0.27 & 0.2354 & 0.2904 & 0.2844 & 0.2071 & 0.2041 & 0.2075 & 0.2096 & 0.3208 & 0.3049 & 0.3289 & 0.3287\\
 & CVM-stdev & 5.7089 & 0.1372 & 0.1389 & 9.6171 & 0.175 & 0.1659 & 0.1692 & 0.1853 & 0.1703 & 0.1602 & 0.1822 & 0.1689 & 0.2293 & 0.2061 & 0.243 & 0.2378\\
 & CVM-rejections@5\% & 8.89\% & 4.67\% & 4.00\% & 18.00\% & 13.33\% & 9.33\% & 16.00\% & 14.67\% & 7.11\% & 6.67\% & 6.00\% & 8.67\% & 18.89\% & 13.33\% & 20.67\% & 22.67\%\\
\hline
Euro/dollar & \multicolumn{17}{c}{Johnson SU}\\
\hline
 & KS-average & 0.0086 & 0.0084 & 0.0086 & 0.0087 & 0.0087 & 0.0085 & 0.0087 & 0.0087 & 0.0086 & 0.0084 & 0.0086 & 0.0087 & 0.0087 & 0.0085 & 0.0088 & 0.0087\\
 & KS-stdev & 0.0025 & 0.0025 & 0.0024 & 0.0027 & 0.0026 & 0.0025 & 0.0025 & 0.0028 & 0.0025 & 0.0025 & 0.0024 & 0.0027 & 0.0026 & 0.0025 & 0.0025 & 0.0028\\
 & KS-rejections@5\% & 4.67\% & 4.67\% & 4.67\% & 4.67\% & 4.44\% & 3.33\% & 4.00\% & 6.00\% & 4.44\% & 4.00\% & 4.67\% & 4.67\% & 4.89\% & 4.00\% & 4.00\% & 6.67\%\\
 & CVM-average & 0.1625 & 0.156 & 0.1617 & 0.1697 & 0.166 & 0.1591 & 0.1645 & 0.1743 & 0.1625 & 0.1561 & 0.1613 & 0.1703 & 0.1663 & 0.159 & 0.1654 & 0.1745\\
 & CVM-stdev & 0.1401 & 0.1381 & 0.1411 & 0.1416 & 0.1423 & 0.1391 & 0.1404 & 0.1478 & 0.1401 & 0.1382 & 0.1402 & 0.1424 & 0.142 & 0.1398 & 0.1392 & 0.1475\\
 & CVM-rejections@5\% & 5.11\% & 4.67\% & 4.67\% & 6.00\% & 4.67\% & 4.00\% & 4.67\% & 5.33\% & 5.11\% & 4.67\% & 4.67\% & 6.00\% & 4.22\% & 3.33\% & 4.00\% & 5.33\%\\
\hline
 & \multicolumn{17}{c}{Edgeworth}\\
\hline
 & KS-average & 0.0086 & 0.0084 & 0.0086 & 0.0087 & 0.009 & 0.0089 & 0.009 & 0.009 & 0.0086 & 0.0084 & 0.0086 & 0.0087 & 0.009 & 0.0089 & 0.0091 & 0.0091\\
 & KS-stdev & 0.0025 & 0.0025 & 0.0024 & 0.0027 & 0.0027 & 0.0025 & 0.0025 & 0.0029 & 0.0025 & 0.0025 & 0.0024 & 0.0027 & 0.0026 & 0.0026 & 0.0025 & 0.0028\\
 & KS-rejections@5\% & 5.11\% & 4.67\% & 4.67\% & 6.00\% & 6.00\% & 6.67\% & 4.00\% & 7.33\% & 4.67\% & 4.67\% & 4.67\% & 4.67\% & 6.44\% & 6.67\% & 4.67\% & 8.00\%\\
 & CVM-average & 0.1626 & 0.1559 & 0.1617 & 0.1703 & 0.1784 & 0.1722 & 0.173 & 0.1899 & 0.1627 & 0.156 & 0.1612 & 0.171 & 0.1793 & 0.172 & 0.1756 & 0.1904\\
 & CVM-stdev & 0.1402 & 0.1382 & 0.1408 & 0.1422 & 0.1474 & 0.1448 & 0.1407 & 0.1567 & 0.1403 & 0.1384 & 0.1395 & 0.1434 & 0.1468 & 0.1458 & 0.1396 & 0.1549\\
 & CVM-rejections@5\% & 5.11\% & 4.67\% & 4.67\% & 6.00\% & 4.89\% & 4.00\% & 4.67\% & 6.00\% & 4.89\% & 4.67\% & 4.67\% & 5.33\% & 4.44\% & 4.00\% & 4.00\% & 5.33\%\\
\hline
3M Bill & \multicolumn{17}{c}{Johnson SU}\\
\hline
 & KS-average & 0.0087 & 0.0084 & 0.0087 & 0.0089 & 0.0155 & 0.0144 & 0.0158 & 0.0165 & 0.0106 & 0.0094 & 0.0112 & 0.0113 & \multicolumn{4}{c}{\multirow{6}{*}{N/A}}\\
 & KS-stdev & 0.0025 & 0.0024 & 0.0024 & 0.0025 & 0.0036 & 0.0042 & 0.0031 & 0.0031 & 0.003 & 0.0026 & 0.003 & 0.0032 & \multicolumn{4}{c}{}\\
 & KS-rejections@5\% & 4.73\% & 4.00\% & 4.67\% & 5.69\% & 72.81\% & 59.33\% & 78.00\% & 82.93\% & 15.37\% & 5.33\% & 18.00\% & 24.39\% & \multicolumn{4}{c}{}\\
 & CVM-average & 0.1614 & 0.152 & 0.1645 & 0.169 & 0.6838 & 0.6 & 0.686 & 0.7834 & 0.2575 & 0.1948 & 0.2889 & 0.2956 & \multicolumn{4}{c}{}\\
 & CVM-stdev & 0.1288 & 0.1233 & 0.1349 & 0.1282 & 0.3202 & 0.3659 & 0.265 & 0.2943 & 0.1931 & 0.1431 & 0.2116 & 0.2044 & \multicolumn{4}{c}{}\\
 & CVM-rejections@5\% & 4.26\% & 4.00\% & 4.67\% & 4.07\% & 76.83\% & 60.67\% & 81.33\% & 91.06\% & 15.60\% & 6.00\% & 18.67\% & 23.58\% & \multicolumn{4}{c}{}\\
\hline
 & \multicolumn{17}{c}{Edgeworth}\\
\hline
 & KS-average & 0.0235 & 0.0179 & 0.0269 & 0.026 & 0.0636 & 0.0511 & 0.0747 & 0.0655 & 0.0289 & 0.0214 & 0.0335 & 0.0326 & 0.0543 & 0.0468 & 0.0627 & 0.0533\\
 & KS-stdev & 0.0054 & 0.003 & 0.0041 & 0.0033 & 0.0123 & 0.0069 & 0.0094 & 0.0032 & 0.0071 & 0.0033 & 0.0054 & 0.004 & 0.0096 & 0.0065 & 0.0089 & 0.0037\\
 & KS-rejections@5\% & 92.44\% & 95.33\% & 100\% & 100\% & 100\% & 100\% & 100\% & 100\% & 100\% & 100\% & 100\% & 100\% & 100\% & 100\% & 100\% & 100\%\\
 & CVM-average & 1.8418 & 0.9387 & 2.3913 & 2.273 & 17.198 & 10.55 & 23.437 & 17.696 & 2.9942 & 1.4219 & 3.9511 & 3.7446 & 11.682 & 8.5839 & 15.458 & 10.857\\
 & CVM-stdev & 0.8681 & 0.3213 & 0.7176 & 0.5385 & 6.6817 & 2.872 & 5.6867 & 1.5538 & 1.4736 & 0.4556 & 1.2011 & 0.8734 & 4.2451 & 2.5548 & 4.32 & 1.248\\
 & CVM-rejections@5\% & 97.64\% & 93.33\% & 100\% & 100\% & 100\% & 100\% & 100\% & 100\% & 100\% & 100\% & 100\% & 100\% & 100\% & 100\% & 100\% & 100\%\\
\hline
\end{tabular}
}

\caption{\small \textbf{Distribution tests for the approximate distributions of 5-day aggregated returns} \\
We report the average KS distance (KS-average) and CVM test statistic (CVM-average), with associated standard deviations (KS-stdev and CVM-stdev, respectively) and the percentage of cases where the test statistics are greater than the asymptotic 5\% CVs(KS-rejections@5\% and CVM-rejections@5\%, respectively) for the 5-day aggregated returns for the S\&P 500, Euro/dollar and 3-month Treasury Bills, respectively. Labels 'low vol', 'high vol' and 'current' refer to the sub-periods: January to August 2006, August 2008 to March 2009 and last 150 observations from 2009, respectively. The label 'total' refers to all three sub-periods, considered together.}
\end{table}
\end{landscape}

\section{Conclusions}\label{sec:Conc}
We have derived analytical expressions for the moments of forward and aggregated returns and variances for an established asymmetric GARCH specification, namely the GJR model, with a generic innovations distribution. Special cases include the normal and Student $t$ GARCH(1,1) and GJR models. We found that the distribution of forward returns is skewed only if the distribution of innovations is skewed, but  the distribution of aggregated returns is skewed even if the innovation distribution is symmetric. The other source of skewness in this case is the asymmetric response of variance to positive and negative shocks (i.e. $\lambda \neq 0$). 

There are two sources of kurtosis in forward returns: the degree of leptokurtosis of the innovation distribution and the uncertainty in forward variance. Since the one-step-ahead variance is deterministic in a GARCH setting, the kurtosis coefficient of the one-step-ahead returns distributions is equal to that of the innovation distribution. However, whenever we forecast $s$-steps ahead (with $s>1$)  using a GARCH(1,1) or GJR model, the $s$-step-ahead returns distribution for $s>1$ will have a higher kurtosis than the one-step-ahead returns distribution, due to the positivity of the conditional variance of the conditional variance, which increases the probability mass in the tails of the forward one-period returns distribution. Also, the time-variability of the conditional variance of the conditional variance introduces dynamics in the higher moments of the forward returns.

Provided the unconditional moments exist (i) the conditional moments of forward returns converge to the corresponding unconditional moments as the time horizon increases, and (ii) the conditional moments of aggregated returns converge to the corresponding moments of a normal distribution. Otherwise, the moments of both the forward returns as well as the aggregated returns generally diverge to (plus or minus) infinity.

An empirical application computed higher moments of the forward and aggregated returns of the S\&P 500 index, the Euro/dollar exchange rate and the 3-month US Treasury bill rate, using our analytic expressions for the first four conditional moments based on four different GARCH processes. Subsequently, we approximated predictive distributions for forward and aggregated returns using these higher moment forecasts and the Johnson SU distribution or the Edgeworth expansion. Using established statistical tests, we evaluated the accuracy of these approximations, relative to the corresponding simulated GARCH returns distributions. The results of these tests are in general very good for the vast majority of the approximate distributions. Hence, our moment expressions may have useful applications to financial problems which, until now, have required GARCH returns distributions to be simulated.


\singlespacing
\section*{References}
\begin{description}
\item Andersen, T.G. \& T. Bollerslev (1998) Answering the Critics: Yes ARCH models Do Provide Good Volatility Forecasts. \textit{International Economic Review} 39, 885-905.

\item Andersen, T.G., T. Bollerslev \& F.X. Diebold (2009) Parametric and Nonparametric Volatility Measurement. In L.P. Hansen \& Y. Ait-Sahalia (eds.), \textit{Handbook of Financial Econometrics}. North-Holland.

\item Anderson, T.W. \& D.A. Darling (1952) Asymptotic Theory of Certain "Goodness of Fit" Criteria Based on Stochastic Processes. \textit{The Annals of Mathematical Statistics} 23, 193-212.

\item Bai, X., J.R. Russell, \& G.C. Tiao (2003) Kurtosis of GARCH and Stochastic Volatility Models with Non-Normal Innovations. \textit{Journal of Econometrics} 114, 349-360.


\item Bauwens, L., S. Laurent \& J.V.K. Rombouts (2006) Multivariate GARCH Models: a Survey. \textit{Journal of Applied Econometrics} 21, 79-109.

\item Bhattacharya, R.N. \&  J.K. Ghosh (1978) On the Validity of the Formal Edgeworth Expansion. \textit{The Annals of Statistics} 6, 435-451.

\item Bollerslev, T. (1986) Generalized Autoregressive Conditional Heteroskedasticity. \textit{Journal of Econometrics} 31, 307-327.

\item Bollerslev, T. (1987) A Conditionally Heteroskedastic Time Series Model for Speculative Prices and Rates of Return. \textit{The Review of Economics and Statistics} 69, 542-547.

\item Breuer, T. \& M. Jandacka (2010) Temporal Aggregation of GARCH Models: Conditional Kurtosis and Optimal Frequency. \textit{Working Paper}, available from http://ssrn.com.



\item Charlier, C.V.L. (1905) Uber das Fehlergesetz. \textit{Arkiv for Matematik, Astronomi Och Fysic}, 8, 1-9.

\item Charlier, C.V.L. (1906) Uber die Darstellung Willkurlicher Funktionen. \textit{Arkiv for Matematik, Astronomi Och Fysic}, 20, 1-35.

\item Chebyshev, P.L. (1860) Sur le Developpment des Fonctions a une Seule Variable. \textit{Bulletin de la Classe Physique-Mathematique de l'Academie Imperiale des Sciences St. Petersbourg} 3, 193-202.

\item Chebyshev, P.L. (1890) Sur Deux Theorems Relatifs aux Probabilites. \textit{Acta Mathematica} 14, 305-315.


\item Christoffersen, P.F., C. Dorion, K. Jacobs \& Y. Wang (2010) Volatility Components: Affine Restrictions and Non-normal Innovations. \textit{Journal of Business and Economic Statistics} 28, 483-502.


\item Christoffersen, P.F., K. Jacobs, C. Ornthanalai \&  Y. Wang (2008) Option Valuation with Long-run and Short-run Volatility Components. \textit{Journal of Financial Economics} 90, 272-297.



\item Demos, A. (2002) Moments and Dynamic Structure of a Time-Varying-Parameter Stochastic Volatility in Mean Model. \textit{The Econometrics Journal} 5, 345-357.

\item Diebold, F.X. (1988) \textit{Empirical Modelling of Exchange Rates}. Springer.


\item Duan, J.C., G. Gauthier, C. Sasseville \& J.G. Simonato (2006) Approximating the GJR-GARCH and EGARCH Option Pricing Models Analytically. \textit{Journal of Computational Finance} 9, 41-69.

\item Duan, J.C., G. Gauthier \& J.G. Simonato (1999) An Analytical Approximation for the GARCH Option Pricing Model. \textit{Journal of Computational Finance} 2, 76-116.

\item Edgeworth, F.Y. (1905) The Law of Error. \textit{Transactions of the Cambridge Philosophical Society} 20, 33-66 and 113 -141.

\item Engle, R.F. (1982) Autoregressive Conditional Heteroskedasticity with Estimates of the Variance of United Kingdom Inflation. \textit{Econometrica} 50, 987-1007.

\item Engle, R.F. (1990) Discussion: Stock Market Volatility and the Crash of '87. \textit{Review of Financial Studies} 3, 103-106.

\item Engle, R.F. \& V.K. Ng (1993) Measuring and Testing the Impact of News on Volatility. \textit{Journal of Finance} 48, 1749-1778.

\item Fama, E. (1965) The Behaviour of Stock Market Prices. \textit{Journal of Business} 38, 34 -105.

\item Glosten, L.R., R. Jagannathan \& D.E. Runkle (1993) On the Relation Between the Expected Value and the Volatility of the Nominal Excess Return on Stocks. \textit{Journal of Finance} 48, 1779-1801.

\item Gram, J.P. (1883) Uber die Entwickelung Reeler Funktionen in Reihen mittelst der Methode der Kleinsten Quadrate. \textit{Creile} 94, 41-73.

\item Haas, M., S. Mittnik, \& M.S. Paolella (2004) Mixed Normal Conditional Heteroskedasticity. \textit{Journal of Financial Econometrics} 2, 493-530.



\item He, C. \& T. Terasvirta (1999a) Properties of Moments of a Family of GARCH Processes. \textit{Journal of Econometrics} 92, 173-192.

\item He, C. \& T. Terasvirta (1999b) Fourth Moment Structure of the GARCH(p,q) process. \textit{Econometric Theory} 15, 824-846.

\item He, C., T. Terasvirta \& H. Malmsten (2002)  Fourth Moments Structure of a Family of First-Order Exponential GARCH Models. \textit{Econometric Theory} 18, 868-885.


\item Heston, S.L. \& S. Nandi (2000) A Closed-Form GARCH Option Pricing Model. \textit{Review of Financial Studies} 13, 281-300.

\item Ishida, I. \& R. F. Engle (2002) Modeling Variance of Variance: The Square-Root, the Affine, and the CEV GARCH Models. \textit{Working Paper}, available from http://www.stern.nyu.edu/rengle/.


\item Jaschke, S. (2002) The Cornish-Fisher-Expansion in the Context of Delta-Gamma-Normal Approximations. \textit{Journal of Risk} 4, 33-52.

\item Johnson, N.L. (1949) Systems of Frequency Curves Generated by Methods of Translation. \textit{Biometrica} 36, 149-76.

\item Karanasos, M. (1999) The Second Moment and the Autocovariance Function of the Squared errors of the GARCH Model. \textit{Journal of Econometrics} 9, 63-76.

\item Karanasos, M. (2001) Prediction in ARMA Models with GARCH-in-Mean Effects. \textit{Journal of Time Series Analysis} 22, 555-78.

\item Karanasos, M. \& J. Kim (2003) Moments of ARMA-EGARCH model. \textit{The Econometrics Journal} 6, 146-166.

\item Karanasos, M., Z. Psaradakis \& M. Sola (2004) On the Autocorrelation Properties of Long-Memory GARCH Processes. \textit{Journal of Time Series Analysis} 25, 265-281.

\item Kolmogorov, A. (1933) Sulla Determinazione Empirica di una Legge di Distribuzione. \textit{Giornale dell'Instituto Italiano degh Attuari} 4, 1-11.

\item Ling, S. \& M. McAleer (2002a) Necessary and Sufficient Moment Conditions for the GARCH(r,s) and the Asymmetric Power GARCH(r,s). \textit{Econometric Theory} 18, 722-729.

\item Ling, S. \& M. McAleer (2002b) Stationarity and the Existence of Moments of a Family of GARCH Processes. \textit{Journal of Econometrics} 106, 109-117.

\item Mandelbrot, B. (1963) The Variations of Certain Speculative Prices. \textit{Journal of Business} 36, 394-419.

\item Marcucci, J. (2005) Forecasting Stock Market Volatility with Regime-Switching GARCH Models. \textit{Studies in Nonlinear Dynamics \& Econometrics} 9, 1-53.

\item Massey, F.J. Jr. (1951) The Kolmogorov-Smirnov Test for Goodness of Fit. \textit{Journal of the American Statistical Association} 46, 68-78.

\item Milhoj, A. (1985) The Moment Structure of ARCH Processes. \textit{Scandinavian Journal of Statistics} 12, 281-292.

\item Nelson, D.B. (1991) Conditional Heteroskedasticity in Asset Returns: a New Approach. \textit{Econometrica} 59, 347-370.

\item Nemec, A.F.L. (1985) Conditionally Heteroskedastic Autoregressions. \textit{Technical Report no. 43}. Department of Statistics, University of Washington.

\item Pearson, E.S. \& H.O. Hartley (1972) \textit{Biometrika Tables for Statisticians}, vol.2. Cambridge University Press.




\item Serfling, R.J. (1980) \textit{Approximation Theorems of Mathematical Statistics}. Wiley.

\item Smirnov, N. (1939) Sur les Ecarts de la Courbe de Distribution Empirique. \textit{Matematicheskii Sbornik} 48, 3-26.


\item Stephens, M.A. (1970) Use of the Kolmogorov-Smirnov, Cramer-von Mises and Related Statistics without Extensive Tables. \textit{Journal of the Royal Statistical Society B} 32, 115-122.


\item Taylor, S. (1986) \textit{Modeling Financial Time Series}. Wiley.

\item Tuenter, H.J.H. (2001) An Algorithm to Determine the Parameters of SU-Curves in the Johnson System of Probability Distributions by Moment Matching. \textit{Journal of Statistical Computation and Simulation} 70, 325-347.

\item Wallace, D.L. (1958) Asymptotic Approximations to Distributions. \textit{The Annals of Mathematical Statistics} 29, 635-654.


\item Wong, C.M. \& M.K.P. So (2003) On Conditional Moments of GARCH Models with Applications to Multiple Period Value at Risk Estimation. \textit{Statistica Sinica} 13, 1015-1044.


\end{description}
\newpage
\section*{Appendices: Conditional Moments for the Generic GJR model}

\doublespacing
\setlength{\parindent}{0pt}
The model's $s$-step-ahead conditional variances are given by:
\vspace{-0.5cm}
\begin{equation}\label{varianceeq}
{h_{t + s}} = \omega  + \left( {\alpha  + \lambda I_{t + s - 1}^ - } \right)\varepsilon _{t + s - 1}^2 + \beta {h_{t + s - 1}}
\vspace{-0.5cm}
\end{equation}
The aim of this appendix is to calculate the conditional moments of the return the forward return ${r_{t+s}}$ and of its conditional variance ${h_{t+s}}$, as well as of the aggregated return and of its conditional variance according to the model. Specifically, for $i$ = 1, 2, 3 and 4, and for $x = r$ and $h$, we compute:
\begin{center}
	$\tilde \mu _{x,s}^{\left( i \right)} = {E_t}\left( {x_{t + s}^i} \right),\quad \mu _{x,s}^{\left( i \right)} = {E_t}\left( {{{\left( {{x_{t + s}} - \tilde \mu _{x,s}^{\left( 1 \right)}} \right)}^i}} \right),$ \\
	$\tilde M_{x,n}^{\left( i \right)} = {E_t}\left[ {{{\left( {\sum\limits_{s = 1}^n {{x_{t + s}}} } \right)}^i}} \right],\quad M_{x,n}^{\left( i \right)} = {E_t}\left( {{{\left( {\sum\limits_{s = 1}^n {\left( {{x_{t + s}} - \tilde \mu _{x,s}^{\left( 1 \right)}} \right)} } \right)}^i}} \right),$
	\vspace{0.5cm}
\end{center}
and the corresponding standardized moments:
\begin{center}
	${\tau _{x,s}} = \frac{{\mu _{x,s}^{\left( 3 \right)}}}{{{{\left( {\mu _{x,s}^{\left( 2 \right)}} \right)}^{3/2}}}},\quad {{\rm T}_{x,n}} = \frac{{M_{x,n}^{\left( 3 \right)}}}{{{{\left( {M_{x,n}^{\left( 2 \right)}} \right)}^{3/2}}}},\quad {\kappa _{x,s}} = \frac{{\mu _{x,s}^{\left( 4 \right)}}}{{{{\left( {\mu _{x,s}^{\left( 2 \right)}} \right)}^2}}},\quad {{\rm K}_{x,n}} = \frac{{M_{x,n}^{\left( 4 \right)}}}{{{{\left( {M_{x,n}^{\left( 2 \right)}} \right)}^2}}}.$
	\vspace{0.5cm}
\end{center}
We focus on calculating the un-centred conditional moments; the centred moments will follow using simple formulae. We denote by ${F_0}$ the distribution function for $D\left( {0,1} \right)$, evaluated at zero, and set:
\vspace{-0.5cm}
\begin{equation}\label{fi}
\varphi  = \alpha  + \lambda {F_0} + \beta
\vspace{-0.5cm}
\end{equation}
and $\bar h = \omega {\left( {1 - \varphi } \right)^{ - 1}}$.
For both the normal and the standardized Student \textit{t}, ${F_0} = \frac{1}{2}$, since the two distributions are symmetric, thus for these two special cases $\varphi$ becomes:
\vspace{-0.5cm}
\begin{equation}\label{fi_1}
\varphi  = \alpha  + \frac{\lambda }{2} + \beta
\vspace{-0.5cm}
\end{equation}
The following notation will also be useful, where $s, u, v, w >0$:
\begin{center}
	$\tilde \mu _{h,su}^{\left( {i,j} \right)} = {E_t}\left( {h_{t + s}^ih_{t + s + u}^j} \right)$ \\
	$\tilde \mu _{h,suv}^{\left( {i,j,k} \right)} = {E_t}\left( {h_{t + s}^ih_{t + s + u}^jh_{t + s + u + v}^k} \right)$ \\
	${\tilde \mu _{h,suvw}} = {E_t}\left( {{h_{t + s}}{h_{t + s + u}}{h_{t + s + u + v}}{h_{t + s + u + v + w}}} \right)$ \\
	$\theta _{su}^{\left( j \right)} = {E_t}\left( {{\varepsilon _{t + s}}h_{t + s + u}^j} \right)$ \\
\end{center}

After deriving the formulae for the generic model, we allow the innovations distribution  $D\left( {0,1} \right)$ to take two specific functional forms, largely used in practice: the standard normal and the (standardized) Student \textit{t}.\footnote{We shall only derive the expressions for the two special cases when a particular expression derived for the generic case differs for one (or both) of the special cases; when no formulae for the special cases are mentioned, the generic formulae apply.} When $D\left( {0,1} \right)$ is the standard normal distribution the moments with odd order are all zero and the even moments are given by $\mu _z^{\left( {2r} \right)} = \prod\limits_{i = 1}^r {\left( {2i - 1} \right)}$. When $D\left( {0,1} \right)$ is the standardized Student \textit{t} distribution, the odd order moments are again all zero (as long as the number of degrees of freedom $\nu >$ the order of the moment) and the even moments are given by: $\mu _z^{\left( {2r} \right)} = {\left( {\nu  - 2} \right)^r}\frac{{\Gamma \left( {r + \frac{1}{2}} \right)\Gamma \left( {\frac{1}{2}\nu  - r} \right)}}{{\Gamma \left( {\frac{1}{2}} \right)\Gamma \left( {\frac{1}{2}\nu } \right)}}$, if $\nu > 2r$.

\subsubsection*{T.A.1: Conditional Moments of Forward and Aggregated Returns}
(a) \textit{First Conditional Moments of Forward and Aggregated Returns}  \\
For $s \ge 1$ and using the notation ${R_{tn}}=\sum\limits_{s=1}^n{{r_{t+s}}}$ for the aggregated return and also using the tower law of expectations, we get:
\begin{center}
	${E_t}\left( {{r_{t + s}}} \right)  = {E_t}\left( {\mu  + {\varepsilon _{t + s}}} \right) = \mu  + {E_t}\left( {\underbrace {{E_{t + s - 1}}\left( {{\varepsilon _{t + s}}} \right)}_0} \right) = \mu$ \\
	${E_t}\left( {{R_{tn}}} \right) = {E_t}\left( {\sum\limits_{s = 1}^n {{r_{t + s}}} } \right) = \sum\limits_{s = 1}^n {{E_t}\left( {{r_{t + s}}} \right)}  = n\mu$
\end{center}
\vspace{0.5cm}
(b) \textit{Second Conditional Moments of Forward and Aggregated Returns}  \\
The second moment of the forward return is:
\begin{center}
	${E_t}\left( {r_{t + s}^2} \right) = {E_t}\left[ {{{\left( {\mu  + {\varepsilon _{t + s}}} \right)}^2}} \right] = {\mu ^2} + \bar h + {\varphi ^{s - 1}}\left( {{h_{t + 1}} - \bar h} \right)$,
\end{center}
where we used the expression for the second moment of variance $\tilde \mu _{h,s}^{(1)} = \mu _{r,s}^{\left( 2 \right)} = \bar h + {\varphi ^{s - 1}}\left( {{h_{t + 1}} - \bar h} \right)$ derived in the Technical Appendix T.A.2 below.
The second moment of aggregated returns is:
\begin{center}
	${E_t}\left( {R_{tn}^2} \right) = {E_t}{\left( {\sum\limits_{s = 1}^n {{r_{t + s}}} } \right)^2} = {E_t}\left( {\sum\limits_{s = 1}^n {r_{t + s}^2 + 2\sum\limits_{s = 1}^n {\sum\limits_{u = 1}^{n - s} {{r_{t + s}}{r_{t + s + u}}} } } } \right).$ \\
	${E_t}\left( {\sum\limits_{s = 1}^n {r_{t + s}^2} } \right) = \sum\limits_{s = 1}^n {\left( {{\mu ^2} + \bar h + {\varphi ^{s - 1}}\left( {{h_{t + 1}} - \bar h} \right)} \right)}  = n\left( {{\mu ^2} + \bar h} \right) + \left( {{h_{t + 1}} - \bar h} \right){\left( {1 - \varphi } \right)^{ - 1}}\left( {1 - {\varphi ^n}} \right)$
\end{center}
${E_t}\left( {\sum\limits_{s = 1}^n {\sum\limits_{u = 1}^{n - s} {{r_{t + s}}{r_{t + s + u}}} } } \right) = \sum\limits_{s = 1}^n {\sum\limits_{u = 1}^{n - s} {{E_t}\left( {{r_{t + s}}{r_{t + s + u}}} \right)} }  = \sum\limits_{s = 1}^n {\sum\limits_{u = 1}^{n - s} {{E_t}\left( {\left( {\mu  + {\varepsilon _{t + s}}} \right)\left( {\mu  + {\varepsilon _{t + s + u}}} \right)} \right)} }  = \raise.5ex\hbox{$\scriptstyle 1$}\kern-.1em/
\kern-.15em\lower.25ex\hbox{$\scriptstyle 2$} \;n\left( {n - 1} \right){\mu ^2}$.\\

Hence, the expression for the second moment of aggregated returns becomes:
\begin{center}
	${E_t}\left( {R_{tn}^2} \right) = {n^2}{\mu ^2} + n\bar h + \left( {{h_{t + 1}} - \bar h} \right){\left( {1 - \varphi } \right)^{ - 1}}\left( {1 - {\varphi ^n}} \right)$
\end{center}

(c) \textit{Third Conditional Moments of Forward and Aggregated Returns} \\
For the third moment of forward returns we write:
\begin{center}
	$\begin{array}{l}
	{E_t}\left( {r_{t + s}^3} \right) = {E_t}\left[ {{{\left( {\mu  + {\varepsilon _{t + s}}} \right)}^3}} \right] = {E_t}\left( {{\mu ^3} + 3{\mu ^2}{\varepsilon _{t + s}} + 3\mu \varepsilon _{t + s}^2 + \varepsilon _{t + s}^3} \right) \\
	\qquad \quad \quad \, = {\mu ^3} + 3\mu \tilde \mu _{h,s}^{\left( 1 \right)} + {E_t}\left( {z_{t + s}^3h_{t + s}^{3/2}} \right) = {\mu ^3} + 3\mu \tilde \mu _{h,s}^{\left( 1 \right)} + {\tau _z}{E_t}\left( {h_{t + s}^{3/2}} \right)\,. \\
	\end{array}$
\end{center}
For both the normal and Student \textit{t} GJR (or rather for any GJR model with a symmetric innovations distribution), the expression for the third moment of returns is given by:
\begin{center}
	${E_t}\left( {r_{t + s}^3} \right) = {\mu ^3} + 3\mu \tilde \mu _{h,s}^{\left( 1 \right)} = {\mu ^3} + 3\mu \left( {\bar h + {\varphi ^{s - 1}}\left( {{h_{t + 1}} - \bar h} \right)} \right)$
\end{center}
However, for the generic, skewed model, we need to approximate ${E_t}\left( {h_{t + s}^{3/2}} \right)$ using a second order Taylor series expansion. In general, for a smooth function $g\left( X \right)$:
\begin{center}
	$g\left( X \right) \approx g\left( {{E_t}\left( X \right)} \right) + g'\left( {{E_t}\left( X \right)} \right)\left( {X - {E_t}\left( X \right)} \right) + \raise.5ex\hbox{$\scriptstyle 1$}\kern-.1em/
	\kern-.15em\lower.25ex\hbox{$\scriptstyle 2$} g''\left( {{E_t}\left( X \right)} \right){\left( {X - {E_t}\left( X \right)} \right)^2}.$
\end{center}
Taking expectations we get:
${E_t}\left( {g\left( X \right)} \right) \approx g\left( {{E_t}\left( X \right)} \right) + \raise.5ex\hbox{$\scriptstyle 1$}\kern-.1em/
\kern-.15em\lower.25ex\hbox{$\scriptstyle 2$} {\rm{ }}g''\left( {{E_t}\left( X \right)} \right){V_t}\left( X \right).$
Setting $g\left( X \right) = {X^{3/2}}$, so
$g'\left( X \right) = \frac{3}{2}{X^{\raise.5ex\hbox{$\scriptstyle 1$}\kern-.1em/
		\kern-.15em\lower.25ex\hbox{$\scriptstyle 2$} {\rm{ }}}}$
and
$g''\left( X \right) = \frac{3}{4}{X^{ - \raise.5ex\hbox{$\scriptstyle 1$}\kern-.1em/
		\kern-.15em\lower.25ex\hbox{$\scriptstyle 2$} {\rm{ }}}}$
and setting $X = {h_{t + s}}$ yields:
\begin{center}
	${E_t}\left( {h_{t + s}^{3/2}} \right) \simeq \frac{5}{8}{\left( {\tilde \mu _{h,s}^{\left( 1 \right)}} \right)^{3/2}} + \frac{3}{8}\tilde \mu _{h,s}^{\left( 2 \right)}{\left( {\tilde \mu _{h,s}^{\left( 1 \right)}} \right)^{ - 1/2}},$
\end{center}
where the expressions for $ {\tilde \mu _{h,s}^{\left( 1 \right)}}$ and $ {\tilde \mu _{h,s}^{\left( 2 \right)}}$ are given in the Technical Appendix T.A.2 below. \\
We now compute the third moment of the aggregated returns: \\
$\begin{array}{l}
{E_t}\left( {R_{tn}^3} \right) = {E_t}{\left( {\sum\limits_{s = 1}^n {{r_{t + s}}} } \right)^3}  = \sum\limits_{s = 1}^n {{E_t}\left( {r_{t + s}^3} \right)}  + 3\sum\limits_{s = 1}^n {\sum\limits_{u = 1}^{n - s} {\left[ {{E_t}\left( {r_{t + s}^2{r_{t + s + u}}} \right) + {E_t}\left( {{r_{t + s}}r_{t + s + u}^2} \right)} \right]} }  \\
\quad \qquad \quad+ 6\sum\limits_{s = 1}^n {\sum\limits_{u = 1}^{n - s} {\sum\limits_{v = 1}^{n - s - u} {{E_t}\left( {{r_{t + s}}{r_{t + s + u}}{r_{t + s + u + v}}} \right)} } }  \\
\end{array}$ \\
$\sum\limits_{s = 1}^n {{E_t}\left( {r_{t + s}^3} \right)}  = n\mu \left( {{\mu ^2} + 3\bar h} \right) + 3\mu {\left( {1 - \varphi } \right)^{ - 1}}\left( {1 - {\varphi ^n}} \right)\left( {{h_{t + 1}} - \bar h} \right) + {\tau_z}\sum\limits_{s = 1}^n {{E_t}\left( {h_{t + s}^{3/2}} \right)}$ \\
$\begin{array}{l}
\sum\limits_{s = 1}^n {\sum\limits_{u = 1}^{n - s} {{E_t}\left( {r_{t + s}^2{r_{t + s + u}}} \right) = } } \sum\limits_{s = 1}^n {\sum\limits_{u = 1}^{n - s} {{E_t}\left[ {\left( {{\mu ^2} + 2\mu {\varepsilon _{t + s}} + \varepsilon _{t + s}^2} \right)\left( {\mu  + {\varepsilon _{t + s + u}}} \right)} \right]} }
= \sum\limits_{s = 1}^n {\sum\limits_{u = 1}^{n - s} {\left( {{\mu ^3} + \mu \mu _{h,s}^{\left( 1 \right)}} \right)} }
\\
\qquad \qquad \quad \quad \quad \quad \quad \;\, = \mu \left[ {\frac{{n\left( {n - 1} \right)}}{2}\left( {{\mu ^2} + \bar h} \right) + {{\left( {1 - \varphi } \right)}^{ - 1}}\left[ {n - {{\left( {1 - \varphi } \right)}^{ - 1}}\left( {1 - {\varphi ^n}} \right)} \right]\left( {{h_{t + 1}} - \bar h} \right)} \right] \\
\end{array}$ \\
$\begin{array}{l}
\sum\limits_{s = 1}^n {\sum\limits_{u = 1}^{n - s} {{E_t}\left( {{r_{t + s}}r_{t + s + u}^2} \right)} }  = \sum\limits_{s = 1}^n {\sum\limits_{u = 1}^{n - s} {{E_t}\left( {\left( {\mu  + {\varepsilon _{t + s}}} \right)\left( {{\mu ^2} + 2\mu {\varepsilon _{t + s + u}} + \varepsilon _{t + s + u}^2} \right)} \right)} }  \\
\quad \quad\quad \quad \quad \quad \quad \quad \quad \;\;\; = \sum\limits_{s = 1}^n {\sum\limits_{u = 1}^{n - s} {\left( {{\mu ^3} + \mu \tilde \mu _{h,s + u}^{\left( 1 \right)} + {E_t}\left( {{\varepsilon _{t + s}}\varepsilon _{t + s + u}^2} \right)} \right)} }  \\
\end{array}$ \\
$\begin{array}{l}
{E_t}\left( {{\varepsilon _{t + s}}\varepsilon _{t + s + u}^2} \right)
= {E_t}\left( {{\varepsilon _{t + s}}{E_{t + s + u - 1}}\left( {\varepsilon _{t + s + u}^2} \right)} \right)
= \theta _{su}^{\left( 1 \right)} \\
\quad \quad \quad \quad \quad \quad \;\;\;= {E_t}\left( {{\varepsilon _{t + s}}\left( {\omega  + \left( {\alpha  + \lambda {I_{t + s + u - 1}}} \right)\varepsilon _{t + s + u - 1}^2 + \beta {h_{t + s + u - 1}}} \right)} \right) = \varphi {E_t}\left( {{\varepsilon _{t + s}}\varepsilon _{t + s + u - 1}^2} \right) \\
\quad \quad \quad \quad \quad \quad \;\;\;
= {\varphi ^{u - 1}}{E_t}\left( {{\varepsilon _{t + s}}{h_{t + s + 1}}} \right)
= {\varphi ^{u - 1}}\left( {\alpha {\tau _{z}} + \lambda \int\limits_{z =  - \infty }^0 {{z^3}f\left( z \right)dz} } \right){E_t}\left( {h_{t + s}^{3/2}} \right)\, \\
\end{array}$ \\
where $f$ is the density of the innovation distribution; the final expression for $\sum\limits_{s = 1}^n {\sum\limits_{u = 1}^{n - s} {{E_t}\left( {{r_{t + s}}r_{t + s + u}^2} \right)} }$ becomes: \\
$\begin{array}{l}
\sum\limits_{s = 1}^n {\sum\limits_{u = 1}^{n - s} {{E_t}\left( {{r_{t + s}}r_{t + s + u}^2} \right)} }  = \frac{{n\left( {n - 1} \right)}}{2}\mu \left( {{\mu ^2} + \bar h} \right) \\
\qquad \qquad \qquad \quad \quad \quad \quad  + {\left( {1 - \varphi } \right)^{ - 1}}\left( \begin{array}{l}
\mu \left[ {\varphi {{\left( {1 - \varphi } \right)}^{ - 1}}\left( {1 - {\varphi ^n}} \right) - n{\varphi ^n}} \right]\left( {{h_{t + 1}} - \bar h} \right) \\
+ \left( {\alpha {\tau _z} + \lambda \int\limits_{z =  - \infty }^0 {{z^3}f\left( z \right)dz} } \right)\sum\limits_{s = 1}^n {\left( {1 - {\varphi ^{n - s}}} \right)} {E_t}\left( {h_{t + s}^{3/2}} \right) \\
\end{array} \right) \\
\end{array}$ \\
For the normal GJR we have ${\tau_z=0}$ and also it can be easily shown that:
\begin{center}
	$\int\limits_{z =  - \infty }^0 {{z^3}f\left( z \right)dz}  = \int\limits_{z =  - \infty }^0 {\frac{1}{{\sqrt {2\pi } }}{z^3}\exp \left( { - \frac{{{z^2}}}{2}} \right)dz =  - \sqrt {\frac{2}{\pi }} }.$
\end{center}
Similarly, for the Student \textit{t} GJR, we have  ${\tau_z=0}$ and easily get:
\begin{center}
	$\int\limits_{z =  - \infty }^0 {{z^3}} f\left( z \right) = \int\limits_{z =  - \infty }^0 {{z^3}\frac{{\Gamma \left( {\frac{{\nu  + 1}}{2}} \right)}}{{\Gamma \left( {\frac{\nu }{2}} \right)\sqrt {\pi \left( {\nu  - 2} \right)} }}} {\left( {1 + \frac{{{z^2}}}{{\nu  - 2}}} \right)^{ - \frac{{\nu  + 1}}{2}}}dz\, =  - \frac{2}{{\sqrt \pi  }}\frac{{{{\left( {\nu  - 2} \right)}^{3/2}}}}{{\left( {\nu  - 1} \right)\left( {\nu  - 3} \right)}}\frac{{\Gamma \left( {\frac{{\nu  + 1}}{2}} \right)}}{{\Gamma \left( {\frac{\nu }{2}} \right)}},$
\end{center}
Repeatedly applying the tower law, we get that: \\
$\sum\limits_{s = 1}^n {\sum\limits_{u = 1}^{n - s} {\sum\limits_{v = 1}^{n - s - u} {{E_t}\left( {{r_{t + s}}{r_{t + s + u}}{r_{t + s + u + v}}} \right)} } }  = \sum\limits_{s = 1}^n {\sum\limits_{u = 1}^{n - s} {\sum\limits_{v = 1}^{n - s - u} {{\mu ^3}} } }  = \frac{{n\left( {n - 1} \right)\left( {n - 2} \right)}}{6}{\mu ^3}$. \\

(d) \textit{Fourth Conditional Moments of Forward and Aggregated Returns} \\
The fourth moment of the forward returns is \\
${E_t}\left( {r_{t + s}^4} \right) = {E_t}\left( {{\mu ^4} + 4{\mu ^3}{\varepsilon _{t + s}} + 6{\mu ^2}\varepsilon _{t + s}^2 + 4\mu \varepsilon _{t + s}^3 + \varepsilon _{t + s}^4} \right) = {\mu ^4} + 6{\mu ^2}\tilde \mu _{h,s}^{\left( 1 \right)} + 4\mu {\tau _ z}{E_t}\left( {h_{t + s}^{3/2}} \right) + {\kappa _z}\tilde \mu _{h,s}^{\left( 2 \right)}$, \\
where $\tilde \mu _{h,s}^{\left( 1 \right)}$ and $\tilde \mu _{h,s}^{\left( 2 \right)}$ are derived in the Technical Appendix T.A.2 below and ${E_t}\left( {h_{t + s}^{3/2}} \right)$ is given above as a function of these first two conditional moments of the forward variance. \\
In the special case that the innovation distribution is the standard normal, \\${E_t}\left( {r_{t + s}^4} \right) = {\mu ^4} + 6{\mu ^2}\tilde \mu _{h,s}^{\left( 1 \right)} + 3\tilde \mu _{h,s}^{\left( 2 \right)}$, while for the Student \textit{t} GJR, ${E_t}\left( {r_{t + s}^4} \right) = {\mu ^4} + 6{\mu ^2}\tilde \mu _{h,s}^{\left( 1 \right)} + 3\frac{{\nu  - 2}}{{\nu  - 4}}\tilde \mu _{h,s}^{\left( 2 \right)}$.

For the fourth moment of aggregated returns, we write: \\
$\begin{array}{l}
{E_t}\left( {R_{tn}^4} \right) = \sum\limits_{s = 1}^n {{E_t}\left( {r_{t + s}^4} \right)}  + \sum\limits_{s = 1}^n {\sum\limits_{u = 1}^{n - s} {\left[ {4\left( {{E_t}\left( {r_{t + s}^3{r_{t + s + u}}} \right) + {E_t}\left( {{r_{t + s}}r_{t + s + u}^3} \right)} \right) + 6{E_t}\left( {r_{t + s}^2r_{t + s + u}^2} \right)} \right]} }  \\
\quad \quad \quad \quad + 12\sum\limits_{s = 1}^n {\sum\limits_{u = 1}^{n - s} {\sum\limits_{v = 1}^{n - s - u} {\left( {{E_t}\left( {r_{t + s}^2{r_{t + s + u}}{r_{t + s + u + v}}} \right) + {E_t}\left( {{r_{t + s}}r_{t + s + u}^2{r_{t + s + u + v}}} \right) + {E_t}\left( {{r_{t + s}}{r_{t + s + u}}r_{t + s + u + v}^3} \right)} \right)} } }  \\
\quad \quad \quad \quad + 24\sum\limits_{s = 1}^n {\sum\limits_{u = 1}^{n - s} {\sum\limits_{v = 1}^{n - s - u} {\sum\limits_{w = 1}^{n - s - u - v} {{E_t}\left( {{r_{t + s}}{r_{t + s + u}}{r_{t + s + u + v}}{r_{t + s + u + v + w}}} \right)} } } }  \\
\end{array}$ \\

$\begin{array}{l}
\sum\limits_{s = 1}^n {{E_t}\left( {r_{t + s}^4} \right)}  = \sum\limits_{s = 1}^n {\left( {{\mu ^4} + 6{\mu ^2}\tilde \mu _{h,s}^{\left( 1 \right)} + 4\mu {\tau _ z}{E_t}\left( {h_{t + s}^{3/2}} \right) + {\kappa _z}\tilde \mu _{h,s}^{\left( 2 \right)}} \right)}  \\
\quad \quad \quad \quad \quad \; \; \,= n{\mu ^2}\left( {{\mu ^2} + 6\bar h} \right) + 6{\mu ^2}{\left( {1 - \varphi } \right)^{ - 1}}\left( {1 - {\varphi ^n}} \right)\left( {{h_{t + 1}} - \bar h} \right) + \sum\limits_{s = 1}^n {\left( {4\mu {\tau _z}{E_t}\left( {h_{t + s}^{3/2}} \right) + {\kappa _z}\tilde \mu _{h,s}^{\left( 2 \right)}} \right)} \\
\end{array}$ \\

For the normal GJR, $\sum\limits_{s = 1}^n {{E_t}\left( {r_{t + s}^4} \right)}  = n{\mu ^2}\left( {{\mu ^2} + 6\bar h} \right) + 6{\mu ^2}{\left( {1 - \varphi } \right)^{ - 1}}\left( {1 - {\varphi ^n}} \right)\left( {{h_{t + 1}} - \bar h} \right) + 3\sum\limits_{s = 1}^n {\tilde \mu _{h,s}^{\left( 2 \right)}}$, while for the Student \textit{t} GJR the sum above becomes:  \\$\sum\limits_{s = 1}^n {{E_t}\left( {r_{t + s}^4} \right)}  = n{\mu ^2}\left( {{\mu ^2} + 6\bar h} \right) + 6{\mu ^2}{\left( {1 - \varphi } \right)^{ - 1}}\left( {1 - {\varphi ^n}} \right)\left( {{h_{t + 1}} - \bar h} \right) + 3\frac{{\nu  - 2}}{{\nu  - 4}}\sum\limits_{s = 1}^n {\tilde \mu _{h,s}^{\left( 2 \right)}}$. \\

$\begin{array}{l}
\sum\limits_{s = 1}^n {\sum\limits_{u = 1}^{n - s} {{E_t}\left( {r_{t + s}^3{r_{t + s + u}}} \right)} }  = \sum\limits_{s = 1}^n {\sum\limits_{u = 1}^{n - s} {{E_t}\left( {\left( {{\mu ^3} + 3\left( {{\mu ^2}{\varepsilon _{t + s}} + \mu \varepsilon _{t + s}^2} \right) + \varepsilon _{t + s}^3} \right)\left( {\mu  + {\varepsilon _{t + s + u}}} \right)} \right)} }  \\
\qquad \qquad \qquad \qquad \qquad \, = \frac{{n\left( {n - 1} \right)}}{2}{\mu ^2}\left( {{\mu ^2} + 3\bar h} \right) + 3{\mu ^2}{\left( {1 - \varphi } \right)^{ - 1}}\left[ {n - {{\left( {1 - \varphi } \right)}^{ - 1}}\left( {1 - {\varphi ^n}} \right)} \right]\left( {{h_{t + 1}} - \bar h} \right) \\
\qquad \qquad \qquad \qquad \qquad \, + \mu {\tau _z}\sum\limits_{s = 1}^n {\left( {n - s} \right)\,} {E_t}\left( {h_{t + s}^{3/2}} \right), \\
\end{array}$ \\
which for the normal and Student \textit{t} GJR models becomes: \\
$\sum\limits_{s = 1}^n {\sum\limits_{u = 1}^{n - s} {{E_t}\left( {r_{t + s}^3{r_{t + s + u}}} \right)} }  = \;\frac{{n\left( {n - 1} \right)}}{2}{\mu ^2}\left( {{\mu ^2} + 3\bar h} \right) + 3{\mu ^2}{\left( {1 - \varphi } \right)^{ - 1}}\left[ {n - {{\left( {1 - \varphi } \right)}^{ - 1}}\left( {1 - {\varphi ^n}} \right)} \right]\left( {{h_{t + 1}} - \bar h} \right)$. \\

$\begin{array}{l}
\sum\limits_{s = 1}^n {\sum\limits_{u = 1}^{n - s} {{E_t}\left( {{r_{t + s}}r_{t + s + u}^3} \right)} }  = \sum\limits_{s = 1}^n {\sum\limits_{u = 1}^{n - s} {{E_t}\left( {\left( {\mu  + {\varepsilon _{t + s}}} \right)\left( {{\mu ^3} + 3\left( {{\mu ^2}{\varepsilon _{t + s + u}} + \mu \varepsilon _{t + s + u}^2} \right) + \varepsilon _{t + s + u}^3} \right)} \right)} }  \\
\qquad \qquad \qquad \qquad \quad \; \;= \sum\limits_{s = 1}^n {\sum\limits_{u = 1}^{n - s} {\left( {{\mu ^4} + 3\,{\mu ^2}\tilde \mu _{h,s + u}^{\left( 1 \right)} + \mu {\tau _z}{E_t}\left( {h_{t + s + u}^{3/2}} \right) + 3\mu {E_t}\left( {{\varepsilon _{t + s}}\varepsilon _{t + s + u}^2} \right) + {E_t}\left( {{\varepsilon _{t + s}}\varepsilon _{t + s + u}^3} \right)} \right)} } . \\
\end{array}$ \\
Now, ${E_t}\left( {{\varepsilon _{t + s}}\varepsilon _{t + s + u}^3} \right) = {E_t}\left( {{\varepsilon _{t + s}}{E_{t + s + u - 1}}\left( {z_{t + s + u}^3h_{t + s + u}^{3/2}} \right)} \right) = {\tau _z}{E_t}\left( {{\varepsilon _{t + s}}h_{t + s + u}^{3/2}} \right)=  {\tau _z}\theta _{su}^{\left( {3/2} \right)}$, which for the normal and Student \textit{t} GJR models reduces to: ${E_t}\left( {{\varepsilon _{t + s}}\varepsilon _{t + s + u}^3} \right) = 0$, since ${\tau_z}=0$.\footnote{Even though  for the normal and Student \textit{t} GJR ${\tau_z}=0$ , $\theta _{su}^{\left( {3/2} \right)} = {E_t}\left( {{\varepsilon _{t + s}}h_{t + s + u}^{3/2}} \right)$ is generally non-zero for these models (and enters the expressions of higher moments computed below) and this is why we still consider the normal and Student \textit{t} special cases in the derivation of $\theta _{su}^{\left( {3/2} \right)}$.} To solve for $\theta _{su}^{\left( {3/2} \right)},$
we are using a second order Taylor expansion around $\tilde \mu _{h,s + u}^{\left( 1 \right)}$
, and obtain: \\
$h_{t + s + u}^{3/2} \simeq {\left( {\tilde \mu _{h,s + u}^{\left( 1 \right)}} \right)^{3/2}} + \frac{3}{2}{\left( {\tilde \mu _{h,s + u}^{\left( 1 \right)}} \right)^{1/2}}\left( {{h_{t + s + u}} - \tilde \mu _{h,s + u}^{\left( 1 \right)}} \right) + \frac{3}{8}{\left( {\tilde \mu _{h,s + u}^{\left( 1 \right)}} \right)^{ - 1/2}}{\left( {{h_{t + s + u}} - \tilde \mu _{h,s + u}^{\left( 1 \right)}} \right)^2}$, \\which yields: \\
$\begin{array}{l}
{E_t}\left( {{\varepsilon _{t + s}}\varepsilon _{t + s + u}^3} \right)
= {\tau _z}\theta _{su}^{\left( {3/2} \right)}
= \frac{3}{4}{\tau _z}{\left( {\tilde \mu _{h,s + u}^{\left( 1 \right)}} \right)^{1/2}}\left( {{E_t}\left( {{\varepsilon _{t + s}}{h_{t + s + u}}} \right) + \frac{1}{2}{{\left( {\tilde \mu _{h,s + u}^{\left( 1 \right)}} \right)}^{ - 1}}{E_t}\left( {{\varepsilon _{t + s}}h_{t + s + u}^2} \right)} \right), \\
\end{array}$ \\
$\begin{array}{l}
\theta _{su}^{\left( 2 \right)}
= {E_t}\left( {{\varepsilon _{t + s}}{{\left( {\omega  + \left( {\alpha  + \lambda I_{t + s + u - 1}^ - } \right)\varepsilon _{t + s + u - 1}^2 + \beta {h_{t + s + u - 1}}} \right)}^2}} \right)
= \gamma \theta _{s\left( {u - 1} \right)}^{\left( 2 \right)} + 2\omega \varphi \theta _{s\left( {u - 1} \right)}^{\left( 1 \right)},  \\
\end{array}$ \\
where we have used that, conditional on the information available at time $t$, the indicator function $I_t^ - $ is independent of all (contemporaneous)   $\varepsilon _t^{2k}$ for any natural number $k$. We also set:\footnote{It can be shown, using the Cauchy -- Buniakowsky -- Schwarz inequality, that the kurtosis is always greater than or equal to 1, hence $\kappa_z\geq 1$.\footnote{See Stuart, A. and J.K. Ord, 1994, Kendall's Advanced Theory of Statistics, Vol 1. Distribution Theory, 6th Edition. Arnold, London. p. 109.} Now it can be easily seen that $\gamma > 0$.} \\
$\gamma= {\left( {{\alpha ^2} + \left( {2\alpha \lambda  + {\lambda ^2}} \right){F_0}} \right){\kappa _z} + {\beta ^2} + 2\beta \left( {\alpha  + \lambda {F_0}} \right)}
={\varphi ^2} + \left( {{\kappa _z} - 1} \right){\left( {\alpha  + \lambda {F_0}} \right)^2} + {\kappa _z}{\lambda ^2}{F_0}\left( {1 - {F_0}} \right)$.
If $D(0,1)$is the standard normal distribution, $\gamma$ becomes:
\begin{center}
	$\gamma  = {\varphi ^2} + 2{\left( {\alpha  + \frac{\lambda }{2}} \right)^2} + \frac{3}{4}{\lambda ^2}$.
\end{center}
When $D(0,1)$ is the standardized Student \textit{t} distribution, $\gamma$ is given by:
\begin{center}
	$\gamma  = {\varphi ^2} + \left( {3\frac{{\nu  - 2}}{{\nu  - 4}} - 1} \right){\left( {\alpha  + \frac{\lambda }{2}} \right)^2} + \frac{3}{4}\left( {\frac{{\nu  - 2}}{{\nu  - 4}}} \right){\lambda ^2}$.
\end{center}
Solving the recursion for  $\theta _{su}^{\left( 2 \right)}$, we get: \\$\theta _{su}^{\left( 2 \right)} = {\gamma ^{u - 1}}\theta _{s1}^{\left( 2 \right)} + 2\omega \varphi \sum\limits_{j = 1}^{u - 1} {{\gamma ^{j - 1}}\theta _{s\left( {u - j} \right)}^{\left( 1 \right)}} $.

$\sum\limits_{j = 1}^{u - 1} {{\gamma ^{j - 1}}\theta _{s\left( {u - j} \right)}^{\left( 1 \right)}}  = {c_9}\sum\limits_{j = 1}^{u - 1} {{\gamma ^{j - 1}}{\varphi ^{u - j - 1}}{E_t}\left( {h_{t + s}^{3/2}} \right)}  = {c_9}{\left( {\varphi  - \gamma } \right)^{ - 1}}\left( {{\varphi ^{u - 1}} - {\gamma ^{u - 1}}} \right){E_t}\left( {h_{t + s}^{3/2}} \right)$, where \\ ${c_9} = \left( {\alpha {\tau _ z} + \lambda \int\limits_{x =  - \infty }^0 {{x^3}f\left( x \right)dx} } \right)$. \\
$\begin{array}{l}
\theta _{s1}^{\left( 2 \right)} = {E_t}\left( {{\varepsilon _{t + s}}h_{t + s + 1}^2} \right) = {E_t}\left( {{\varepsilon _{t + s}}{{\left( {\omega  + \left( {\alpha  + \lambda I_{t + s}^ - } \right)\varepsilon _{t + s}^2 + \beta {h_{t + s}}} \right)}^2}} \right) \\
\quad \; \; \,  = \left( {\alpha \left( {\alpha \mu _z^{\left( 5 \right)} + 2\beta {\tau _z}} \right) + \lambda \left( {2\alpha  + \lambda  } \right)\int\limits_{x =  - \infty }^0 {{x^5}f\left( x \right)dx}+ 2 \lambda\beta\int\limits_{x =  - \infty }^0 {{x^3}f\left( x \right)dx} } \right){E_t}\left( {h_{t + s}^{5/2}} \right) \\
\quad \; \; \,  + 2\left( {\omega \alpha {\tau _z} + \lambda \omega \int\limits_{x =  - \infty }^0 {{x^3}f\left( x \right)dx} } \right){E_t}\left( {h_{t + s}^{3/2}} \right) \\
\end{array}$ \\
For the normal GJR, ${\tau_z=\mu_z^{(5)} =0}$, $f\left( z \right) = \varphi \left( z \right)$, and it can be easily shown that
$\int\limits_{z =  - \infty }^0 {{z^5}\varphi \left( z \right)dz}  = - 4\sqrt {\frac{2}{\pi }}$.
Similarly, for the Student \textit{t} GJR, we again have have ${\tau_z=\mu_z^{(5)} =0}$ and $f\left( z \right) = {f_\nu }\left( z \right)$. After some algebraic manipulation, we get that: $\int\limits_{z =  - \infty }^0 {{z^5}} {f_v}\left( z \right) =  - \frac{8}{{\sqrt \pi  }}\frac{{{{\left( {\nu  - 2} \right)}^{5/2}}}}{{\left( {\nu  - 1} \right)\left( {\nu  - 3} \right)\left( {\nu  - 5} \right)}}\frac{{\Gamma \left( {\frac{{\nu  + 1}}{2}} \right)}}{{\Gamma \left( {\frac{\nu }{2}} \right)}}$.

Thus, the final expression for $\theta _{su}^{\left( {3/2} \right)}$ becomes: \\
$\theta _{su}^{\left( {3/2} \right)} = \frac{3}{4}{\left( {\tilde \mu _{h,s + u}^{\left( 1 \right)}} \right)^{1/2}}\left( \begin{array}{l}
{c_9}{\varphi ^{u - 1}}{E_t}\left( {h_{t + s}^{3/2}} \right) + \frac{1}{2}{\left( {\tilde \mu _{h,s + u}^{\left( 1 \right)}} \right)^{ - 1}} \\
\left( {{c_{10}}{\gamma ^{u - 1}}{E_t}\left( {h_{t + s}^{5/2}} \right) + 2\omega {c_9}\left( {\varphi {{\left( {\varphi  - \gamma } \right)}^{ - 1}}\left( {{\varphi ^{u - 1}} - {\gamma ^{u - 1}}} \right) + {\gamma ^{u - 1}}} \right){E_t}\left( {h_{t + s}^{3/2}} \right)} \right) \\
\end{array} \right),$ \\
with ${c_{10}} = \alpha \left( {\alpha \mu _ z^{\left( 5 \right)} + 2\beta {\tau _z}} \right) + \lambda \left( {2\alpha  + \lambda} \right)\int\limits_{x =  - \infty }^0 {{x^5}f\left( x \right)dx} + 2\beta\lambda\int\limits_{x =  - \infty }^0 {{x^3}f\left( x \right)dx}$ and ${E_t}\left( {h_{t + s}^{5/2}} \right)$ is given approximately using a second order Taylor expansion for ${h_{t + s}^{5/2}}$ around ${E_t}\left( {h_{t + s}} \right)$: \\${E_t}\left( {h_{t + s}^{5/2}} \right) \simeq \frac{1}{8}{\left( {\tilde \mu _{h,s}^{\left( 1 \right)}} \right)^{1/2}}\left( {15\tilde \mu _{h,s}^{\left( 2 \right)} - 7{{\left( {\tilde \mu _{h,s}^{\left( 1 \right)}} \right)}^2}} \right)$. \\
$\begin{array}{l}
\sum\limits_{s = 1}^n {\sum\limits_{u = 1}^{n - s} {{E_t}\left( {r_{t + s}^2r_{t + s + u}^2} \right) = } } \sum\limits_{s = 1}^n {\sum\limits_{u = 1}^{n - s} {{E_t}\left[ {{{\left( {\mu  + {\varepsilon _{t + s}}} \right)}^2}{{\left( {\mu  + {\varepsilon _{t + s + u}}} \right)}^2}} \right]} }  \\
\qquad \qquad \qquad \qquad \quad \; \;\, = \sum\limits_{s = 1}^n {\sum\limits_{u = 1}^{n - s} {\left[ {{\mu ^4} + {\mu ^2}\left( {\tilde \mu _{h,s}^{\left( 1 \right)} + \tilde \mu _{h,s + u}^{\left( 1 \right)}} \right) + 2\mu {E_t}\left( {{\varepsilon _{t + s}}\varepsilon _{t + s + u}^2} \right) + {E_t}\left( {\varepsilon _{t + s}^2\varepsilon _{t + s + u}^2} \right)} \right]} } , \\
\end{array}$ \\
$\begin{array}{l}
{E_t}\left( {\varepsilon _{t + s}^2\varepsilon _{t + s + u}^2} \right)
= {E_t}\left( {\varepsilon _{t + s}^2{h_{t + s + u}}} \right)
= \omega \tilde \mu _{h,s}^{\left( 1 \right)} + \varphi {E_t}\left( {\varepsilon _{t + s}^2\varepsilon _{t + s + u - 1}^2} \right) \\
\qquad \qquad \qquad \; \; \, \, = \omega \tilde \mu _{h,s}^{\left( 1 \right)}\left( {{{\left( {1 - \varphi } \right)}^{ - 1}}\left( {1 - {\varphi ^u}} \right) - {\varphi ^{u - 1}}} \right) + {\varphi ^{u - 1}}{E_t}\left( {\varepsilon _{t + s}^2\varepsilon _{t + s + 1}^2} \right), \mathrm{with} \\
\end{array}$ \\
$\begin{array}{l}
{E_t}\left( {\varepsilon _{t + s}^2\varepsilon _{t + s + 1}^2} \right) = {E_t}\left( {\varepsilon _{t + s}^2{h_{t + s + 1}}} \right) = {E_t}\left( {\varepsilon _{t + s}^2\left( {\omega  + \left( {\alpha  + \lambda I_{t + s}^ - } \right)\varepsilon _{t + s}^2 + \beta {h_{t + s}}} \right)} \right) \\
\qquad \quad \quad \quad \quad \; \; \, = \omega \tilde \mu _{h,s}^{\left( 1 \right)} + {\kappa _z}\left( {\alpha  + \lambda {F_0} + \kappa _z^{ - 1}\beta } \right)\tilde \mu _{h,s}^{\left( 2 \right)} \\
\end{array}$ \\
The final expression for ${E_t}\left( {\varepsilon _{t + s}^2\varepsilon _{t + s + u}^2} \right)$ becomes: \\${E_t}\left( {\varepsilon _{t + s}^2\varepsilon _{t + s + u}^2} \right) = \bar h\left( {1 - {\varphi ^u}} \right)\tilde \mu _{h,s}^{\left( 1 \right)} + {\varphi ^{u - 1}}{\kappa _z}\left( {\alpha  + \lambda {F_0} + \kappa _z^{ - 1}\beta } \right)\tilde \mu _{h,s}^{\left( 2 \right)}$. \\The expressions for the normal and Student \textit{t} GJR are obtained by replacing ${\kappa_z=3}$ and ${\kappa_z=3\frac{\left(\nu-2\right)}{\left(\nu-4\right)}}$, respectively, and ${F_0=\frac{1}{2}}$ in the expression above. \\
$\begin{array}{l}
{E_t}\left( {r_{t + s}^2{r_{t + s + u}}{r_{t + s + u + v}}} \right) = {E_t}\left( {r_{t + s}^2{r_{t + s + u}}{E_{t + s + u + v - 1}}\left( {{r_{t + s + u + v}}} \right)} \right) = \mu {E_t}\left( {r_{t + s}^2{r_{t + s + u}}} \right) = {\mu ^2}{E_t}\left( {r_{t + s}^2} \right) \\
\qquad \qquad \qquad \quad \quad \quad \quad \;  = {\mu ^2}\left( {{\mu ^2} + \tilde \mu _{h,s}^{\left( 1 \right)}} \right) \\
\end{array}$

$ {E_t}\left( {{r_{t + s}}r_{t + s + u}^2{r_{t + s + u + v}}} \right)
= \mu {E_t}\left( {\left( {\mu  + {\varepsilon _{t + s}}} \right)\left( {{\mu ^2} + {h_{t + s + u}}} \right)} \right) = {\mu ^4} + {\mu ^2}\tilde \mu _{h,s}^{\left( 1 \right)} + \mu \theta _{su}^{\left( 1 \right)} $

$\begin{array}{l}
{E_t}\left( {{r_{t + s}}{r_{t + s + u}}r_{t + s + u + v}^2} \right) = {E_t}\left( {{r_{t + s}}{r_{t + s + u}}{E_{t + s + u + v - 1}}\left( {{\mu ^2} + 2\mu {\varepsilon _{t + s + u + v}} + \varepsilon _{t + s + u + v}^2} \right)} \right) \\
\qquad \qquad \qquad \qquad \qquad \; = {E_t}\left( {{r_{t + s}}{r_{t + s + u}}\left( {{\mu ^2} + {h_{t + s + u + v}}} \right)} \right) \\
\qquad \qquad \qquad \qquad \qquad \; = {\mu ^4} + {\mu ^2}\tilde \mu _{h,\left( {s + u + v} \right)}^{\left( 1 \right)} + \mu \theta _{s\left( {u + v} \right)}^{\left( 1 \right)} + \mu {E_t}\left( {{\varepsilon _{t + s + u}}{h_{t + s + u + v}}} \right) + {E_t}\left( {{\varepsilon _{t + s}}{\varepsilon _{t + s + u}}{h_{t + s + u + v}}} \right) \\
\end{array}$ \\

${E_t}\left( {{\varepsilon _{t + s + u}}{h_{t + s + u + v}}} \right) = {E_t}\left( {{E_{t + s}}\left( {{\varepsilon _{t + s + u}}{h_{t + s + u + v}}} \right)} \right) = {E_t}\left( {{E_{{t_1}}}\left( {{\varepsilon _{{t_1} + u}}{h_{{t_1} + u + v}}} \right)} \right)$, where ${t_1} = t + s$.

We showed that ${E_t}\left( {{\varepsilon _{t + s}}{h_{t + s + u}}} \right) = \theta _{su}^{\left( 1 \right)} = {c_9}{\varphi ^{u - 1}}{E_t}\left( {h_{t + s}^{3/2}} \right)$. Hence \\
${E_{{t_1}}}\left( {{\varepsilon _{{t_1} + u}}{h_{{t_1} + u + v}}} \right)  = {c_9}{\varphi ^{v - 1}}{E_{{t_1}}}\left( {h_{{t_1} + u}^{3/2}} \right)$ and thus we get: \\
${E_t}\left( {{\varepsilon _{t + s + u}}{h_{t + s + u + v}}} \right) = {c_9}{\varphi ^{v - 1}}{E_t}\left( {{E_{{t_1}}}\left( {h_{{t_1} + u}^{3/2}} \right)} \right) = {c_9}{\varphi ^{v - 1}}{E_t}\left( {h_{t + s + u}^{3/2}} \right)$. \\
$\begin{array}{l}
{E_t}\left( {{\varepsilon _{t + s}}{\varepsilon _{t + s + u}}\varepsilon _{t + s + u + v}^2} \right) = {E_t}\left( {{\varepsilon _{t + s}}{\varepsilon _{t + s + u}}{h_{t + s + u + v}}} \right) = {E_t}\left( {{\varepsilon _{t + s}}{E_{{t_1}}}\left( {{\varepsilon _{{t_1} + u}}{h_{{t_1} + u + v}}} \right)} \right) \\
\qquad \qquad \qquad \qquad \quad \quad \,\, = {c_9}{\varphi ^{v - 1}}{E_t}\left( {{\varepsilon _{t + s}}{E_{{t_1}}}\left( {h_{{t_1} + u}^{3/2}} \right)} \right) = {c_9}{\varphi ^{v - 1}}{E_t}\left( {{\varepsilon _{t + s}}h_{t + s + u}^{3/2}} \right) = {c_9}{\varphi ^{v - 1}}\theta _{su}^{\left( {3/2} \right)} \\
\end{array}$

Finally, repeatedly applying the tower law, we get that: ${E_t}\left( {{r_{t + s}}{r_{t + s + u}}{r_{t + s + u + v}}{r_{t + s + u + v + w}}} \right) = {\mu ^4}.$ \\

(e) \textit{Centred Conditional Moments of Forward and Aggregated Returns} \\
%
The second conditional centred moment of the forward returns (i.e. the conditional variance of the forward return) is:
\begin{center}
	$\mu _{r,s}^{\left( 2 \right)} = {E_t}\left( {\varepsilon _{t + s}^2} \right) = {E_t}\left( {{h_{t + s}}} \right) = \tilde \mu _{h,s}^{\left( 1 \right)} = \bar h + {\varphi ^{s - 1}}\left( {{h_{t + 1}} - \bar h} \right),$
\end{center}
and for the aggregated returns it is:
\begin{center}
	$M_{r,n}^{\left( 2 \right)} = {E_t}\left( {{{\left( {\sum\limits_{s = 1}^n {{\varepsilon _{t + s}}} } \right)}^2}} \right) = \sum\limits_{s = 1}^n {\tilde \mu _{h,s}^{\left( 1 \right)}}  + 2\sum\limits_{s = 1}^n {\sum\limits_{u = 1}^{n - s} {\underbrace {{E_t}\left( {{\varepsilon _{t + s}}{\varepsilon _{t + s + u}}} \right)}_0} }  = \sum\limits_{s = 1}^n {\tilde \mu _{h,s}^{\left( 1 \right)}}$
\end{center}
Hence, \qquad \qquad \qquad \qquad
$ M_{r,n}^{\left( 2 \right)} = n\bar h + {\left( {1 - \varphi } \right)^{ - 1}}\left( {1 - {\varphi ^n}} \right)\left( {{h_{t + 1}} - \bar h} \right).$ \\
The third conditional centred moment of the forward returns is:
\begin{center}
	$\mu _{r,s}^{\left( 3 \right)} = {E_t}\left( {\varepsilon _{t + s}^3} \right) = {\tau _z}{E_t}\left( {h_{t + s}^{3/2}} \right) \simeq \frac{1}{8}{\tau _z}\left( {5{{\left( {\mu _{h,s}^{\left( 1 \right)}} \right)}^{3/2}} + 3\,\mu _{h,s}^{\left( 2 \right)}{{\left( {\mu _{h,s}^{\left( 1 \right)}} \right)}^{ - 1/2}}} \right),$
\end{center}
so in the special cases when the innovation distribution is either the standard normal, or the standardized Student \textit{t}, $\mu _{r,s}^{\left( 3 \right)} = 0$.
The third conditional centred moment of the aggregated returns is:
\begin{center}
	$\begin{array}{l}
	M_{r,n}^{\left( 3 \right)} = {E_t}\left( {{{\left( {\sum\limits_{s = 1}^n {{\varepsilon _{t + s}}} } \right)}^3}} \right) = \sum\limits_{s = 1}^n {{E_t}\left( {\varepsilon _{t + s}^3} \right) + 3} \sum\limits_{s = 1}^n {\sum\limits_{u = 1}^{n - s} {\left[ {{E_t}\left( {\varepsilon _{t + s}^2{\varepsilon _{t + s + u}}} \right) + {E_t}\left( {{\varepsilon _{t + s}}\varepsilon _{t + s + u}^2} \right)} \right]} }  \\
	\quad \;\;\; + 6\sum\limits_{s = 1}^n {\sum\limits_{u = 1}^{n - s} {\sum\limits_{v = 1}^{n - s - u} {{E_t}\left( {{\varepsilon _{t + s}}{\varepsilon _{t + s + u}}{\varepsilon _{t + s + u + v}}} \right)} } }  \\
	\quad \;\;\; = {\tau _z}\sum\limits_{s = 1}^n {{E_t}\left( {h_{t + s}^{3/2}} \right) + } 3\sum\limits_{s = 1}^n {\sum\limits_{u = 1}^{n - s} {{E_t}\left( {{\varepsilon _{t + s}}\varepsilon _{t + s + u}^2} \right)} }  \\
	\quad \;\;\; \simeq \frac{1}{8}{\tau _z}\sum\limits_{s = 1}^n {\left( {5{{\left( {\tilde \mu _{h,s}^{\left( 1 \right)}} \right)}^{3/2}} + 3\tilde \mu _{h,s}^{\left( 2 \right)}{{\left( {\tilde \mu _{h,s}^{\left( 1 \right)}} \right)}^{ - 1/2}}} \right) + } 3\sum\limits_{s = 1}^n {\sum\limits_{u = 1}^{n - s} {{E_t}\left( {{\varepsilon _{t + s}}\varepsilon _{t + s + u}^2} \right)} }; \\
	\end{array}$
\end{center}
for the normal and Student \textit{t} GJR models, this expression simplifies to: $M_{r,n}^{\left( 3 \right)} = 3\sum\limits_{s = 1}^n {\sum\limits_{u = 1}^{n - s} {{E_t}\left( {{\varepsilon _{t + s}}\varepsilon _{t + s + u}^2} \right)} } $.
The fourth conditional centred moment of the forward returns is:
\begin{center}
	$\mu _{r,s}^{\left( 4 \right)} = {E_t}\left( {\varepsilon _{t + s}^4} \right) = {\kappa _z}\tilde \mu _{h,s}^{\left( 2 \right)},$
\end{center}
so in the special case that the innovation distribution is the standard normal, $\mu _{r,s}^{\left( 4 \right)} = 3\tilde \mu _{h,s}^{\left( 2 \right)}$, while for the Student \textit{t} GJR we get: $\mu _{r,s}^{\left( 4 \right)} = 3\frac{{\nu  - 2}}{{\nu  - 4}}\tilde \mu _{h,s}^{\left( 2 \right)}$.
\\
The fourth conditional centred moment of the aggregated returns is: \\
$\begin{array}{l}
M_{r,n}^{\left( 4 \right)} = {E_t}\left( {{{\left( {\sum\limits_{s = 1}^n {{\varepsilon _{t + s}}} } \right)}^4}} \right) \\
\quad \;\;\; = \sum\limits_{s = 1}^n {\varepsilon _{t + s}^4 + } \sum\limits_{s = 1}^n {\sum\limits_{u = 1}^{n - s} {\left( {4\left( {{E_t}\left( {\varepsilon _{t + s}^3{\varepsilon _{t + s + u}}} \right) + {E_t}\left( {{\varepsilon _{t + s}}\varepsilon _{t + s + u}^3} \right)} \right) + 6{E_t}\left( {\varepsilon _{t + s}^2\varepsilon _{t + s + u}^2} \right)} \right)} }  \\
\quad \;\;\; + 12\sum\limits_{s = 1}^n {\sum\limits_{u = 1}^{n - s} {\sum\limits_{v = 1}^{n - s - u} {{E_t}} } } \left( {\varepsilon _{t + s}^2{\varepsilon _{t + s + u}}{\varepsilon _{t + s + u + v}}} \right) + {E_t}\left( {{\varepsilon _{t + s}}\varepsilon _{t + s + u}^2{\varepsilon _{t + s + u + v}}} \right) + {E_t}\left( {{\varepsilon _{t + s}}{\varepsilon _{t + s + u}}\varepsilon _{t + s + u + v}^2} \right) \\
\quad \;\;\; + 24\sum\limits_{s = 1}^n {\sum\limits_{u = 1}^{n - s} {\sum\limits_{v = 1}^{n - s - u} {\sum\limits_{w = 1}^{n - s - u - v} {{E_t}} } } } \left( {{\varepsilon _{t + s}}{\varepsilon _{t + s + u}}{\varepsilon _{t + s + u + v}}{\varepsilon _{t + s + u + v + w}}} \right) \\
\quad \;\;\; = {\kappa _z}\sum\limits_{s = 1}^n {\tilde \mu _{h,s}^{\left( 2 \right)} + \sum\limits_{s = 1}^n {\sum\limits_{u = 1}^{n - s} {\left( {4{E_t}\left( {{\varepsilon _{t + s}}\varepsilon _{t + s + u}^3} \right) + 6{E_t}\left( {\varepsilon _{t + s}^2\varepsilon _{t + s + u}^2} \right)} \right)} } }  + 12\sum\limits_{s = 1}^n {\sum\limits_{u = 1}^{n - s} {\sum\limits_{v = 1}^{n - s - u} {{E_t}\left( {{\varepsilon _{t + s}}{\varepsilon _{t + s + u}}\varepsilon _{t + s + u + v}^2} \right)} } }. \\
\end{array}$ \\
In the special case that the conditional distribution is the standard normal, \\
$M_{r,n}^{\left( 4 \right)} = 3\sum\limits_{s = 1}^n {\tilde \mu _{h,s}^{\left( 2 \right)} + 6\sum\limits_{s = 1}^n {\sum\limits_{u = 1}^{n - s} {{E_t}\left( {\varepsilon _{t + s}^2\varepsilon _{t + s + u}^2} \right)} } }  + 12\sum\limits_{s = 1}^n {\sum\limits_{u = 1}^{n - s} {\sum\limits_{v = 1}^{n - s - u} {{E_t}\left( {{\varepsilon _{t + s}}{\varepsilon _{t + s + u}}\varepsilon _{t + s + u + v}^2} \right)} } }$, \\ while for a Student \textit{t} GJR we obtain, \\
$M_{r,n}^{\left( 4 \right)} = 3\frac{{\nu  - 2}}{{\nu  - 4}}\sum\limits_{s = 1}^n {\tilde \mu _{h,s}^{\left( 2 \right)} + 6\sum\limits_{s = 1}^n {\sum\limits_{u = 1}^{n - s} {{E_t}\left( {\varepsilon _{t + s}^2\varepsilon _{t + s + u}^2} \right)} } }  + 12\sum\limits_{s = 1}^n {\sum\limits_{u = 1}^{n - s} {\sum\limits_{v = 1}^{n - s - u} {{E_t}\left( {{\varepsilon _{t + s}}{\varepsilon _{t + s + u}}\varepsilon _{t + s + u + v}^2} \right)} } }$. \\

(f) \textit{Standardized Conditional Moments of Forward and Aggregated Returns} \\
The skewness of the forward returns is: \\
$\begin{array}{l}
{\tau _{r,s}} = \mu _{r,s}^{\left( 3 \right)}{\left( {\mu _{r,s}^{\left( 2 \right)}} \right)^{ - 3/2}} = {\tau _z}{E_t}\left( {h_{t + s}^{3/2}} \right){\left( {\tilde \mu _{h,s}^{\left( 1 \right)}} \right)^{ - 3/2}} \simeq \frac{1}{8}{\tau _z}\left( {5{{\left( {\tilde \mu _{h,s}^{\left( 1 \right)}} \right)}^{3/2}} + 3\tilde \mu _{h,s}^{\left( 2 \right)}{{\left( {\tilde \mu _{h,s}^{\left( 1 \right)}} \right)}^{ - 1/2}}} \right){\left( {\tilde \mu _{h,s}^{\left( 1 \right)}} \right)^{ - 3/2}} \\
\quad \; \;\,= \frac{1}{8}{\tau _z}\left( {5 + 3\tilde \mu _{h,s}^{\left( 2 \right)}{{\left( {\tilde \mu _{h,s}^{\left( 1 \right)}} \right)}^{ - 2}}} \right). \\
\end{array}$

It can be easily observed that if we used only a first order Taylor series expansion we would obtain ${\tau _{r,s}} \approx {\tau _z}$ and that ${\tau _{r,s}} = 0$ for both the normal and Student \textit{t} GJR. \\
%
The kurtosis of the forward returns is:
\begin{center}
	${\kappa _{r,s}} = \mu _{r,s}^{\left( 4 \right)}{\left( {\mu _{r,s}^{\left( 2 \right)}} \right)^{ - 2}} = {\kappa _z}\tilde \mu _{h,s}^{\left( 2 \right)}{\left( {\tilde \mu _{h,s}^{\left( 1 \right)}} \right)^{ - 2}},$
\end{center}
so in the special cases that the conditional distribution is standard normal ${\kappa _{h,s}} = 3\tilde \mu _{h,s}^{\left( 2 \right)}{\left( {\tilde \mu _{h,s}^{\left( 1 \right)}} \right)^{ - 2}},$ while for the Student \textit{t} GJR we obtain: ${\kappa _{h,s}} = 3\frac{{\nu  - 2}}{{\nu  - 4}}\tilde \mu _{h,s}^{\left( 2 \right)}{\left( {\tilde \mu _{h,s}^{\left( 1 \right)}} \right)^{ - 2}}$.\\

Finally, the skewness and kurtosis of the aggregated returns are:
\begin{center}
	${{\rm T}_{r,n}} = M_{r,n}^{\left( 3 \right)}{\left( {M_{r,n}^{\left( 2 \right)}} \right)^{ - 3/2}}$ and ${{\rm K}_{r,n}} = M_{r,n}^{\left( 4 \right)}{\left( {M_{r,n}^{\left( 2 \right)}} \right)^{ - 2}}$.
\end{center}

\subsection*{T.A.2: Conditional Moments of Forward and Aggregated Variances}
(a) \textit{First Conditional Moments of Forward and Aggregated Variances} \\
Applying the conditional expectation operator to (\ref{varianceeq}), the first un-centred conditional moment of the forward variance may be written:
\begin{center}
	$\tilde \mu _{h,s}^{(1)} = \mu _{r,s}^{\left( 2 \right)} = \bar h + {\varphi ^{s - 1}}\left( {{h_{t + 1}} - \bar h} \right)$.
\end{center}
Similarly, the first un-centred conditional moment of the aggregated variance becomes:
\begin{center}
	$\tilde M_{h,n}^{\left( 1 \right)} = \sum\limits_{s = 1}^n {\tilde \mu _{h,s}^{\left( 1 \right)}}  = n\bar h + {\left( {1 - \varphi } \right)^{ - 1}}\left( {1 - {\varphi ^n}} \right)\left( {{h_{t + 1}} - \bar h} \right)$
\end{center}
or equivalently, in recursive form: $\tilde M_{h,n}^{\left( 1 \right)} = \tilde M_{h,n - 1}^{\left( 1 \right)} + \bar h + {\varphi ^{n - 1}}\left( {{h_{t + 1}} - \bar h} \right)$. \\

(b) \textit{Second Conditional Moments of Forward and Aggregated Variances} \\
The second moment of the forward variance is:
\begin{center}
	$\begin{array}{l}
	\tilde \mu _{h,s}^{\left( 2 \right)} = {E_t}\left( {{{\left( {\omega  + \left( {\alpha  + \lambda I_{t + s - 1}^ - } \right)\varepsilon _{t + s - 1}^2 + \beta {h_{t + s - 1}}} \right)}^2}} \right) \\
	\qquad  = {\omega ^2} + 2\omega \varphi \tilde \mu _{h,s - 1}^{\left( 1 \right)} + \left( {{\varphi ^2} + \left( {{\kappa _z} - 1} \right){{\left( {\alpha  + \lambda {F_0}} \right)}^2} + {\kappa _z}{\lambda ^2}{F_0}\left( {1 - {F_0}} \right)} \right)\tilde \mu _{h,s - 1}^{\left( 2 \right)} \\
	\qquad  = \sum\limits_{i = 1}^{s - 1} {{\gamma ^{i - 1}}} \left( {{\omega ^2} + 2\omega \varphi \left( {\bar h + {\varphi ^{s - i - 1}}\left( {{h_{t + 1}} - \bar h} \right)} \right)} \right) + {\gamma ^{s - 1}}h_{t + 1}^2.
	\end{array}$
\end{center}
When $\gamma = 1$, the expression for the second moment of the forward variance becomes:
\begin{center}
	$\tilde\mu _{h,s}^{\left( 2 \right)} = \left( {s - 1} \right)\left( {{\omega ^2} + 2\omega \varphi \bar h} \right) + 2\varphi \bar h\left( {1 - {\varphi ^{s - 1}}} \right)\left( {{h_{t + 1}} - \bar h} \right) + h_{t + 1}^2$.
\end{center}
For $\gamma\neq 1$  (and $\gamma\neq \varphi$ ), we introduce the following additional notation:
\begin{center}
	${c_1} = \left( {{\omega ^2} + 2\omega \varphi \bar h} \right){\left( {1 - \gamma } \right)^{ - 1}}{\rm{,}}\quad {c_2} = 2\omega \varphi \left( {{h_{t + 1}} - \bar h} \right){\left( {\varphi  - \gamma } \right)^{ - 1}}{\rm{     and    }} \quad {c_3} = {c_1} + {c_2}{\rm{ }}$.
\end{center}
Now the expression for the second moment of variance may be written:
\begin{center}
	$\tilde \mu _{h,s}^{\left( 2 \right)} = {c_1} + \left( {h_{t + 1}^2 - {c_3}} \right){\gamma ^{s - 1}} + {c_2}{\varphi ^{s - 1}}.$
\end{center}
The second moment of the aggregated variance is given by:
\begin{center}
	$\tilde M_{h,n}^{\left( 2 \right)} = {E_t}\left[ {{{\left( {\sum\limits_{s = 1}^n {{h_{t + s}}} } \right)}^2}} \right] = \sum\limits_{s = 1}^n {\tilde \mu _{h,s}^{\left( 2 \right)}}  + 2\sum\limits_{s = 1}^n {\sum\limits_{u = 1}^{n - s} {\tilde \mu _{h,su}^{\left( {1,1} \right)}} }$
\end{center}
$\begin{array}{l}
\tilde \mu _{h,su}^{\left( {1,1} \right)} = {E_t}\left( {{h_{t + s}}\left( {\omega  + \left( {\alpha  + \lambda I_{t + s + u - 1}^ - } \right)\varepsilon _{t + s + u - 1}^2 + \beta {h_{t + s + u - 1}}} \right)} \right) = \omega \tilde \mu _{h,s}^{(1)} + \varphi \tilde \mu _{h,s\left( {u - 1} \right)}^{\left( {1,1} \right)} \\
\quad \quad \; \,
= \bar h\tilde \mu _{h,s}^{(1)} + {\varphi ^u}\left( {\tilde \mu _{h,s}^{(2)} - \bar h\tilde \mu _{h,s}^{(1)}} \right), \\
\end{array}$ \\
hence
\vspace{-0.5cm}
\begin{equation}\label{momagggvar2}
\tilde M_{h,n}^{\left( 2 \right)} = \sum\limits_{s = 1}^n {\left( {\tilde \mu _{h,s}^{\left( 2 \right)} + 2\bar h\left( {n - s} \right)\tilde \mu _{h,s}^{(1)}} \right)}  + 2\sum\limits_{s = 1}^n {\sum\limits_{u = 1}^{n - s} {\left( {{\varphi ^u}\left( {\tilde \mu _{h,s}^{(2)} - \bar h\tilde \mu _{h,s}^{(1)}} \right)} \right)}}.
\end{equation}
Consider the first sum in (\ref{momagggvar2}). For $\gamma \neq 1$ and $\gamma \neq \varphi$ , we have:
\vspace{-0.5cm}
\begin{equation}
\begin{array}{l}
\sum\limits_{s = 1}^n {\tilde \mu _{h,s}^{\left( 2 \right)}}  = \sum\limits_{s = 1}^n {\left( {{c_1} + \left( {h_{t + 1}^2 - {c_3}} \right){\gamma ^{s - 1}} + {c_2}{\varphi ^{s - 1}}} \right)}  \\
\qquad \;\;\;\;{\rm{ }} = n{c_1} + \left( {h_{t + 1}^2 - {c_3}} \right){\left( {1 - \gamma } \right)^{ - 1}}\left( {1 - {\gamma ^n}} \right) + {c_2}{\left( {1 - \varphi } \right)^{ - 1}}\left( {1 - {\varphi ^n}} \right), \\
\end{array}
\vspace{-0.5cm}
\end{equation}
and
\begin{equation}
\begin{array}{l}
\sum\limits_{s = 1}^n {\left( {n - s} \right)\;} \tilde \mu _{h,s}^{(1)} = \sum\limits_{s = 1}^n {\left( {n - s} \right)} \left( {\bar h + {\varphi ^{s - 1}}\left( {{h_{t + 1}} - \bar h} \right)} \right) \\
\qquad \qquad \quad \quad \;\;\, = \frac{{n\left( {n - 1} \right)}}{2}\bar h + {\left( {1 - \varphi } \right)^{ - 1}}\left[ {n - {{\left( {1 - \varphi } \right)}^{ - 1}}\left( {1 - {\varphi ^n}} \right)} \right]\left( {{h_{t + 1}} - \bar h} \right). \\
\end{array}
\label{eq:3.55}
\end{equation}
Next we evaluate the double sum term in (\ref{momagggvar2}). We have: \\
$\begin{array}{l}
\sum\limits_{s = 1}^n {\sum\limits_{u = 1}^{n - s} {{\varphi ^u}\tilde \mu _{h,s}^{\left( 2 \right)}} }  = \varphi {\left( {1 - \varphi } \right)^{ - 1}}\sum\limits_{s = 1}^n {\tilde \mu _{h,s}^{\left( 2 \right)}\left( {1 - {\varphi ^{n - s}}} \right)}  \\
\qquad \qquad \quad \;\;\;\, = \varphi {\left( {1 - \varphi } \right)^{ - 1}}\left[ \begin{array}{l}
n\left( {{c_1} - {c_2}{\varphi ^{n - 1}}} \right) + \left( {{c_2} - {c_1}} \right){\left( {1 - \varphi } \right)^{ - 1}}\left( {1 - {\varphi ^n}} \right) + \left( {h_{t + 1}^2 - {c_3}} \right) \\
\left[ {{{\left( {1 - \gamma } \right)}^{ - 1}}\left( {1 - {\gamma ^n}} \right) - {{\left( {\varphi  - \gamma } \right)}^{ - 1}}\left( {{\varphi ^n} - {\gamma ^n}} \right)} \right] \\
\end{array} \right] \\
\end{array}$ \\
and
\vspace{-0.5cm}
\begin{equation}
\begin{array}{l}
\sum\limits_{s = 1}^n {\sum\limits_{u = 1}^{n - s} {{\varphi ^u}\tilde \mu _{h,s}^{(1)}} }  = \varphi {\left( {1 - \varphi } \right)^{ - 1}}\sum\limits_{s = 1}^n {\left( {1 - {\varphi ^{n - s}}} \right)} \left( {\bar h + {\varphi ^{s - 1}}\left( {{h_{t + 1}} - \bar h} \right)} \right) \\
\qquad \qquad \quad \;\,{\rm{   }}\, = \varphi {\left( {1 - \varphi } \right)^{ - 1}}\left( {n\left( {\bar h - {\varphi ^{n - 1}}\left( {{h_{t + 1}} - \bar h} \right)} \right) + {{\left( {1 - \varphi } \right)}^{ - 1}}\left( {{h_{t + 1}} - 2\bar h} \right)\left( {1 - {\varphi ^n}} \right)} \right). \\
\end{array}
\label{eq:3.56}
\vspace{-0.5cm}
\end{equation}
Thus the final expression for $\tilde M_{h,n}^{\left( 2 \right)}$ is:
\vspace{-0.5cm}
\begin {equation}
\begin{array}{l}
	\tilde M_{h,n}^{\left( 2 \right)} = n{c_1} + \left( {h_{t + 1}^2 - {c_3}} \right){\left( {1 - \gamma } \right)^{ - 1}}\left( {1 - {\gamma ^n}} \right) + {c_2}{\left( {1 - \varphi } \right)^{ - 1}}\left( {1 - {\varphi ^n}} \right) \\
	\qquad\;\, + 2\bar h\left( {\frac{{n\left( {n - 1} \right)}}{2}\bar h + {{\left( {1 - \varphi } \right)}^{ - 1}}\left[ {n - {{\left( {1 - \varphi } \right)}^{ - 1}}\left( {1 - {\varphi ^n}} \right)} \right]\left( {{h_{t + 1}} - \bar h} \right)} \right) \\
	\qquad\;\,  + 2\varphi {\left( {1 - \varphi } \right)^{ - 1}}\left[ \begin{array}{l}
		n\left( {{c_1} - {c_2}{\varphi ^{n - 1}}} \right) + \left( {{c_2} - {c_1}} \right){\left( {1 - \varphi } \right)^{ - 1}}\left( {1 - {\varphi ^n}} \right) + \left( {h_{t + 1}^2 - {c_3}} \right) \\
		\left[ {{{\left( {1 - \gamma } \right)}^{ - 1}}\left( {1 - {\gamma ^n}} \right) - {{\left( {\varphi  - \gamma } \right)}^{ - 1}}\left( {{\varphi ^n} - {\gamma ^n}} \right)} \right] \\
	\end{array} \right] \\
	\qquad\;\,  - 2\bar h\varphi {\left( {1 - \varphi } \right)^{ - 1}}\left( {n\left( {\bar h - {\varphi ^{n - 1}}\left( {{h_{t + 1}} - \bar h} \right)} \right) + {{\left( {1 - \varphi } \right)}^{ - 1}}\left( {{h_{t + 1}} - 2\bar h} \right)\left( {1 - {\varphi ^n}} \right)} \right)\,. \\
\end{array}
\label{eq:3.41}
\end{equation}

(c) \textit{Third Conditional Moments of Forward and Aggregated Variances}
\\
We now consider the third moment of the forward variance: \\
\begin{eqnarray*}
\tilde \mu _{h,s}^{\left( 3 \right)} &=& E_t\left[ \left( \omega  + \left( \alpha  + \lambda I_{t + s - 1}^-  \right)\varepsilon_{t + s - 1}^2 + \beta h_{t + s - 1} \right)^3 \right] \\
&=& \omega^3 + 3\omega^2\varphi \tilde \mu _{h,s - 1}^{(1)} + 3\omega \underbrace {\left[ {{\kappa _z}\left( {{\alpha ^2} + \lambda \left( {2\alpha  + \lambda } \right){F_0}} \right) + {\beta ^2} + 2\beta \left( {\alpha  + \lambda {F_0}} \right)} \right]}_\gamma \tilde \mu _{h,s - 1}^{(2)} \\
&+& \left[ {\mu _z^{\left( 6 \right)}\left( {{\alpha ^3} + 3\alpha \lambda \left( {\alpha  + \lambda } \right){F_0} + {\lambda ^3}{F_0}} \right) + 3\beta {\kappa _z}\left( {{\alpha ^2} + \lambda \left( {2\alpha  + \lambda } \right){F_0}} \right) + 3{\beta ^2}\left( {\alpha  + \lambda {F_0}} \right) + {\beta ^3}} \right]\tilde \mu _{h,s - 1}^{(3)} \\
\end{eqnarray*}
Hence $\quad \tilde \mu _{h,s}^{(3)} = \sum\limits_{i = 0}^{s - 2} {c_4^i} \left( {{\omega ^3} + 3{\omega ^2}\varphi \tilde \mu _{h,s - i - 1}^{(1)} + 3\omega \gamma \tilde \mu _{h,s - i - 1}^{(2)}} \right) + c_4^{s - 1}h_{t + 1}^3,$ where

 ${c_4} = \mu _z^{\left( 6 \right)}\left( {{\alpha ^3} + 3\alpha \lambda \left( {\alpha  + \lambda } \right){F_0} + {\lambda ^3}{F_0}} \right) + 3\beta \gamma  - {\beta ^2}\left( {2\beta  + 3\left( {\alpha  + \lambda {F_0}} \right)} \right)$.\\

For the special case when innovations are normally distributed, $F_0=\frac{1}{2}$ and $ \mu _z^{\left( 6 \right)}$ = 15. Similarly, when innovations are Student \textit{t} distributed, $F_0=\frac{1}{2}$ still and $\mu _z^{\left( 6 \right)} = 15\frac{{{{\left( {\nu  - 2} \right)}^2}}}{{\left( {\nu  - 4} \right)\left( {\nu  - 6} \right)}}$.\\

For ${c_4} \ne 1$ (and ${c_4} \ne \gamma$, ${c_4} \ne \varphi$, and $\gamma \ne 1$), we get that: \\
$ \begin{array}{l}
\tilde \mu _{h,s}^{\left( 3 \right)} = \omega \left( {{\omega ^2} + 3\omega \varphi \bar h + 3\gamma {c_1}} \right){\left( {1 - {c_4}} \right)^{ - 1}} + \left( {3{\omega ^2}\varphi \left( {{h_{t + 1}} - \bar h} \right) + 3\omega \gamma {c_2}} \right){\left( {\varphi  - {c_4}} \right)^{ - 1}}{\varphi ^{s - 1}} \\
\qquad  + {c_{18}}c_4^{s - 1} + 3\omega \gamma \left( { - {c_3} + h_{t + 1}^2} \right){\left( {\gamma  - {c_4}} \right)^{ - 1}}{\gamma ^{s - 1}}\,. \\
\end{array}$ \\
where \\
${c_{18}} = h_{t + 1}^3 - \omega \left( {{\omega ^2} + 3\omega \varphi \bar h + 3\gamma {c_1}} \right){\left( {1 - {c_4}} \right)^{ - 1}} - \left( {3{\omega ^2}\varphi \left( {{h_{t + 1}} - \bar h} \right) + 3\omega \gamma \left( { - {c_1} + h_{t + 1}^2} \right)} \right){\left( {\varphi  - {c_4}} \right)^{ - 1}}$. \\

For $c_4 = 1$ and $\gamma \ne 1$, the expression for $\tilde \mu _{h,s}^{\left( 3 \right)}$
becomes: \\
$\begin{array}{l}
\tilde \mu _{h,s}^{\left( 3 \right)} = \sum\limits_{i = 0}^{s - 2} {\left( {{\omega ^3} + 3{\omega ^2}\varphi \tilde \mu _{h,s - i - 1}^{\left( 1 \right)} + 3\omega \gamma \tilde \mu _{h,s - i - 1}^{\left( 2 \right)}} \right)}  + h_{t + 1}^3 \\
\quad \;\; = h_{t + 1}^3 + \left( {s - 1} \right)\omega \left( {{\omega ^2} + 3\bar h\varphi \omega  + 3\gamma {c_1}} \right) + 3\omega \left[ {\omega \varphi \left( {{h_{t + 1}} - \bar h} \right) + \gamma {c_2}} \right]{\left( {1 - \varphi } \right)^{ - 1}}\left( {1 - {\varphi ^{s - 1}}} \right) \\
\quad \;\; + 3\omega \gamma \left( {h_{t + 1}^2 - {c_3}} \right){\left( {1 - \gamma } \right)^{ - 1}}\left( {1 - {\gamma ^{s - 1}}} \right)\,. \\
\end{array}$

For ${c_4} \ne 1$ and $\gamma  = 1$, we get that: \\
$\begin{array}{l}
\tilde \mu _{h,s}^{\left( 3 \right)} = {\omega ^3}{\left( {1 - {c_4}} \right)^{ - 1}}\left( {1 - c_4^{s - 1}} \right) + 3{\omega ^2}\varphi \sum\limits_{i = 0}^{s - 2} {\left( {c_4^i\left( {\bar h + {\varphi ^{s - i - 2}}\left( {{h_{t + 1}} - \bar h} \right)} \right)} \right)}  \\
\quad  + 3\omega \gamma \sum\limits_{i = 0}^{s - 2} {\left( {c_4^i\left( {\left( {s - i - 2} \right)\left( {{\omega ^2} + 2\omega \varphi \bar h} \right) + 2\varphi \bar h\left( {1 - {\varphi ^{s - i - 2}}} \right)\left( {{h_{t + 1}} - \bar h} \right) + h_{t + 1}^2} \right)} \right)}  + c_4^{s - 1}h_{t + 1}^3 \\
\quad  = \left( {{\omega ^3} + 3{\omega ^2}\varphi \bar h + 6\omega \gamma \varphi \bar h\left( {{h_{t + 1}} - \bar h} \right) + 3\omega \gamma h_{t + 1}^2} \right){\left( {1 - {c_4}} \right)^{ - 1}}\left( {1 - c_4^{s - 1}} \right) \\
\quad  + \left[ {3{\omega ^2}\varphi \left( {{h_{t + 1}} - \bar h} \right) - 6\omega \gamma \varphi \bar h\left( {{h_{t + 1}} - \bar h} \right)} \right]{\left( {\varphi  - {c_4}} \right)^{ - 1}}\left( {{\varphi ^{s - 1}} - c_4^{s - 1}} \right) \\
\quad  + 3\omega \gamma \left( {{\omega ^2} + 2\omega \varphi \bar h} \right)\left( {{{\left( {1 - {c_4}} \right)}^{ - 1}}\left( {s - 2} \right) - {c_4}{{\left( {1 - {c_4}} \right)}^{ - 2}}\left( {1 - c_4^{s - 2}} \right)} \right) + c_4^{s - 1}h_{t + 1}^3. \\
\end{array}$

For the third moment of aggregated variance we write: \\
$\begin{array}{l}
\tilde M_{h,n}^{\left( 3 \right)} = {E_t}\left( {{{\left( {\sum\limits_{s = 1}^n {{h_{t + s}}} } \right)}^3}} \right) = \sum\limits_{s = 1}^n {{E_t}\left( {h_{t + s}^3} \right)} + 3 \sum\limits_{s = 1}^n {\sum\limits_{u = 1}^{n - s} {\left[ {{E_t}\left( {h_{t + s}^2{h_{t + s + u}}} \right) + {E_t}\left( {{h_{t + s}}h_{t + s + u}^2} \right)} \right]} }  \\
\qquad \; + 6\sum\limits_{s = 1}^n {\sum\limits_{u = 1}^{n - s} {\sum\limits_{v = 1}^{n - s - u} {{E_t}\left( {{h_{t + s}}{h_{t + s + u}}{h_{t + s + u + v}}} \right)} } }  \\
\qquad \; = \sum\limits_{s = 1}^n {\tilde \mu _{h,s}^{\left( 3 \right)} + 3} \sum\limits_{s = 1}^n {\sum\limits_{u = 1}^{n - s} {\left( {\tilde \mu _{h,su}^{\left( {1,2} \right)} + \tilde \mu _{h,su}^{\left( {2,1} \right)}} \right)} }  + 6\sum\limits_{s = 1}^n {\sum\limits_{u = 1}^{n - s} {\sum\limits_{v = 1}^{n - s - u} {\tilde \mu _{h,suv}^{\left( {1,1,1} \right)}} } }  \\
\end{array}$ \\
\begin{equation}
\begin{array}{l}
\tilde \mu _{h,su}^{\left( {2,1} \right)}= {E_t}\left( {h_{t + s}^2\left( {\omega  + \left( {\alpha  + \lambda I_{t + s + u - 1}^ - } \right)\varepsilon _{t + s + u - 1}^2 + \beta {h_{t + s + u - 1}}} \right)} \right) = \omega \tilde \mu _{h,s}^{\left( 2 \right)} + \varphi \tilde \mu _{h,s\left( {u - 1} \right)}^{\left( {2,1} \right)} \\
\qquad \; = \bar h\tilde \mu _{h,s}^{\left( 2 \right)} + {\varphi ^u}\left( {\tilde \mu _{h,s}^{\left( 3 \right)} - \bar h\tilde \mu _{h,s}^{\left( 2 \right)}} \right) \\
\end{array} \\
\label{eq:2.41}
\end{equation}
\begin{equation}
\begin{array}{l}
\tilde \mu _{h,su}^{\left( {1,2} \right)} = {E_t}\left( {{h_{t + s}}\left( \begin{array}{l}
{\omega ^2} + \left( {{\alpha ^2} + 2\alpha \lambda I_{t + s + u - 1}^ -  + {\lambda ^2}I_{t + s + u - 1}^ - } \right)\varepsilon _{t + s + u - 1}^4 + {\beta ^2}h_{t + s + u - 1}^2 \\
+ 2\omega \left( {\alpha  + \lambda I_{t + s + u - 1}^ - } \right)\varepsilon _{t + s + u - 1}^2 + 2\omega \beta {h_{t + s + u - 1}} \\ +2\beta \left( {\alpha  + \lambda I_{t + s + u - 1}^ - } \right)\varepsilon _{t + s + u - 1}^2{h_{t + s + u - 1}} \\
\end{array} \right)} \right) \\
\qquad \;\, = {\omega ^2}\tilde \mu _{h,s}^{(1)} + 2\omega \varphi \tilde \mu _{h,s\left( {u - 1} \right)}^{\left( {1,1} \right)} + \gamma \tilde \mu _{h,s\left( {u - 1} \right)}^{\left( {1,2} \right)}.
= \sum\limits_{j = 0}^{u - 1} {{\gamma ^j}} \left( {{\omega ^2}\tilde \mu _{h,s}^{(1)} + 2\omega \varphi \tilde \mu _{h,s\left( {u - j - 1} \right)}^{\left( {1,1} \right)}} \right) + {\gamma ^u}\tilde \mu _{h,s}^{(3)}  \\
\end{array}
\label{eq:2.42}
\end{equation}
$\tilde \mu _{h,suv}^{\left( {1,1,1} \right)} = {E_t}\left( {{h_{t + s}}{E_{t + s}}\left( {{h_{t + s + u}}{h_{t + s + u + v}}} \right)} \right) = {E_t}\left( {{h_{t + s}}{E_{{t_1}}}\left( {{h_{{t_1} + u}}{h_{{t_1} + u + v}}} \right)} \right)$, where $t_1 = t+s$. \\
We have already shown that $\tilde \mu _{h,su}^{\left( {1,1} \right)} = {E_t}\left( {{h_{t + s}}{h_{t + s + u}}} \right) = \bar h\tilde \mu _{h,s}^{(1)} + {\varphi ^u}\left( {\tilde \mu _{h,s}^{(2)} - \bar h\tilde \mu _{h,s}^{(1)}} \right)$, thus:
\\
${E_{{t_1}}}\left( {{h_{{t_1} + u}}{h_{{t_1} + u + v}}} \right) = \bar h\tilde {\tilde \mu} _{h,u}^{\left( 1 \right)} + {\varphi ^v}\left( {\tilde {\tilde \mu} _{h,u}^{\left( 2 \right)} - \bar h\tilde {\tilde \mu} _{h,u}^{\left( 1 \right)}} \right)$, where $\tilde {\tilde \mu }_{h,u}^{\left( 1 \right)} = {E_{{t_1}}}\left( {{h_{{t_1} + u}}} \right)$ and $\tilde {\tilde \mu} _{h,u}^{\left( 2 \right)} = {E_{{t_1}}}\left( {h_{{t_1} + u}^2} \right)$. Since
$
\bar h\tilde \mu _{h,s}^{(1)} + {\varphi ^u}\left( {\tilde \mu _{h,s}^{(2)} - \bar h\tilde \mu _{h,s}^{(1)}} \right)
= \left( {{\varphi ^u}{\gamma ^{s - 1}}} \right)h_{t + 1}^2 + \left( {\bar h{\varphi ^{s - 1}}\left( {1 - {\varphi ^u}} \right) + 2\omega {\varphi ^{u + 1}}{{\left( {\varphi  - \gamma } \right)}^{ - 1}}\left( {{\varphi ^{s - 1}} - {\gamma ^{s - 1}}} \right)} \right){h_{t + 1}} + {{\bar h}^2}\left( {1 - {\varphi ^{s - 1}}} \right)
+ {\varphi ^u}\left( {{c_1} + \left( {2\omega \varphi \bar h{{\left( {\varphi  - \gamma } \right)}^{ - 1}} - {c_1}} \right){\gamma ^{s - 1}} - 2\omega {\varphi ^s}\bar h{{\left( {\varphi  - \gamma } \right)}^{ - 1}} - {{\bar h}^2}\left( {1 - {\varphi ^{s - 1}}} \right)} \right)
$, we get: \\
$\begin{array}{l}
{E_{{t_1}}}\left( {{h_{{t_1} + u}}{h_{{t_1} + u + v}}} \right) = \left( {{\varphi ^v}{\gamma ^{u - 1}}} \right)h_{t + s + 1}^2 + \left( {\bar h{\varphi ^{u - 1}}\left( {1 - {\varphi ^v}} \right) + 2\omega {\varphi ^{v + 1}}{{\left( {\varphi  - \gamma } \right)}^{ - 1}}\left( {{\varphi ^{u - 1}} - {\gamma ^{u - 1}}} \right)} \right){h_{t + s + 1}} \\
\qquad \qquad \quad \quad \quad \quad  + {{\bar h}^2}\left( {1 - {\varphi ^{u - 1}}} \right) + {\varphi ^v}\left( \begin{array}{l}
{c_1} + \left( {2\omega \varphi \bar h{{\left( {\varphi  - \gamma } \right)}^{ - 1}} - {c_1}} \right){\gamma ^{u - 1}} \\
- 2\omega {\varphi ^u}\bar h{\left( {\varphi  - \gamma } \right)^{ - 1}} - {{\bar h}^2}\left( {1 - {\varphi ^{u - 1}}} \right) \\
\end{array} \right), \\
\end{array}$
and the expression for $\tilde \mu _{h,suv}^{\left( {1,1,1} \right)}$ now becomes:
\begin{center}
$\begin{array}{l}
\tilde \mu _{h,suv}^{\left( {1,1,1} \right)} = \left( {{\varphi ^v}{\gamma ^{u - 1}}} \right){E_t}\left( {{h_{t + s}}h_{t + s + 1}^2} \right) \\
\qquad \quad + \left( {\bar h{\varphi ^{u - 1}}\left( {1 - {\varphi ^v}} \right) + 2\omega {\varphi ^{v + 1}}{{\left( {\varphi  - \gamma } \right)}^{ - 1}}\left( {{\varphi ^{u - 1}} - {\gamma ^{u - 1}}} \right)} \right){E_t}\left( {{h_{t + s}}{h_{t + s + 1}}} \right) \\
\qquad \quad + \left[ {{{\bar h}^2}\left( {1 - {\varphi ^{u - 1}}} \right) + {\varphi ^v}\left( \begin{array}{l}
	{c_1} + \left( {2\omega \varphi \bar h{{\left( {\varphi  - \gamma } \right)}^{ - 1}} - {c_1}} \right){\gamma ^{u - 1}} \\
	- 2\omega {\varphi ^u}\bar h{\left( {\varphi  - \gamma } \right)^{ - 1}} - {{\bar h}^2}\left( {1 - {\varphi ^{u - 1}}} \right) \\
\end{array} \right)} \right]\tilde \mu _{h,s}^{(1)}. \\
\end{array}.$
\end{center}
But
$
{E_t}\left( {{h_{t + s}}h_{t + s + 1}^2} \right) = {E_t}\left( {{h_{t + s}}{{\left( {\omega  + \left( {\alpha  + \lambda I_{t + s}^ - } \right)\varepsilon _{t + s}^2 + \beta {h_{t + s}}} \right)}^2}} \right)
= {\omega ^2}\tilde \mu _{h,s}^{(1)} + 2\omega \varphi \tilde \mu _{h,s}^{(2)} + \gamma \tilde \mu _{h,s}^{(3)}$ \\
and
${E_t}\left( {{h_{t + s}}{h_{t + s + 1}}} \right) = {E_t}\left( {{h_{t + s}}\left( {\omega  + \left( {\alpha  + \lambda I_{t + s}^ - } \right)\varepsilon _{t + s}^2 + \beta {h_{t + s}}} \right)} \right) = \omega \tilde \mu _{h,s}^{(1)} + \varphi \tilde \mu _{h,s}^{(2)}$. Hence the final expression for $\tilde \mu _{h,suv}^{\left( {1,1,1} \right)}$ is:
\begin{center}
$\begin{array}{l}
\tilde \mu _{h,suv}^{\left( {1,1,1} \right)} = \left( {{\varphi ^v}{\gamma ^{u - 1}}} \right)\left( {{\omega ^2}\tilde \mu _{h,s}^{(1)} + 2\omega \varphi \tilde \mu _{h,s}^{(2)} + \gamma \tilde \mu _{h,s}^{(3)}} \right) \\
\quad \quad \; \;\;+ \left( {\bar h{\varphi ^{u - 1}}\left( {1 - {\varphi ^v}} \right) + 2\omega {\varphi ^{v + 1}}{{\left( {\varphi  - \gamma } \right)}^{ - 1}}\left( {{\varphi ^{u - 1}} - {\gamma ^{u - 1}}} \right)} \right)\left( {\omega \tilde \mu _{h,s}^{(1)} + \varphi \tilde \mu _{h,s}^{(2)}} \right) \\
\quad \quad \;\;\; + \left[ {{{\bar h}^2}\left( {1 - {\varphi ^{u - 1}}} \right) + {\varphi ^v}\left( \begin{array}{l}
	{c_1} + \left( {2\omega \varphi \bar h{{\left( {\varphi  - \gamma } \right)}^{ - 1}} - {c_1}} \right){\gamma ^{u - 1}} \\
	- 2\omega {\varphi ^u}\bar h{\left( {\varphi  - \gamma } \right)^{ - 1}} - {{\bar h}^2}\left( {1 - {\varphi ^{u - 1}}} \right) \\
\end{array} \right)} \right]\tilde \mu _{h,s}^{(1)}. \\
\end{array}$
\end{center}
%
\vspace{0.5cm}
(d) \textit{Fourth Conditional Moments of Forward and Aggregated Variances} \\
For the fourth moment of the forward variance we write: \\
$\begin{array}{l}
\tilde \mu _{h,s}^{(4)} = {E_t}\left[ {{{\left( {\omega  + \left( {\alpha  + \lambda I_{t + s - 1}^ - } \right)\varepsilon _{t + s - 1}^2 + \beta {h_{t + s - 1}}} \right)}^4}} \right] = {\omega ^4} + 4{\omega ^3}\varphi \tilde \mu _{h,s - 1}^{(1)} + {c_5}\tilde \mu _{h,s - 1}^{(2)} + {c_6}\tilde \mu _{h,s - 1}^{(3)} + {c_7}\tilde \mu _{h,s - 1}^{(4)} \\
\qquad= \sum\limits_{j = 0}^{s - 2} {c_7^j} \left( {{\omega ^4} + 4{\omega ^3}\varphi \tilde \mu _{h,s - j - 1}^{(1)} + {c_5}\tilde \mu _{h,s - j - 1}^{(2)} + {c_6}\tilde \mu _{h,s - j - 1}^{(3)}} \right) + c_7^{s - 1}h_{t + 1}^4 \\
\end{array}$
where \\
${c_5} = 6{\omega ^2}\left( {{\kappa _z}\left( {{\alpha ^2} + \lambda {F_0}\left( {\lambda  + 2\alpha } \right)} \right) + {\beta ^2}} \right) + 12{\omega ^2}\beta \left( {\alpha  + \lambda {F_0}} \right) = 6{\omega ^2}\gamma $ \\
${c_6} = 4\omega \left[ \begin{array}{l}
\mu _z^{\left( 6 \right)}\left( {{\alpha ^3} + {F_0}\left( {{\lambda ^3} + 3\left( {{\alpha ^2}\lambda  + \alpha {\lambda ^2}} \right)} \right)} \right) \\
+ 3{\kappa _z}\beta \left( {{\alpha ^2} + {\lambda ^2}{F_0} + 2\alpha \lambda {F_0}} \right) + 3{\beta ^2}\left( {\alpha  + \lambda {F_0}} \right) + {\beta ^3} \\
\end{array} \right] = 4\omega {c_4}$ \\
$\begin{array}{l}
{c_7} = \mu _z^{\left( 8 \right)}\left( {{\alpha ^4} + {F_0}\left( {{\lambda ^4} + 4\left( {{\alpha ^3}\lambda  + \alpha {\lambda ^3}} \right) + 6{\alpha ^2}{\lambda ^2}} \right)} \right) + {\beta ^4} \\
\quad  + 4\left[ {\mu _z^{\left( 6 \right)}\beta \left( {{\alpha ^3} + {F_0}\left( {{\lambda ^3} + 3\left( {{\alpha ^2}\lambda  + \alpha {\lambda ^2}} \right)} \right)} \right) + {\beta ^3}\left( {\alpha  + \lambda {F_0}} \right)} \right] \\
\quad  + 6{\kappa _z}{\beta ^2}\left( {{\alpha ^2} + {\lambda ^2}{F_0} + 2\alpha \lambda {F_0}} \right). \\
\end{array}$ \\
When the innovations are normally distributed, $\mu _z^{\left( 8 \right)}=105$, while when they are Student \textit{t} distributed $\mu _z^{\left( 8 \right)}=105\frac{{{{\left( {\nu  - 2} \right)}^3}}}{{\left( {\nu  - 4} \right)\left( {\nu  - 6} \right)\left( {\nu  - 8} \right)}}$.
Finally, for the fourth moment of aggrregated variance we write: \\
$\begin{array}{l}
\tilde M_{h,n}^{\left( 4 \right)} = \sum\limits_{s = 1}^n {\tilde \mu _{h,s}^{\left( 4 \right)} + } \sum\limits_{s = 1}^n {\sum\limits_{u = 1}^{n - s} {\left( {4\left( {\tilde \mu _{h,su}^{\left( {3,1} \right)} + \tilde \mu _{h,su}^{\left( {1,3} \right)}} \right) + 6\tilde \mu _{h,su}^{\left( {2,2} \right)}} \right)} }  \\
\qquad \; + 12\sum\limits_{s = 1}^n {\sum\limits_{u = 1}^{n - s} {\sum\limits_{v = 1}^{n - s - u} {\left( {\tilde \mu _{h,suv}^{\left( {2,1,1} \right)} + \tilde \mu _{h,suv}^{\left( {1,2,1} \right)} + \tilde \mu _{h,suv}^{\left( {1,1,2} \right)}} \right)} } }  + 24\sum\limits_{s = 1}^n {\sum\limits_{u = 1}^{n - s} {\sum\limits_{v = 1}^{n - s - u} {\sum\limits_{w = 1}^{n - s - u - v} {\tilde \mu _{h,suvw}^{\left( {1,1,1,1} \right)}} } } }  \\
\end{array}$ \\
$\tilde \mu _{h,su}^{\left( {3,1} \right)} = {E_t}\left( {h_{t + s}^3\left( {\omega  + \left( {\alpha  + \lambda I_{t + s + u - 1}^ - } \right)\varepsilon _{t + s + u - 1}^2 + \beta {h_{t + s + u - 1}}} \right)} \right)=\bar h\tilde \mu _{h,s}^{\left( 3 \right)} + {\varphi ^u}\left( {\tilde \mu _{h,s}^{\left( 4 \right)} - \bar h\tilde \mu _{h,s}^{\left( 3 \right)}} \right)$ \\
$\begin{array}{l}
\tilde \mu _{h,su}^{\left( {1,3} \right)}
= {E_t}\left( {{h_{t + s}}{{\left( {\omega  + \left( {\alpha  + \lambda I_{t + s + u - 1}^ - } \right)\varepsilon _{t + s + u - 1}^2 + \beta {h_{t + s + u - 1}}} \right)}^3}} \right) \\
\quad \quad \;\, = {\omega ^3}\tilde \mu _{h,s}^{\left( 1 \right)} + 3\omega \left( {\omega \varphi \tilde \mu _{h,s\left( {u - 1} \right)}^{\left( {1,1} \right)} + \gamma \tilde \mu _{h,s\left( {u - 1} \right)}^{\left( {1,2} \right)}} \right) + {c_4}\tilde \mu _{h,s\left( {u - 1} \right)}^{\left( {1,3} \right)} \\
\quad \quad \;\, =\sum\limits_{j = 0}^{u - 1} {c_4^j} \left( {{\omega ^3}\tilde \mu _{h,s}^{\left( 1 \right)} + 3\omega \left( {\omega \varphi \tilde \mu _{h,s\left( {u - j - 1} \right)}^{\left( {1,1} \right)} + \gamma \tilde \mu _{h,s\left( {u - j - 1} \right)}^{\left( {1,2} \right)}} \right)} \right) + c_4^u\tilde \mu _{h,s}^{\left( 4 \right)} \\
\end{array}$ \\
$\begin{array}{l}
\tilde \mu _{h,su}^{\left( {2,2} \right)} = {E_t}\left( {h_{t + s}^2{{\left( {\omega  + \left( {\alpha  + \lambda I_{t + s + u - 1}^ - } \right)\varepsilon _{t + s + u - 1}^2 + \beta {h_{t + s + u - 1}}} \right)}^2}} \right) \\
\quad \quad \;  = {\omega ^2}\tilde \mu _{h,s}^{\left( 2 \right)} + 2\omega \varphi \tilde \mu _{h,s\left( {u - 1} \right)}^{\left( {2,1} \right)} + \gamma \tilde \mu _{h,s\left( {u - 1} \right)}^{\left( {2,2} \right)} = \sum\limits_{j = 0}^{u - 1} {{\gamma ^j}\left( {{\omega ^2}\tilde \mu _{h,s}^{\left( 2 \right)} + 2\omega \varphi \tilde \mu _{h,s\left( {u - j - 1} \right)}^{\left( {2,1} \right)}} \right)}  + {\gamma ^u}\tilde \mu _{h,s}^{\left( 4 \right)}. \\
\end{array}$ \\
$\tilde \mu _{h,suv}^{\left( {2,1,1} \right)} = {E_t}\left( {h_{t + s}^2{E_{t + s}}\left( {{h_{t + s + u}}{h_{t + s + u + v}}} \right)} \right) = {E_t}\left( {h_{t + s}^2{E_{{t_1}}}\left( {{h_{{t_1} + u}}{h_{{t_1} + u + v}}} \right)} \right)$, where $t_1=t+s$.

By analogy with $\tilde \mu _{h,suv}^{\left( {1,1,1} \right)}$, we obtain the following expression for $\tilde \mu _{h,suv}^{\left( {2,1,1} \right)}$: \\
$\begin{array}{l}
\tilde \mu _{h,suv}^{\left( {2,1,1} \right)} = \left( {{\varphi ^v}{\gamma ^{u - 1}}} \right)\left( {{\omega ^2}\tilde \mu _{h,s}^{\left( 2 \right)} + 2\omega \varphi \tilde \mu _{h,s}^{\left( 3 \right)} + \gamma \tilde \mu _{h,s}^{\left( 4 \right)}} \right) \\
\qquad \quad \; + \left( {\bar h{\varphi ^{u - 1}}\left( {1 - {\varphi ^v}} \right) + 2\omega {\varphi ^{v + 1}}{{\left( {\varphi  - \gamma } \right)}^{ - 1}}\left( {{\varphi ^{u - 1}} - {\gamma ^{u - 1}}} \right)} \right)\left( {\omega \tilde \mu _{h,s}^{\left( 2 \right)} + \varphi \tilde \mu _{h,s}^{\left( 3 \right)}} \right) \\
\qquad \quad \; + \left[ {{{\bar h}^2}\left( {1 - {\varphi ^{u - 1}}} \right) + {\varphi ^v}\left( \begin{array}{l}
	{c_1} + \left( {2\omega \varphi \bar h{{\left( {\varphi  - \gamma } \right)}^{ - 1}} - {c_1}} \right){\gamma ^{u - 1}} \\
	- 2\omega {\varphi ^u}\bar h{\left( {\varphi  - \gamma } \right)^{ - 1}} - {{\bar h}^2}\left( {1 - {\varphi ^{u - 1}}} \right) \\
\end{array} \right)} \right]\tilde \mu _{h,s}^{\left( 2 \right)}. \\
\end{array}$ \\
Using the same idea (the tower law), we can also solve for $\tilde \mu _{h,suv}^{\left( {1,2,1} \right)}$, $\tilde \mu _{h,suv}^{\left( {1,1,2} \right)}$ and $\tilde \mu _{h,suv}^{\left( {1,1,1,1} \right)}$.

(e) \textit{Centred Conditional Moments of Forward and Aggregated Variances}
\\
The second centred moment of the forward variance, i.e. the conditional variance of the forward conditional variance is:
\begin{center}
$\mu _{h,s}^{\left( 2 \right)} = {E_t}\left( {{{\left( {{h_{t + s}} - \tilde \mu _{h,s}^{\left( 1 \right)}} \right)}^2}} \right) = \tilde \mu _{h,s}^{\left( 2 \right)} - {\left( {\tilde \mu _{h,s}^{\left( 1 \right)}} \right)^2}$.
\end{center}
The second centred moment of the aggregated variance, i.e. the conditional variance of the aggregated conditional variance is:
\vspace{-0.5cm}
\begin{equation}
\begin{array}{l}
M_{h,n}^{\left( 2 \right)} = {E_t}\left( {{{\left( {\sum\limits_{s = 1}^n {\left( {{h_{t + s}} - \tilde \mu _{h,s}^{\left( 1 \right)}} \right)} } \right)}^2}} \right)
= \sum\limits_{s = 1}^n {\left( {\tilde \mu _{h,s}^{\left( 2 \right)} - {{\left( {\tilde \mu _{h,s}^{\left( 1 \right)}} \right)}^2}} \right)}  + 2\sum\limits_{s = 1}^n {\sum\limits_{u = 1}^{n - s} {\left( {\tilde \mu _{h,su}^{\left( {1,1} \right)} - \tilde \mu _{h,s}^{\left( 1 \right)}\tilde \mu _{h,s + u}^{\left( 1 \right)}} \right)} }  \\
\quad \quad \;\, = \sum\limits_{s = 1}^n {\tilde \mu _{h,s}^{\left( 2 \right)}}  + 2\sum\limits_{s = 1}^n {\sum\limits_{u = 1}^{n - s} {{\varphi ^u}\tilde \mu _{h,s}^{\left( 2 \right)}} }  - \sum\limits_{s = 1}^n {{{\left( {\tilde \mu _{h,s}^{\left( 1 \right)}} \right)}^2}}  + 2\sum\limits_{s = 1}^n {\sum\limits_{u = 1}^{n - s} {\left( {1 - {\varphi ^u}} \right)\bar h\tilde \mu _{h,s}^{\left( 1 \right)}} }  - 2\sum\limits_{s = 1}^n {\sum\limits_{u = 1}^{n - s} {\tilde \mu _{h,s}^{\left( 1 \right)}\tilde \mu _{h,s + u}^{\left( 1 \right)}} }
\end{array}
\label{eq:3.54}
\end{equation}
or
\vspace{-0.5cm}
\begin{equation}
M_{h,n}^{\left( 2 \right)} = \tilde M_{h,n}^{\left( 2 \right)} - {\sum\limits_{s = 1}^n {\left( {\tilde \mu _{h,s}^{\left( 1 \right)}} \right)} ^2} - 2\sum\limits_{s = 1}^n {\sum\limits_{u = 1}^{n - s} {\tilde \mu _{h,s}^{\left( 1 \right)}\tilde \mu _{h,s + u}^{\left( 1 \right)}}}.
\label{eq:3.40}
\end{equation}
For $\gamma \neq 1$, the expression for the second moment of the aggregated variance is given by (\ref{eq:3.41}).
\vspace{-0.5cm}
\begin{equation}
\begin{array}{l}
{\sum\limits_{s = 1}^n {\left( {\tilde \mu _{h,s}^{\left( 1 \right)}} \right)} ^2} = \sum\limits_{s = 1}^n {{{\left( {\bar h + {\varphi ^{s - 1}}\left( {{h_{t + 1}} - \bar h} \right)} \right)}^2}}
\\
\quad \quad \quad \;\; = n{{\bar h}^2} + {\left( {{h_{t + 1}} - \bar h} \right)^2}{\left( {1 - {\varphi ^2}} \right)^{ - 1}}\left( {1 - {\varphi ^{2n}}} \right) + 2\bar h\left( {{h_{t + 1}} - \bar h} \right){\left( {1 - \varphi } \right)^{ - 1}}\left( {1 - {\varphi ^n}} \right) \\
\end{array}
\label{eq:3.42}
\end{equation}
\begin{equation}
\begin{array}{l}
\sum\limits_{s = 1}^n {\sum\limits_{u = 1}^{n - s} {\tilde \mu _{h,s}^{\left( 1 \right)}\tilde \mu _{h,s + u}^{\left( 1 \right)}} }  = \sum\limits_{s = 1}^n {\sum\limits_{u = 1}^{n - s} {\left( {\bar h + {\varphi ^{s - 1}}\left( {{h_{t + 1}} - \bar h} \right)} \right)} } \left( {\bar h + {\varphi ^{s + u - 1}}\left( {{h_{t + 1}} - \bar h} \right)} \right) \\
\quad \quad \quad \quad  = \frac{1}{2}n\left( {n - 1} \right){{\bar h}^2} + \bar h\left( {{h_{t + 1}} - \bar h} \right){\left( {1 - \varphi } \right)^{ - 1}}\left[ {n - {{\left( {1 - \varphi } \right)}^{ - 1}}\left( {1 - {\varphi ^n}} \right)} \right] \\
\quad \quad \quad \quad  + \bar h\left( {{h_{t + 1}} - \bar h} \right){\left( {1 - \varphi } \right)^{ - 1}}\left[ {\varphi {{\left( {1 - \varphi } \right)}^{ - 1}}\left( {1 - {\varphi ^n}} \right) - n{\varphi ^n}} \right] \\
\quad \quad \quad \quad  + {\left( {{h_{t + 1}} - \bar h} \right)^2}{\left( {1 - \varphi } \right)^{ - 1}}\left[ {\varphi {{\left( {1 - {\varphi ^2}} \right)}^{ - 1}}\left( {1 - {\varphi ^{2n}}} \right) - {{\left( {1 - \varphi } \right)}^{ - 1}}{\varphi ^n}\left( {1 - {\varphi ^n}} \right)} \right] \\
\end{array}
\label{eq:3.43}
\end{equation}
For $\gamma = 1$, consider the formula in (\ref{eq:3.54}). The expressions for the last three sums do not depend on $\gamma$ and hence remain the same as in the $\gamma \neq 1$ case (see (\ref{eq:3.55}), (\ref{eq:3.56}), (\ref{eq:3.42}) and (\ref{eq:3.43})), while  $\sum\limits_{s = 1}^n {\tilde \mu _{h,s}^{\left( 2 \right)}}$ and $\sum\limits_{s = 1}^n {\sum\limits_{u = 1}^{n - s} {{\varphi ^u}\tilde \mu _{h,s}^{\left( 2 \right)}} }$ become:
\begin{equation}
\begin{array}{l}
\sum\limits_{s = 1}^n {\tilde \mu _{h,s}^{\left( 2 \right)}}  = \sum\limits_{s = 1}^n {\left[ {\left( {s - 1} \right)\left( {{\omega ^2} + 2\omega \varphi \bar h} \right) + 2\varphi \bar h\left( {1 - {\varphi ^{s - 1}}} \right)\left( {{h_{t + 1}} - \bar h} \right) + h_{t + 1}^2} \right]}  \\
\quad \quad \; = \frac{1}{2}n\left( {n - 1} \right)\left( {{\omega ^2} + 2\omega \varphi \bar h} \right) + 2\varphi \bar h\left( {{h_{t + 1}} - \bar h} \right)\left( {n - {{\left( {1 - \varphi } \right)}^{ - 1}}\left( {1 - {\varphi ^n}} \right)} \right) + nh_{t + 1}^2 \\
\end{array}
\label{eq:3.51}
\end{equation}
\begin{equation}
\sum\limits_{s = 1}^n {\sum\limits_{u = 1}^{n - s} {{\varphi ^u}\tilde \mu _{h,s}^{\left( 2 \right)}} }  = \sum\limits_{s = 1}^n {\tilde \mu _{h,s}^{\left( 2 \right)}} \sum\limits_{u = 1}^{n - s} {{\varphi ^u}}
= \varphi {\left( {1 - \varphi } \right)^{ - 1}}\left[ {\sum\limits_{s = 1}^n {\tilde \mu _{h,s}^{\left( 2 \right)}}  - \sum\limits_{s = 1}^n {{\varphi ^{n - s}}\tilde \mu _{h,s}^{\left( 2 \right)}} } \right]
\label{eq:3.52}
\end{equation}
\begin{equation}
\begin{array}{l}
\sum\limits_{s = 1}^n {{\varphi ^{n - s}}\tilde \mu _{h,s}^{\left( 2 \right)} = \sum\limits_{s = 1}^n {{\varphi ^{n - s}}\left[ {\left( {s - 1} \right)\left( {{\omega ^2} + 2\omega \varphi \bar h} \right) + 2\varphi \bar h\left( {1 - {\varphi ^{s - 1}}} \right)\left( {{h_{t + 1}} - \bar h} \right) + h_{t + 1}^2} \right]} }  \\
= \left( {{\omega ^2} + 2\omega \varphi \bar h} \right){\left( {1 - \varphi } \right)^{ - 1}}\left[ {n - {{\left( {1 - \varphi } \right)}^{ - 1}}\left( {1 - {\varphi ^n}} \right)} \right] - 2n\bar h\left( {{h_{t + 1}} - \bar h} \right){\varphi ^n} \\
+ \left[ {2\varphi \bar h\left( {{h_{t + 1}} - \bar h} \right) + h_{t + 1}^2} \right]{\left( {1 - \varphi } \right)^{ - 1}}\left( {1 - {\varphi ^n}} \right)\,. \\
\end{array}
\label{eq:3.53}
\end{equation}
The third centred moment of the forward variance is:
\vspace{-0.5cm}
\begin{equation}
\mu _{h,s}^{\left( 3 \right)} = {E_t}\left( {{{\left( {{h_{t + s}} - \tilde \mu _{h,s}^{\left( 1 \right)}} \right)}^3}} \right)= \tilde \mu _{h,s}^{\left( 3 \right)} - 3\tilde \mu _{h,s}^{\left( 2 \right)}\tilde \mu _{h,s}^{\left( 1 \right)} + 2{\left( {\tilde \mu _{h,s}^{\left( 1 \right)}} \right)^3}
\label{third_centr_mom_fwd_var}
\end{equation}
and the third centred moment of the aggregated variance is: \\
$\begin{array}{l}
M_{h,n}^{\left( 3 \right)} = {E_t}\left( {{{\left( {\sum\limits_{s = 1}^n {\left( {{h_{t + s}} - \tilde \mu _{h,s}^{\left( 1 \right)}} \right)} } \right)}^3}} \right) \\
\quad \quad \;\,= \sum\limits_{s = 1}^n {\left( {\tilde \mu _{h,s}^{\left( 3 \right)} - 3\tilde \mu _{h,s}^{\left( 2 \right)}\tilde \mu _{h,s}^{\left( 1 \right)} + 2{{\left( {\tilde \mu _{h,s}^{\left( 1 \right)}} \right)}^3}} \right)}  + 3\sum\limits_{s = 1}^n {\sum\limits_{u = 1}^{n - s} {\left( \begin{array}{l}
		\tilde \mu _{h,su}^{\left( {2,1} \right)} + \tilde \mu _{h,su}^{\left( {1,2} \right)} + 2\left( {\tilde \mu _{h,s}^{\left( 1 \right)} + \tilde \mu _{h,s + u}^{\left( 1 \right)}} \right) \\
		\left( {\tilde \mu _{h,s}^{\left( 1 \right)}\tilde \mu _{h,s + u}^{\left( 1 \right)} - \tilde \mu _{h,su}^{\left( {1,1} \right)}} \right) - \tilde \mu _{h,s}^{\left( 1 \right)}\tilde \mu _{h,s + u}^{\left( 2 \right)} - \tilde \mu _{h,s + u}^{\left( 1 \right)}\tilde \mu _{h,s}^{\left( 2 \right)} \\
	\end{array} \right)} }  \\
\quad \quad\;\, + 6\sum\limits_{s = 1}^n {\sum\limits_{u = 1}^{n - s} {\sum\limits_{v = 1}^{n - s - u} {\left( \begin{array}{l}
			\tilde \mu _{h,suv}^{\left( {1,1,1} \right)} - \tilde \mu _{h,s}^{\left( 1 \right)}\tilde \mu _{h,\left( {s + u} \right)v}^{\left( {1,1} \right)} - \tilde \mu _{h,\left( {s + u} \right)}^{\left( 1 \right)}\tilde \mu _{h,s\left( {u + v} \right)}^{\left( {1,1} \right)} \\
			- \tilde \mu _{h,\left( {s + u + v} \right)}^{\left( 1 \right)}\tilde \mu _{h,su}^{\left( {1,1} \right)} + 2\tilde \mu _{h,s}^{\left( 1 \right)}\tilde \mu _{h,\left( {s + u} \right)}^{\left( 1 \right)}\tilde \mu _{h,\left( {s + u + v} \right)}^{\left( 1 \right)} \\
		\end{array} \right)} } }.  \\
\end{array}$ \\
The fourth centred moment of the forward variance is:
\begin{center}
$ \mu _{h,s}^{\left( 4 \right)} = \tilde \mu _{h,s}^{\left( 4 \right)} - 4\tilde \mu _{h,s}^{\left( 1 \right)}\tilde \mu _{h,s}^{\left( 3 \right)} + 6{\left( {\tilde \mu _{h,s}^{\left( 1 \right)}} \right)^2}\tilde \mu _{h,s}^{\left( 2 \right)} - 3{\left( {\tilde \mu _{h,s}^{\left( 1 \right)}} \right)^4}.$
\end{center}
Finally, the fourth centred moment of the aggregated variance is: \\
$\begin{array}{l}
M_{h,n}^{\left( 4 \right)} = {E_t}\left( {{{\left( {\sum\limits_{s = 1}^n {\left( {{h_{t + s}} - \tilde \mu _{h,s}^{\left( 1 \right)}} \right)} } \right)}^4}} \right) = \sum\limits_{s = 1}^n {{E_t}\left( {{{\left( {{h_{t + s}} - \tilde \mu _{h,s}^{\left( 1 \right)}} \right)}^4}} \right)}  \\
\quad \quad  + \sum\limits_{s = 1}^n {\sum\limits_{u = 1}^{n - s} {\left( \begin{array}{l}
		4\left( {{E_t}\left( {{{\left( {{h_{t + s}} - \tilde \mu _{h,s}^{\left( 1 \right)}} \right)}^3}\left( {{h_{t + s + u}} - \tilde \mu _{h,s + u}^{\left( 1 \right)}} \right)} \right) + {E_t}\left( {\left( {{h_{t + s}} - \tilde \mu _{h,s}^{\left( 1 \right)}} \right){{\left( {{h_{t + s + u}} - \tilde \mu _{h,s + u}^{\left( 1 \right)}} \right)}^3}} \right)} \right) \\
		+ 6{E_t}\left( {{{\left( {{h_{t + s}} - \tilde \mu _{h,s}^{\left( 1 \right)}} \right)}^2}{{\left( {{h_{t + s + u}} - \tilde \mu _{h,s + u}^{\left( 1 \right)}} \right)}^2}} \right) \\
	\end{array} \right)} }  \\
\quad \quad  + 12\sum\limits_{s = 1}^n {\sum\limits_{u = 1}^{n - s} {\sum\limits_{v = 1}^{n - s - u} {\left[ \begin{array}{l}
			{E_t}\left( {{{\left( {{h_{t + s}} - \tilde \mu _{h,s}^{\left( 1 \right)}} \right)}^2}\left( {{h_{t + s + u}} - \tilde \mu _{h,s + u}^{\left( 1 \right)}} \right)\left( {{h_{t + s + u + v}} - \tilde \mu _{h,s + u + v}^{\left( 1 \right)}} \right)} \right) \\
			+ {E_t}\left( {\left( {{h_{t + s}} - \tilde \mu _{h,s}^{\left( 1 \right)}} \right){{\left( {{h_{t + s + u}} - \tilde \mu _{h,s + u}^{\left( 1 \right)}} \right)}^2}\left( {{h_{t + s + u + v}} - \tilde \mu _{h,s + u + v}^{\left( 1 \right)}} \right)} \right) \\
			+ {E_t}\left( {\left( {{h_{t + s}} - \tilde \mu _{h,s}^{\left( 1 \right)}} \right)\left( {{h_{t + s + u}} - \tilde \mu _{h,s + u}^{\left( 1 \right)}} \right){{\left( {{h_{t + s + u + v}} - \tilde \mu _{h,s + u + v}^{\left( 1 \right)}} \right)}^2}} \right) \\
		\end{array} \right]} } }  \\
\quad \quad  + 24\sum\limits_{s = 1}^n {\sum\limits_{u = 1}^{n - s} {\sum\limits_{v = 1}^{n - s - u} {\sum\limits_{w = 1}^{n - s - u - v} {{E_t}} } } } \left[ \begin{array}{l}
\left( {{h_{t + s}} - \tilde \mu _{h,s}^{\left( 1 \right)}} \right)\left( {{h_{t + s + u}} - \tilde \mu _{h,s + u}^{\left( 1 \right)}} \right) \\
\left( {{h_{t + s + u + v}} - \tilde \mu _{h,s + u + v}^{\left( 1 \right)}} \right)\left( {{h_{t + s + u + v + w}} - \tilde \mu _{h,s + u + v + w}^{\left( 1 \right)}} \right) \\
\end{array} \right]. \\
\end{array}$
Performing the necessary calculations yields the formula for $M_{h,n}^{\left( 4 \right)}$ from Theorem 2. The standardized moments (i.e. skewness and kurtosis) of the forward and aggregated variance distributions are now easily obtained from the central moments, as defined in Section 2.1.

%

\subsection*{T.A.3: Limits of the Moments of Returns}

This appendix derives the limits of the conditional moments of the forward and aggregated returns of the generic GJR model as the time horizon increases. We also outline the results for two important special cases of the generic framework, namely the normal GJR (i.e. $D(0,1)$ is now the standard normal) and also the normal GARCH(1,1) (i.e. $D(0,1)$ is the standard normal and $\lambda = 0$). We only specify the results for these special cases when they differ from the results obtained for the generic model.\\

In what follows we use the notation defined in the beginning of this Appendix.
For the two special cases where $D(0,1)$ is the standard normal, $\tau_z=0$, $F_0 = \frac{1}{2}$  and $\kappa_z = 3$ and  $\varphi$ and $\gamma$ become (for the normal GJR):
\vspace{-0.5cm}
\begin{equation}
\varphi  = \alpha  + \frac{\lambda }{2} + \beta \;{\rm{and}}\;\gamma  = {\varphi ^2} + 2{\left( {\alpha  + \frac{\lambda }{2}} \right)^2} + \frac{3}{4}{\lambda ^2}\,.
\label{eq:3.21}
\vspace{-0.5cm}
\end{equation}
Moreover, for the normal GARCH(1,1), $\lambda = 0$ and the two constants above simplify further:
\vspace{-0.5cm}
\begin{equation}
\vspace{-0.5cm}
\varphi  = \alpha  + \beta \;{\rm{and}}\;\gamma  = {\varphi ^2} + 2{\alpha ^2} = {\left( {\alpha  + \beta } \right)^2} + 2{\alpha ^2}\,.
\label{eq:3.22}
\vspace{-0.5cm}
\end{equation}

We assume $\varphi  \in \left( {0,1} \right)$ and $\varphi \neq \gamma$.

(a) \textit{Limits of the Forward and Aggregated Conditional Variance}
\\
Both the forward variance and the aggregated variance limit, expressed in daily units, are equal to the long term variance, which we have denoted by $\bar h$. That is,
\begin{center}
$\mathop {\lim }\limits_{s \to \infty } \mu _{r,s}^{\left( 2 \right)} =\mathop {\lim }\limits_{s \to \infty } \tilde \mu _{h,s}^{\left( 1 \right)}= \mathop {\lim }\limits_{s \to \infty } \left( {\bar h + {\varphi ^{s - 1}}\left( {{h_{t + 1}} - \bar h} \right)} \right) = \bar h$ \\
$\mathop {\lim }\limits_{n \to \infty } \frac{{M_{r,n}^{\left( 2 \right)}}}{n} = \mathop {\lim }\limits_{n \to \infty } \left[ {\frac{{n\bar h + {{\left( {1 - \varphi } \right)}^{ - 1}}\left( {1 - {\varphi ^n}} \right)\left( {{h_{t + 1}} - \bar h} \right)}}{n}} \right] = \bar h.$
\end{center}
\vspace{0.5cm}
(b) \textit{Limits of the Forward and Aggregated Conditional Skewness} \\
The forward skewness limit is: \\
$\mathop {\lim }\limits_{s \to \infty } {\tau _{r,s}} = \mathop {\lim }\limits_{s \to \infty } \left[ {\frac{1}{8}{\tau _z}\left( {5 + 3\tilde \mu _{h,s}^{\left( 2 \right)}{{\left( {\tilde \mu _{h,s}^{\left( 1 \right)}} \right)}^{ - 2}}} \right)} \right] = \frac{1}{8}{\tau _z}\left( {5 + 3\mathop {\lim }\limits_{s \to \infty } \tilde \mu _{h,s}^{\left( 2 \right)}{{\left( {\mathop {\lim }\limits_{s \to \infty } \tilde \mu _{h,s}^{\left( 1 \right)}} \right)}^{ - 2}}} \right)$ \\
where $\mathop {\lim }\limits_{s \to \infty } \tilde \mu _{h,s}^{\left( 2 \right)} == \left\{ \begin{array}{l}
{c_1}\quad {\rm{if }}\;\gamma  \in \left( {0,1} \right), \\
\infty \quad {\rm{if }}\;\gamma  \in \left[ {1,\infty } \right). \\
\end{array} \right. $ Hence: \\
$\mathop {\lim }\limits_{s \to \infty } {\tau _{r,s}} = \left\{ \begin{array}{l}
\frac{1}{8}{\tau _z}\left( {5 + 3\left( {{\omega ^2} + 2\omega \varphi \bar h} \right){{\left( {1 - \gamma } \right)}^{ - 1}}{{\left( {\bar h} \right)}^{ - 2}}} \right)\quad {\rm{if }}\;\gamma  \in \left( {0,1} \right), \\
{\mathop{\rm sgn}} \left( {{\tau _z}} \right)\infty \qquad \qquad \qquad \qquad \qquad \qquad \qquad  \, \,{\rm{if }}\;\gamma  \in \left[ {1,\infty } \right). \\
\end{array} \right.$

For the normal GJR and the normal GARCH(1,1),
${\tau _{r,s}} = \mathop {\lim }\limits_{s \to \infty } {\tau _{r,s}} = 0$. \\
For the limit of the aggregated skewness set:
\begin{center}
${c_{12}} = \frac{1}{8}\left( {{\tau _z} + 3\left( {\alpha {\tau _z} + \lambda \int\limits_{x =  - \infty }^0 {{x^3}f\left( x \right)dx} } \right){{\left( {1 - \varphi } \right)}^{ - 1}}} \right)$ \\
${c_{13}} = \frac{3}{8}\left( {\left( {\alpha {\tau _z} + \lambda \int\limits_{x =  - \infty }^0 {{x^3}f\left( x \right)dx} } \right){{\left( {1 - \varphi } \right)}^{ - 1}}} \right) = {c_{12}} - \frac{{{\tau _z}}}{8}$.
\end{center}
We have:
\begin{center}
$\begin{array}{l}
\mathop {\lim }\limits_{n \to \infty } {{\rm T}_{r,n}}
= \mathop {\lim }\limits_{n \to \infty } \frac{{M_{r,n}^{\left( 3 \right)}}}{{{n^{3/2}}}}{\left( {\frac{{M_{r,n}^{\left( 2 \right)}}}{n}} \right)^{ - 3/2}} = \mathop {\lim }\limits_{n \to \infty } \frac{{M_{r,n}^{\left( 3 \right)}}}{{{n^{3/2}}}}{\left[ {\mathop {\lim }\limits_{n \to \infty } \left( {\frac{{M_{r,n}^{\left( 2 \right)}}}{n}} \right)} \right]^{ - 3/2}} \\
\qquad \qquad {\rm{            }} = {c_{13}}{{\bar h}^{ - 3/2}}\mathop {\lim }\limits_{n \to \infty } \left[ {{n^{ - 3/2}}\sum\limits_{s = 1}^n {\left( {\frac{{{c_{12}}}}{{{c_{13}}}} - {\varphi ^{n - s}}} \right)} \left( {5{{\left( {\tilde \mu _{h,s}^{\left( 1 \right)}} \right)}^{3/2}} + 3\tilde \mu _{h,s}^{\left( 2 \right)}{{\left( {\tilde \mu _{h,s}^{\left( 1 \right)}} \right)}^{ - 1/2}}} \right)} \right]. \\
\end{array}$
\end{center}
Write	
${S_n} = \frac{{{c_{13}}\sum\limits_{s = 1}^n {\left( {\frac{{{c_{12}}}}{{{c_{13}}}} - {\varphi ^{n - s}}} \right)} \left( {5{{\left( {\tilde \mu _{h,s}^{\left( 1 \right)}} \right)}^{3/2}} + 3\tilde \mu _{h,s}^{\left( 2 \right)}{{\left( {\tilde \mu _{h,s}^{\left( 1 \right)}} \right)}^{ - 1/2}}} \right)}}{{{n^{3/2}}}}$
and
$L = \mathop {\lim }\limits_{n \to \infty } {S_n}$

${S_{1,n}} = \frac{{{c_{13}}\sum\limits_{s = 1}^n {\left( {\frac{{{c_{12}}}}{{{c_{13}}}} - {\varphi ^{n - s}}} \right)\left( {5{{\left( {\mathop {\max }\limits_{1 \le s \le n} \left( {\tilde \mu _{r,s}^{\left( 1 \right)}} \right)} \right)}^{3/2}} + 3\tilde \mu _{r,s}^{\left( 2 \right)}{{\left( {\mathop {\min }\limits_{1 \le s \le n} \left( {\tilde \mu _{r,s}^{\left( 1 \right)}} \right)} \right)}^{ - 1/2}}} \right)} }}{{{n^{3/2}}}}$
and
${L_1} = \mathop {\lim }\limits_{n \to \infty } {S_{1,n}}$

${S_{2,n}} = \frac{{{c_{13}}\sum\limits_{s = 1}^n {\left( {\frac{{{c_{12}}}}{{{c_{13}}}} - {\varphi ^{n - s}}} \right)\left( {5{{\left( {\mathop {\min }\limits_{1 \le s \le n} \left( {\tilde \mu _{r,s}^{\left( 1 \right)}} \right)} \right)}^{3/2}} + 3\tilde \mu _{r,s}^{\left( 2 \right)}{{\left( {\mathop {\max }\limits_{1 \le s \le n} \left( {\tilde \mu _{r,s}^{\left( 1 \right)}} \right)} \right)}^{ - 1/2}}} \right)} }}{{{n^{3/2}}}}$
and
${L_2} = \mathop {\lim }\limits_{n \to \infty } {S_{2,n}}$.\\

\underline{Case 1} \\
${c_{14}} = \frac{{{c_{12}}}}{{{c_{13}}}} \le 0$, then ${{c_{14}} - {\varphi ^{n - s}}} \le 0$ for any $s$. Also, if ${\tau _z} < 0$, then ${S_{1,n}} \le {S_n} \le {S_{2,n}}$ for any $s$.\footnote{If ${\tau _z} > 0$, then ${S_{2,n}} \le {S_n} \le {S_{1,n}}$. However, the limit does not change: the proof above still applies, only that $S_{2,n}$ and $S_{1,n}$ swap place above.} Hence, if ${L_1} = {L_2}$ then $L = {L_1} = {L_2}$, by the squeeze theorem. We now prove that
${L_1} = {L_2}$. Setting
${L_{\max }} = \mathop {\lim }\limits_{n \to \infty } \left( {\mathop {\max }\limits_{1 \le s \le n} \;\tilde \mu _{h,s}^{\left( 1 \right)}} \right) = \max \left( {\bar h,{h_{t + 1}}} \right)$ and ${L_{\min }} = \mathop {\lim }\limits_{n \to \infty } \left( {\mathop {\min }\limits_{1 \le s \le n} \;\tilde \mu _{h,s}^{\left( 1 \right)}} \right) = \min \left( {\bar h,{h_{t + 1}}} \right)$
we may write: \\
${L_1} = {c_{13}}\left[ {5L_{\max }^{3/2}\mathop {\lim }\limits_{n \to \infty } \left( {{n^{ - 1/2}}{c_{14}} - {n^{ - 3/2}}{{\left( {1 - \varphi } \right)}^{ - 1}}\left( {1 - {\varphi ^n}} \right)} \right) + 3L_{\min }^{ - 1/2}\mathop {\lim }\limits_{n \to \infty } {n^{ - 3/2}}\sum\limits_{s = 1}^n {\left( {{c_{14}} - {\varphi ^{n - s}}} \right)\tilde \mu _{h,s}^{\left( 2 \right)}} } \right]$
and the first term above is zero. For $\gamma  \ne 1$, \\
$\begin{array}{l}
\sum\limits_{s = 1}^n {\left( {{c_{14}} - {\varphi ^{n - s}}} \right)\tilde \mu _{h,s}^{\left( 2 \right)}}  = \sum\limits_{s = 1}^n {\left( {{c_{14}} - {\varphi ^{n - s}}} \right)\left( {{c_1} + \left( {h_{t + 1}^2 - {c_3}} \right){\gamma ^{s - 1}} + {c_2}{\varphi ^{s - 1}}} \right)}  \\
\qquad \qquad \qquad \qquad \;\;\; = n{c_1}{c_{14}} + {c_{14}}\left( {h_{t + 1}^2 - {c_3}} \right){\left( {1 - \gamma } \right)^{ - 1}}\left( {1 - {\gamma ^n}} \right) + \left( {{c_2}{c_{14}} - {c_1}} \right){\left( {1 - \varphi } \right)^{ - 1}}\left( {1 - {\varphi ^n}} \right) \\
\qquad \qquad \qquad \qquad \;\;\;  - \left( {h_{t + 1}^2 - {c_3}} \right){\left( {\varphi  - \gamma } \right)^{ - 1}}\left( {{\varphi ^n} - {\gamma ^n}} \right) - n{c_2}{\varphi ^{n - 1}}\,. \\
\end{array}$ \\
For $\gamma  = 1$, \\
$\begin{array}{l}
\sum\limits_{s = 1}^n {\left( {{c_{14}} - {\varphi ^{n - s}}} \right)\tilde \mu _{h,s}^{\left( 2 \right)}}  = \sum\limits_{s = 1}^n {\left( {{c_{14}} - {\varphi ^{n - s}}} \right)\left[ \begin{array}{l}
\left( {s - 1} \right)\left( {{\omega ^2} + 2\omega \varphi \bar h} \right) \\
+ 2\varphi \bar h\left( {1 - {\varphi ^{s - 1}}} \right)\left( {{h_{t + 1}} - \bar h} \right) + \tilde \mu _{h,1}^{\left( 2 \right)} \\
\end{array} \right]}  \\
\quad \quad \quad \quad \quad \quad  = \raise.5ex\hbox{$\scriptstyle 1$}\kern-.1em/
\kern-.15em\lower.25ex\hbox{$\scriptstyle 2$} {\rm{ }}n\left( {n - 1} \right){c_{14}}\left( {{\omega ^2} + 2\omega \varphi {h_0}} \right)\bar h - \varphi \left( {{\omega ^2} + 2\omega \varphi \bar h} \right){\left( {1 - \varphi } \right)^{ - 1}} \\
\quad \quad \quad \quad \quad \quad \left[ {{{\left( {1 - \varphi } \right)}^{ - 1}}\left( {1 - {\varphi ^n}} \right) - {\varphi ^{ - 1}}\left( {n - 1} \right)} \right] + 2{c_{14}}\varphi {h_0}\left( {{h_{t + 1}} - \bar h} \right)\left( {n - {{\left( {1 - \varphi } \right)}^{ - 1}}\left( {1 - {\varphi ^n}} \right)} \right) \\
\quad \quad \quad \quad \quad \quad  - 2\varphi \bar h\left( {{h_{t + 1}} - \bar h} \right)\left[ {{{\left( {1 - \varphi } \right)}^{ - 1}}\left( {1 - {\varphi ^n}} \right) - n{\varphi ^{n - 1}}} \right] + \tilde \mu _{h,1}^{\left( 2 \right)}\left( {n{c_{14}} - {{\left( {1 - \varphi } \right)}^{ - 1}}\left( {1 - {\varphi ^n}} \right)} \right). \\
\end{array}$ \\
Hence	
\begin{center}
${L_1} = \left\{ \begin{array}{l}
0\qquad \qquad \qquad \qquad \qquad \qquad \qquad \qquad \qquad \qquad \qquad \qquad \; \,{\rm{if }}\;\gamma  \in \left( {0,1} \right), \\
{\mathop{\rm sgn}} \left( {{c_{13}}} \right){\mathop{\rm sgn}} \left( {{c_{14}}\left( {{\omega ^2} + 2\omega \varphi \bar h} \right)} \right)\infty \qquad \qquad \qquad \qquad \qquad \; \; \, {\rm{if }}\;\gamma  = 1, \\
{\mathop{\rm sgn}} \left( {{c_{13}}} \right){\mathop{\rm sgn}} \left[ {\left( {{{\left( {\varphi  - \gamma } \right)}^{ - 1}} - {c_{14}}{{\left( {1 - \gamma } \right)}^{ - 1}}} \right)\left( {h_{t + 1}^2 - {c_3}} \right)} \right]\infty \quad \; \, {\rm{if }}\;\gamma  \in \left( {1,\infty } \right). \\
\end{array} \right.$
\end{center}

Now, for $\gamma  \in \left[ {1,\infty } \right)$,\; ${\mathop{\rm sgn}} \left( {h_{t + 1}^2 - {c_3}} \right) = 1$, \; ${\mathop{\rm sgn}} \left( {{{\left( {\varphi  - \gamma } \right)}^{ - 1}} - {c_{14}}{{\left( {1 - \gamma } \right)}^{ - 1}}} \right) =  - 1$ and \\${\mathop{\rm sgn}} \left( {{\omega ^2} + 2\omega \varphi {h_0}} \right) = 1$, so
${L_1} = \left\{ \begin{array}{l}
0 \qquad \quad \quad\quad \quad \;{\rm{if }}\gamma  \in \left( {0,1} \right), \\
{\mathop{\rm sgn}} \left( {{c_{12}}} \right)\infty \quad \;\;\;\;{\rm{if }}\gamma  \in \left[ {1,\infty } \right). \\
\end{array} \right.$

Analogously it can be shown that ${L_2}$ is the same limit, hence ${L_1} = {L_2}$ and finally:
\begin{center}
$\mathop {\lim }\limits_{n \to \infty } {{\rm T}_{r,n}} = \left\{ \begin{array}{l}
0\quad \quad \quad \quad \quad \; \; \; \;{\rm{if }}\gamma  \in \left( {0,1} \right), \\
{\mathop{\rm sgn}} \left( {{c_{12}}} \right)\infty \quad \;\;\;{\rm{if }}\gamma  \in \left[ {1,\infty } \right). \\
\end{array} \right.$
\end{center}
\underline{Case 2} \\
${c_{14}} > 0$: In this case $\exists $ an integer $\tilde s \ge 1$ such that ${{c_{14}} - {\varphi ^{n - s}}} > 0 \; {\rm{ for }} \; \forall s > \tilde s$ and \\ $ {{c_{14}} - {\varphi ^{n - s}}}  \le 0 \; {\rm{ for }} \; \forall \;\tilde s \ge s \ge 1$. It can be easily seen that if
${c_{14}} > \varphi $, then $\tilde s = 0.$

Write: ${S_n} = S_n^{\left( 1 \right)} + S_n^{\left( 2 \right)}$, ${\tilde S_{1,n}} = S_{1,n}^{\left( 1 \right)} + S_{2,n}^{\left( 2 \right)}$,
${\tilde S_{2,n}} = S_{2,n}^{\left( 1 \right)} + S_{1,n}^{\left( 2 \right)}$ where \\
$S_n^{\left( 1 \right)} = \frac{{{c_{13}}\sum\limits_{s = 1}^{\tilde s} {\left( {{c_{14}} - {\varphi ^{n - s}}} \right)\left( {5{{\left( {\tilde \mu _{h,s}^{\left( 1 \right)}} \right)}^{3/2}} + 3\tilde \mu _{h,s}^{\left( 2 \right)}{{\left( {\tilde \mu _{h,s}^{\left( 1 \right)}} \right)}^{ - 1/2}}} \right)} }}{{{n^{3/2}}}}$,
$S_n^{\left( 2 \right)} = \frac{{{c_{13}}\sum\limits_{s = \tilde s + 1}^n {\left( {{c_{14}} - {\varphi ^{n - s}}} \right)\left( {5{{\left( {\tilde \mu _{h,s}^{\left( 1 \right)}} \right)}^{3/2}} + 3\tilde \mu _{h,s}^{\left( 2 \right)}{{\left( {\tilde \mu _{h,s}^{\left( 1 \right)}} \right)}^{ - 1/2}}} \right)} }}{{{n^{3/2}}}}$ \\
$S_{1,n}^{\left( 1 \right)} = \frac{{{c_{13}}\sum\limits_{s = 1}^{\tilde s} {\left( {{c_{14}} - {\varphi ^{n - s}}} \right)\left( {5{{\left( {\mathop {\max }\limits_{1 \le s \le n} \left( {\tilde \mu _{r,s}^{\left( 1 \right)}} \right)} \right)}^{3/2}} + 3\tilde \mu _{r,s}^{\left( 2 \right)}{{\left( {\mathop {\min }\limits_{1 \le s \le n} \left( {\tilde \mu _{r,s}^{\left( 1 \right)}} \right)} \right)}^{ - 1/2}}} \right)} }}{{{n^{3/2}}}}$, \\
$S_{2,n}^{\left( 2 \right)} = \frac{{{c_{13}}\sum\limits_{s = \tilde s + 1}^n {\left( {{c_{14}} - {\varphi ^{n - s}}} \right)\left( {5{{\left( {\mathop {\min }\limits_{1 \le s \le n} \left( {\tilde \mu _{r,s}^{\left( 1 \right)}} \right)} \right)}^{3/2}} + 3\tilde \mu _{r,s}^{\left( 2 \right)}{{\left( {\mathop {\max }\limits_{1 \le s \le n} \left( {\tilde \mu _{r,s}^{\left( 1 \right)}} \right)} \right)}^{ - 1/2}}} \right)} }}{{{n^{3/2}}}}$ \\
$S_{2,n}^{\left( 1 \right)} = \frac{{{c_{13}}\sum\limits_{s = 1}^{\tilde s} {\left( {{c_{14}} - {\varphi ^{n - s}}} \right)\left( {5{{\left( {\mathop {\min }\limits_{1 \le s \le n} \left( {\tilde \mu _{r,s}^{\left( 1 \right)}} \right)} \right)}^{3/2}} + 3\tilde \mu _{r,s}^{\left( 2 \right)}{{\left( {\mathop {\max }\limits_{1 \le s \le n} \left( {\tilde \mu _{r,s}^{\left( 1 \right)}} \right)} \right)}^{ - 1/2}}} \right)} }}{{{n^{3/2}}}}$ and \\
$S_{1,n}^{\left( 2 \right)} = \frac{{{c_{13}}\sum\limits_{s = \tilde s + 1}^n {\left( {{c_{14}} - {\varphi ^{n - s}}} \right)\left( {5{{\left( {\mathop {\max }\limits_{1 \le s \le n} \left( {\tilde \mu _{r,s}^{\left( 1 \right)}} \right)} \right)}^{3/2}} + 3\tilde \mu _{r,s}^{\left( 2 \right)}{{\left( {\mathop {\min }\limits_{1 \le s \le n} \left( {\tilde \mu _{r,s}^{\left( 1 \right)}} \right)} \right)}^{ - 1/2}}} \right)} }}{{{n^{3/2}}}}\,.$

If we solve further for $S_{1,n}^{\left( 1 \right)}$, we get: \\
$S_{1,n}^{\left( 1 \right)} = {c_{13}}\left[ \begin{array}{l}
5{\left( {\mathop {\max }\limits_{1 \le s \le n} \left( {\tilde \mu _{h,s}^{\left( 1 \right)}} \right)} \right)^{3/2}}{n^{ - 3/2}}\left( {\tilde s{c_{14}} - {\varphi ^{n - \tilde s}}\sum\limits_{s = 1}^{\tilde s} {{\varphi ^{\tilde s - s}}} } \right) \\    + 3{\left( {\mathop {\min }\limits_{1 \le s \le n} \left( {\tilde \mu _{h,s}^{\left( 1 \right)}} \right)} \right)^{ - 1/2}}{n^{ - 3/2}}\left( {{c_{14}}\sum\limits_{s = 1}^{\tilde s} {\tilde \mu _{h,s}^{\left( 2 \right)}}  - {\varphi ^{n - \tilde s}}\sum\limits_{s = 1}^{\tilde s} {{\varphi ^{\tilde s - s}}\tilde \mu _{h,s}^{\left( 2 \right)}} } \right) \\
\end{array} \right].$ \\
Define: $f_1^{a,b} = \sum\limits_{s = a}^b {{\varphi ^{b - s}}}$,\quad
$f_2^{a,b} = \sum\limits_{s = a}^b {\tilde \mu _{h,s}^{\left( 2 \right)}}$ \quad
and \quad $f_3^{a,b} = \sum\limits_{s = a}^b {{\varphi ^{b - s}}} \tilde \mu _{h,s}^{\left( 2 \right)}$. $S_{1,n}^{\left( 1 \right)}$ becomes: \\
$S_{1,n}^{\left( 1 \right)} = {c_{13}}\left[ {5{{\left( {\mathop {\max }\limits_{1 \le s \le n} \left( {\tilde \mu _{h,s}^{\left( 1 \right)}} \right)} \right)}^{3/2}}{n^{ - 3/2}}\left[ {\tilde s{c_{14}} - {\varphi ^{n - \tilde s}}f_1^{1,\tilde s}} \right] + 3{{\left( {\mathop {\min }\limits_{1 \le s \le n} \left( {\tilde \mu _{h,s}^{\left( 1 \right)}} \right)} \right)}^{ - 1/2}}{n^{ - 3/2}}\left[ {f_2^{1,\tilde s} - {\varphi ^{n - \tilde s}}f_3^{1,\tilde s}} \right]} \right],$
where $f_j^{1,\tilde s}$ $j = 1, 2, 3$ are all constant w.r.t. $n$. Thus
$\mathop {\lim }\limits_{n \to \infty } S_{1,n}^{\left( 1 \right)} = 0$.
Also \\
$S_{2,n}^{\left( 2 \right)} = {c_{13}}\left[ \begin{array}{l}
5{\left( {\mathop {\min }\limits_{1 < s \le n} \left( {\tilde \mu _{h,s}^{\left( 1 \right)}} \right)} \right)^{3/2}}{n^{ - 3/2}}\left( {\left( {n - \tilde s} \right){c_{14}} - f_1^{\tilde s + 1,n}} \right) \\
+ 3{\left( {\mathop {\max }\limits_{1 < s \le n} \left( {\tilde \mu _{h,s}^{\left( 1 \right)}} \right)} \right)^{ - 1/2}}{n^{ - 3/2}}\left( {{c_{14}}f_2^{\tilde s + 1,n} - f_3^{\tilde s + 1,n}} \right) \\
\end{array} \right].$ \\
Now $f_1^{\tilde s + 1,n} = {\left( {1 - \varphi } \right)^{ - 1}}\left( {1 - {\varphi ^{n - \tilde s}}} \right)$. Also, for $\gamma \ne 1$,

$\begin{array}{l}
f_2^{\tilde s + 1,n} = \sum\limits_{s = \tilde s + 1}^n {\left( {{c_1} + \left( {h_{t + 1}^2 - {c_3}} \right){\gamma ^{s - 1}} + {c_2}{\varphi ^{s - 1}}} \right)}  \\
\qquad \; \; {\rm{        }} = \left( {n - \tilde s} \right){c_1} + \left( {h_{t + 1}^2 - {c_3}} \right){\gamma ^{\tilde s}}{\left( {1 - \gamma } \right)^{ - 1}}\left( {1 - {\gamma ^{n - \tilde s}}} \right) + {c_2}{\varphi ^{\tilde s}}{\left( {1 - \varphi } \right)^{ - 1}}\left( {1 - {\varphi ^{n - \tilde s}}} \right), \\
\end{array}$ \\
$\begin{array}{l}
f_3^{\tilde s + 1,n} = \sum\limits_{s = \tilde s + 1}^n {{\varphi ^{n - s}}\left( {{c_1} + \left( {h_{t + 1}^2 - {c_3}} \right){\gamma ^{s - 1}} + {c_2}{\varphi ^{s - 1}}} \right)}  \\
\quad \quad \; \; \; = {c_1}{\left( {1 - \varphi } \right)^{ - 1}}\left( {1 - {\varphi ^{n - \tilde s}}} \right) + \left( {h_{t + 1}^2 - {c_3}} \right){\gamma ^{\tilde s}}{\left( {\varphi  - \gamma } \right)^{ - 1}}\left( {{\varphi ^{n - \tilde s}} - {\gamma ^{n - \tilde s}}} \right) + {c_2}\left( {n - \tilde s} \right){\varphi ^{n - 1}}\,. \\
\end{array}$ \\

For $\gamma = 1$, \\
$\begin{array}{l}
f_2^{\tilde s + 1,n} = \sum\limits_{s = \tilde s + 1}^n {\left( {\left( {s - 1} \right)\left( {{\omega ^2} + 2\omega \varphi \bar h} \right) + 2\varphi \bar h\left( {1 - {\varphi ^{s - 1}}} \right)\left( {{h_{t + 1}} - \bar h} \right) + h_{t + 1}^2} \right)}  \\
\quad \quad \; \; \;  = \frac{1}{2}\left( {{\omega ^2} + 2\omega \varphi \bar h} \right)\left( {n - \tilde s} \right)\left( {n - \tilde s - 1} \right) - 2\varphi \bar h\left( {{h_{t + 1}} - \bar h} \right){\varphi ^{\tilde s}}{\left( {1 - \varphi } \right)^{ - 1}}\left( {1 - {\varphi ^{n - \tilde s}}} \right) \\
\quad \quad \; \; \;  + \left( {2\varphi \bar h\left( {{h_{t + 1}} - \bar h} \right) + h_{t + 1}^2} \right)\left( {n - \tilde s} \right)\,, \\
\end{array}$ \\
$\begin{array}{l}
f_3^{\tilde s + 1,n} = \sum\limits_{s = \tilde s + 1}^n {{\varphi ^{n - s}}} \tilde \mu _{h,s}^{\left( 2 \right)} = \sum\limits_{s = \tilde s + 1}^n {{\varphi ^{n - s}}} \left( {\left( {s - 1} \right)\left( {{\omega ^2} + 2\omega \varphi \bar h} \right) + 2\varphi {h_0}\left( {1 - {\varphi ^{s - 1}}} \right)\left( {{h_{t + 1}} - \bar h} \right) + h_{t + 1}^2} \right) \\
\quad \quad \; \; \;  = \left( {{\omega ^2} + 2\omega \varphi \bar h} \right){\left( {1 - \varphi } \right)^{ - 1}}\left( {\varphi {{\left( {1 - \varphi } \right)}^{ - 1}}\left( {1 - {\varphi ^{n - \tilde s}}} \right) - \tilde s{\varphi ^{n - \tilde s}}} \right) \\
\quad \quad \; \; \;   + \left[ {2\varphi \bar h\left( {{h_{t + 1}} - \bar h} \right) + h_{t + 1}^2} \right]{\left( {1 - \varphi } \right)^{ - 1}}\left( {1 - {\varphi ^{n - \tilde s}}} \right) - \left( {2\varphi \bar h\left( {{h_{t + 1}} - \bar h} \right)} \right)\left( {n - \tilde s} \right){\varphi ^{n - 1}}\,. \\
\end{array}$

Hence: \\
$\begin{array}{l}
\mathop {\lim }\limits_{n \to \infty } {\tilde S_{1,n}} = \mathop {\lim }\limits_{n \to \infty } S_{2,n}^{\left( 2 \right)} = \left\{ \begin{array}{l}
0\qquad \qquad \qquad \qquad \qquad \qquad \qquad \qquad \qquad \qquad \qquad \;\; \, \,{\rm{if }} \; \gamma  \in \left( {0,1} \right), \\
{\mathop{\rm sgn}} \left( {{c_{13}}} \right)\infty \qquad \qquad \qquad \qquad \qquad \qquad \qquad \qquad \qquad \; \; \; {\rm{if }} \; \gamma  = 1, \\
{\mathop{\rm sgn}} \left[ {{c_{13}}\left( {h_{t + 1}^2 - {c_3}} \right)\left( { - {c_{14}}{{\left( {1 - \gamma } \right)}^{ - 1}} + {{\left( {\varphi  - \gamma } \right)}^{ - 1}}} \right)} \right]\infty \quad \; \; {\rm{if }} \; \gamma  \in \left( {1,\infty } \right), \\
\end{array} \right. \\
\qquad \qquad = \left\{ \begin{array}{l}
0\qquad \qquad \qquad \qquad \qquad \qquad \qquad \qquad \qquad \qquad \qquad \;\; \, \,{\rm{if }} \; \gamma  \in \left( {0,1} \right), \\
{\mathop{\rm sgn}} \left( {{\tau _z}\left( {\alpha  + \frac{{\gamma  - \varphi }}{3}} \right) + \lambda \int\limits_{x =  - \infty }^0 {{x^3}f\left( x \right)dx} } \right)\infty \quad \qquad \quad \quad \; {\rm{if}}\;\gamma  \in \left[ {1,\infty } \right). \\
\end{array} \right. \\
\end {array} $
Analogously it can be shown that $\mathop {\lim }\limits_{n \to \infty } {\tilde S_{2,n}}$ is identical to $\mathop {\lim }\limits_{n \to \infty } {\tilde S_{1,n}}$. Using $\min \left( {{{\tilde S}_{1,n}},{{\tilde S}_{2,n}}} \right) \le {S_n} \le \max \left( {{{\tilde S}_{1,n}},{{\tilde S}_{2,n}}} \right)$\footnote{If ${c_{13}} > 0$, then ${\tilde S_{1,n}} \le {S_n} \le {\tilde S_{2,n}}$, whereas if ${c_{13}} < 0$, the inequality is reversed. However, this does not change the proof above.} and the squeeze theorem,\footnote{It can be easily noticed that
${\mathop{\rm sgn}} \left( { - {c_{12}}\left( {\varphi  - \gamma } \right) + {c_{13}}\left( {1 - \gamma } \right)} \right) = {\mathop{\rm sgn}} \left( {{c_{12}}} \right)$, for ${c_{14}} < 0$. Hence the limits of aggregated skewness are the same, regardless of $c_{14}$ being greater than or less than 0.}
\begin{center}
$\mathop {\lim }\limits_{n \to \infty } {{\rm T}_{r,n}} = \left\{ \begin{array}{l}
0\qquad \qquad \qquad \qquad \qquad \qquad \quad \quad \quad \quad \quad \quad \quad \quad \quad \quad \;{\rm{if }} \; \gamma  \in \left( {0,1} \right), \\
{\mathop{\rm sgn}} \left( {{\tau _z}\left( {\alpha  + \frac{{\gamma  - \varphi }}{3}} \right) + \lambda \int\limits_{x =  - \infty }^0 {{x^3}f\left( x \right)dx} } \right)\infty \qquad \qquad \;\; {\rm{if}}\;\gamma  \in \left[ {1,\infty } \right). \\
\end{array} \right.$
\end{center}
For the normal GJR, $\tau_z = 0$ and$\int\limits_{x =  - \infty }^0 {{x^3}f\left( x \right)dx}  =  - \sqrt {\frac{2}{\pi }} $ hence:
$\mathop {\lim }\limits_{n \to \infty } {{\rm T}_{r,n}} = \left\{ \begin{array}{l}
\quad \quad \quad 0\quad \quad \;{\rm{if }} \; \gamma  \in \left( {0,1} \right), \\
- {\mathop{\rm sgn}} \left( \lambda  \right)\infty  \quad {\rm{if }}\; \gamma  \in \left[ {1,\infty } \right). \\
\end{array} \right.$
For the normal GARCH(1,1) $\tau_z = \lambda = 0$ and thus ${{\rm T}_{r,n}} = \mathop {\lim }\limits_{n \to \infty } {{\rm T}_{r,n}} = 0$. \\

(c) \textit{Limits of Forward and Aggregated Conditional Kurtosis} \\
The forward kurtosis limit is:
\begin{center}
$\mathop {\lim }\limits_{s \to \infty } {\kappa _{r,s}}
= {\kappa _z}\mathop {\lim }\limits_{s \to \infty } {\left( {\tilde \mu _{h,s}^{\left( 1 \right)}} \right)^{ - 2}}\mathop {\lim }\limits_{s \to \infty } \tilde \mu _{h,s}^{\left( 2 \right)}
= \left\{ \begin{array}{l}
{\kappa _z}\omega {{\bar h}^{ - 2}}\left( {\omega  + 2\varphi \bar h} \right){\left( {1 - \gamma } \right)^{ - 1}}\quad {\rm{if }} \; \gamma  \in \left( {0,1} \right), \\
\infty \qquad \qquad \qquad \;\;\quad \quad \;\;\quad \quad \;\;\;{\rm{if }}\;\gamma  \in \left[ {1,\infty } \right). \\
\end{array} \right.$
\end{center}
For the normal GJR and normal GARCH(1,1) the limit of the kurtosis of forward returns becomes:
\begin{center}
$\mathop {\lim }\limits_{s \to \infty } {\kappa _{r,s}} = \left\{ \begin{array}{l}
3\omega {{\bar h}^{ - 2}}\left( {\omega  + 2\varphi \bar h} \right){\left( {1 - \gamma } \right)^{ - 1}}\quad \; {\rm{if }}\;\gamma  \in \left( {0,1} \right), \\
\infty \qquad \qquad \qquad \;\;\quad \quad \;\;\quad \quad \;\;\,{\rm{if }}\;\gamma  \in \left[ {1,\infty } \right), \\
\end{array} \right.$
\end{center}
where $\varphi$ and $\gamma$ are now given by (\ref{eq:3.21}) and (\ref{eq:3.22}) for the normal GJR and normal GARCH(1,1).\\

For the limit of the aggregated kurtosis, we write: \\
$\mathop {\lim }\limits_{n \to \infty } {{\rm K}_{r,n}} = \mathop {\lim }\limits_{n \to \infty } {\left( {M_{r,n}^{\left( 2 \right)}} \right)^{ - 2}}M_{r,n}^{\left( 4 \right)} = \mathop {\lim }\limits_{n \to \infty } {\left( {{n^{ - 1}}M_{r,n}^{\left( 2 \right)}} \right)^{ - 2}}\left( {{n^{ - 2}}M_{r,n}^{\left( 4 \right)}} \right) = {\bar h^{ - 2}}\mathop {\lim }\limits_{n \to \infty } \left( {{n^{ - 2}}M_{r,n}^{\left( 4 \right)}} \right)$. \\
\begin{equation}\label{eq:3.23}
\mathop {\lim }\limits_{n \to \infty } \left( {{n^{ - 2}}M_{r,n}^{\left( 4 \right)}} \right) = {\kappa _z}{A_1} + 6{A_2} + 4{A_3},
\end{equation}	
where \\
$\begin{array}{l}
{A_1} = \mathop {\lim }\limits_{n \to \infty } \left( {{n^{ - 2}}\sum\limits_{s = 1}^n {\tilde \mu _{h,s}^{\left( 2 \right)}} } \right),\;\quad {A_2} = \mathop {\lim }\limits_{n \to \infty } \left( {{n^{ - 2}}\sum\limits_{s = 1}^n {\sum\limits_{u = 1}^{n - s} {{E_t}\left( {\varepsilon _{t + s}^2\varepsilon _{t + s + u}^2} \right)} } } \right), \\
{A_3} = \mathop {\lim }\limits_{n \to \infty } \left( {{n^{ - 2}}\sum\limits_{s = 1}^n {\sum\limits_{u = 1}^{n - s} {{E_t}\left( {{\varepsilon _{t + s}}\varepsilon _{t + s + u}^3} \right)} } } \right)+3 \mathop {\lim }\limits_{n \to \infty } \left( {{n^{ - 2}}\sum\limits_{s = 1}^n {\sum\limits_{u = 1}^{n - s} {\sum\limits_{v = 1}^{n - s - u} {{E_t}\left( {{\varepsilon _{t + s}}{\varepsilon _{t + s + u}}\varepsilon _{t + s + u + v}^2} \right)} } } } \right). \\
\end{array}$

Now, if $\gamma  \ne 1$, \\
$\begin{array}{l}
{A_1} = \mathop {\lim }\limits_{n \to \infty } {n^{ - 2}}\left( {n{c_1} + \left( {\tilde \mu _{h,1}^{\left( 2 \right)} - {c_3}} \right){{\left( {1 - \gamma } \right)}^{ - 1}}\left( {1 - {\gamma ^n}} \right) + {c_2}{{\left( {1 - \varphi } \right)}^{ - 1}}\left( {1 - {\varphi ^n}} \right)} \right) \\
\quad \; \,  = {c_1}\mathop {\lim }\limits_{n \to \infty } {n^{ - 1}} + \left( {h_{t + 1}^2 - {c_3}} \right){\left( {1 - \gamma } \right)^{ - 1}}\mathop {\lim }\limits_{n \to \infty } {n^{ - 2}}\left( {1 - {\gamma ^n}} \right) + {c_2}{\left( {1 - \varphi } \right)^{ - 1}}\mathop {\lim }\limits_{n \to \infty } {n^{ - 2}}\left( {1 - {\varphi ^n}} \right). \\
\end{array}$ \\

Otherwise, when $\gamma = 1$: \\
$\begin{array}{l}
\sum\limits_{s = 1}^n {\tilde \mu _{h,s}^{\left( 2 \right)} = } \sum\limits_{s = 1}^n {\left( {\left( {s - 1} \right)\left( {{\omega ^2} + 2\omega \varphi \bar h} \right) + 2\varphi {h_0}\left( {1 - {\varphi ^{s - 1}}} \right)\left( {{h_{t + 1}} - \bar h} \right) + h_{t + 1}^2} \right)}  \\
\qquad \quad \; = \raise.5ex\hbox{$\scriptstyle 1$}\kern-.1em/
\kern-.15em\lower.25ex\hbox{$\scriptstyle 2$} \left( {{\omega ^2} + 2\omega \varphi \bar h} \right){n^2} + \left( {2\varphi \bar h\left( {{h_{t + 1}} - \bar h} \right) - \raise.5ex\hbox{$\scriptstyle 1$}\kern-.1em/
\kern-.15em\lower.25ex\hbox{$\scriptstyle 2$}\left( {{\omega ^2} + 2\omega \varphi \bar h} \right) + h_{t + 1}^2} \right)n \\
\qquad \quad \; - 2\varphi \bar h\left( {{h_{t + 1}} - \bar h} \right){\left( {1 - \varphi } \right)^{ - 1}}\left( {1 - {\varphi ^n}} \right). \\
\end{array}$\\

So for $\gamma  = 1$, we obtain that: \\
$\begin{array}{l}
{A_1} = \mathop {\lim }\limits_{n \to \infty } \raise.5ex\hbox{$\scriptstyle 1$}\kern-.1em/
\kern-.15em\lower.25ex\hbox{$\scriptstyle 2$} \left( {{\omega ^2} + 2\omega \varphi \bar h} \right) + \mathop {\lim }\limits_{n \to \infty } {n^{ - 1}}\left( {2\varphi \bar h\left( {{h_{t + 1}} - \bar h} \right) - \raise.5ex\hbox{$\scriptstyle 1$}\kern-.1em/
\kern-.15em\lower.25ex\hbox{$\scriptstyle 2$} \left( {{\omega ^2} + 2\omega \varphi \bar h} \right) + h_{t + 1}^2} \right) \\
\quad \;\, - \mathop {\lim }\limits_{n \to \infty } {n^{ - 2}}2\varphi \bar h\left( {{h_{t + 1}} - \bar h} \right){\left( {1 - \varphi } \right)^{ - 1}}\left( {1 - {\varphi ^n}} \right). \\
\end{array}$\\

Hence
\begin{equation}\label{eq:3.25}
{A_1} = \left\{ \begin{array}{l}
0\qquad \qquad \qquad  \quad \quad \;{\rm{if }} \; \gamma  \in \left( {0,1} \right), \\
\raise.5ex\hbox{$\scriptstyle 1$}\kern-.1em/
\kern-.15em\lower.25ex\hbox{$\scriptstyle 2$} \left( {{\omega ^2} + 2\omega \varphi \bar h} \right)\quad \quad {\rm{if }} \; \gamma  = 1, \\
\infty \qquad \qquad \quad \quad \quad \quad {\rm{if }} \; \gamma  \in \left( {1,\infty } \right). \\
\end{array} \right.\\
\end{equation}

For $A_2$, using the derivations from the Technical Appendix T.A.1, we can write: \\
$\begin{array}{l}
{A_2} = \mathop {\lim }\limits_{n \to 0} \left( {{n^{ - 2}}\sum\limits_{s = 1}^n {\sum\limits_{u = 1}^{n - s} {{E_t}\left( {\varepsilon _{t + s}^2\varepsilon _{t + s + u}^2} \right)} } } \right) \\
\quad  = \mathop {\lim }\limits_{n \to 0} \left( {{n^{ - 2}}\left[ {\sum\limits_{s = 1}^n {\sum\limits_{u = 1}^{n - s} {\left[ {\bar h\left( {1 - {\varphi ^u}} \right)\tilde \mu _{h,s}^{\left( 1 \right)}} \right]} }  + {\kappa _z}\left( {\alpha  + \lambda {F_0} + \kappa _z^{ - 1}\beta } \right)\sum\limits_{s = 1}^n {\sum\limits_{u = 1}^{n - 2} {{\varphi ^{u - 1}}\tilde \mu _{h,s}^{\left( 2 \right)}} } } \right]} \right). \\
\end{array}$\\

For $\gamma  \ne 1$, the expressions for the two double sums in the expressions above were derived in the Technical Appendix T.A.1. Using those results, we have: \\

${A_2} = \mathop {\lim }\limits_{n \to \infty } {n^{ - 2}}\left[ \begin{array}{l}
\raise.5ex\hbox{$\scriptstyle 1$}\kern-.1em/
\kern-.15em\lower.25ex\hbox{$\scriptstyle 2$} n\left( {n - 1} \right){{\bar h}^2} + {\left( {1 - \varphi } \right)^{ - 1}}\bar h\left( {{h_{t + 1}} - \bar h} \right) \\
\left( {\left( {n - {{\left( {1 - \varphi } \right)}^{ - 1}}\left( {1 - {\varphi ^n}} \right)} \right) - \varphi \left( \begin{array}{l}
n\bar h{\left( {{h_{t + 1}} - \bar h} \right)^{ - 1}} + {\left( {1 - \varphi } \right)^{ - 1}}\left( {1 - {\varphi ^n}} \right) - {h_0} \\
{\left( {{h_{t + 1}} - {h_0}} \right)^{ - 1}}{\left( {1 - \varphi } \right)^{ - 1}}\left( {1 - {\varphi ^n}} \right) - n{\varphi ^{n - 1}} \\
\end{array} \right)} \right) \\
+ {\kappa _z}\left( {\alpha  + \lambda {F_0} + {\kappa _z}^{ - 1}\beta } \right){\left( {1 - \varphi } \right)^{ - 1}} \\
\left( \begin{array}{l}
n{c_1} + \left( {\tilde \mu _{h,1}^{\left( 2 \right)} - {c_3}} \right){\left( {1 - \gamma } \right)^{ - 1}}\left( {1 - {\gamma ^n}} \right) + {c_2}{\left( {1 - \varphi } \right)^{ - 1}}\left( {1 - {\varphi ^n}} \right) \\
- {c_1}{\left( {1 - \varphi } \right)^{ - 1}}\left( {1 - {\varphi ^n}} \right) - \left( {h_{t + 1}^2 - {c_3}} \right){\left( {\varphi  - \gamma } \right)^{ - 1}}\left( {{\varphi ^n} - {\gamma ^n}} \right) - n{c_2}{\varphi ^{n - 1}} \\
\end{array} \right) \\
\end{array} \right]. $ \\

For $\gamma = 1$, \\
$ {A_2} = \mathop {\lim }\limits_{n \to \infty } {n^{ - 2}}\left[ \begin{array}{l}
\raise.5ex\hbox{$\scriptstyle 1$}\kern-.1em/
\kern-.15em\lower.25ex\hbox{$\scriptstyle 2$} n\left( {n - 1} \right){{\bar h}^2} + {\left( {1 - \varphi } \right)^{ - 1}}\bar h\left( {{h_{t + 1}} - \bar h} \right) \\
\left( {\left( {n - {{\left( {1 - \varphi } \right)}^{ - 1}}\left( {1 - {\varphi ^n}} \right)} \right) - \varphi \left( \begin{array}{l}
n\bar h{\left( {{h_{t + 1}} - \bar h} \right)^{ - 1}} + {\left( {1 - \varphi } \right)^{ - 1}}\left( {1 - {\varphi ^n}} \right) - {h_0} \\
{\left( {{h_{t + 1}} - {h_0}} \right)^{ - 1}}{\left( {1 - \varphi } \right)^{ - 1}}\left( {1 - {\varphi ^n}} \right) - n{\varphi ^{n - 1}} \\
\end{array} \right)} \right) \\
+ {\kappa _z}\left( {\alpha  + \lambda {F_0} + {\kappa _z}^{ - 1}\beta } \right){\left( {1 - \varphi } \right)^{ - 1}} \\
\left( \begin{array}{l}
\raise.5ex\hbox{$\scriptstyle 1$}\kern-.1em/
\kern-.15em\lower.25ex\hbox{$\scriptstyle 2$} \left( {{\omega ^2} + 2\omega \varphi \bar h} \right){n^2}  + \left( {2\varphi \bar h\left( {{h_{t + 1}} - \bar h} \right) - \raise.5ex\hbox{$\scriptstyle 1$}\kern-.1em/
\kern-.15em\lower.25ex\hbox{$\scriptstyle 2$}\left( {{\omega ^2} + 2\omega \varphi \bar h} \right) + h_{t+1}^2} \right)n \\
- 2\varphi \bar h\left( {{h_{t + 1}} - \bar h} \right){\left( {1 - \varphi } \right)^{ - 1}}\left( {1 - {\varphi ^n}} \right) \\
+ \varphi \left( {{\omega ^2} + 2\omega \varphi \bar h} \right){\left( {1 - \varphi } \right)^{ - 1}}\left[ {{{\left( {1 - \varphi } \right)}^{ - 1}}\left( {1 - {\varphi ^n}} \right) - {\varphi ^{ - 1}}\left( {n - 1} \right)} \right] \\
- 2\varphi \bar h\left( {{h_{t + 1}} - \bar h} \right)\left[ {{{\left( {1 - \varphi } \right)}^{ - 1}}\left( {1 - {\varphi ^n}} \right) - n{\varphi ^{n - 1}}} \right] - \tilde \mu _{h,1}^{\left( 2 \right)}{\left( {1 - \varphi } \right)^{ - 1}}\left( {1 - {\varphi ^n}} \right) \\
\end{array} \right) \\
\end{array} \right].$ \\
Thus:
\begin{equation}\label{eq:3.33}
{A_2} = \left\{ \begin{array}{l}
\raise.5ex\hbox{$\scriptstyle 1$}\kern-.1em/
\kern-.15em\lower.25ex\hbox{$\scriptstyle 2$} \,{{\bar h}^2}\quad \quad \qquad \qquad \qquad \qquad \qquad \quad \quad \quad \quad \quad \quad \quad \quad \quad \;\;\;\;{\rm{if }}\; \gamma  \in \left( {0,1} \right), \\
\raise.5ex\hbox{$\scriptstyle 1$}\kern-.1em/
\kern-.15em\lower.25ex\hbox{$\scriptstyle 2$} \left[ {{{\bar h}^2} + {\kappa _z}\left( {\alpha  + \lambda {F_0} + \kappa _z^{ - 1}\beta } \right){{\left( {1 - \varphi } \right)}^{ - 1}}\left( {{\omega ^2} + 2\omega \varphi \bar h} \right)} \right]\quad \; {\rm{if }}\;\gamma  = 1, \\
\infty \qquad \qquad \qquad \qquad \qquad \quad \quad \quad \quad \quad \quad \quad \quad \quad \quad \quad \quad \;\;\;\;\,{\rm{if }}\;\gamma  \in \left( {1,\infty } \right). \\
\end{array} \right.
\end{equation}
${A_3} = \left[ {{\tau _z} + 3{{\left( {1 - \varphi } \right)}^{ - 1}}{c_9}} \right]\mathop {\lim }\limits_{n \to \infty } {n^{ - 2}}\sum\limits_{s = 1}^n {\sum\limits_{u = 1}^{n - s} {\theta _{su}^{\left( {3/2} \right)}} }  - 3{\left( {1 - \varphi } \right)^{ - 1}}{c_9}\mathop {\lim }\limits_{n \to \infty } {n^{ - 2}}\sum\limits_{s = 1}^n {\sum\limits_{u = 1}^{n - s} {{\varphi ^{n - s - u}}\theta _{su}^{\left( {3/2} \right)}} }$ where

$\begin{array}{l}
\theta _{su}^{\left( {3/2} \right)} = \frac{3}{4}{c_9}\left[ {{{\left( {\tilde \mu _{h,s + u}^{\left( 1 \right)}} \right)}^{1/2}} + \omega \varphi {{\left( {\varphi  - \gamma } \right)}^{ - 1}}{{\left( {\tilde \mu _{h,s + u}^{\left( 1 \right)}} \right)}^{ - 1/2}}} \right]{\varphi ^{u - 1}}{E_t}\left( {h_{t + s}^{3/2}} \right) \\
\qquad \;\; + \frac{3}{8}{\left( {\tilde \mu _{h,s + u}^{\left( 1 \right)}} \right)^{ - 1/2}}{\gamma ^{u - 1}}\left( {{c_{10}}{E_t}\left( {h_{t + s}^{5/2}} \right) + 2\omega \gamma {{\left( {\gamma  - \varphi } \right)}^{ - 1}}{c_9}{E_t}\left( {h_{t + s}^{3/2}} \right)} \right), \\
\end{array}$ \\
$\begin{array}{l}
{\varphi ^{n - s - u}}\theta _{su}^{\left( {3/2} \right)} = \frac{3}{4}{c_9}\left[ {{{\left( {\tilde \mu _{h,s + u}^{\left( 1 \right)}} \right)}^{1/2}} + \omega \varphi {{\left( {\varphi  - \gamma } \right)}^{ - 1}}{{\left( {\tilde \mu _{h,s + u}^{\left( 1 \right)}} \right)}^{ - 1/2}}} \right]{\varphi ^{n - s - 1}}{E_t}\left( {h_{t + s}^{3/2}} \right) \\
\qquad \qquad \quad \;\; + \frac{3}{8}{\left( {\tilde \mu _{h,s + u}^{\left( 1 \right)}} \right)^{ - 1/2}}{\varphi ^{n - s - 1}}{\left( {\gamma /\varphi } \right)^{u - 1}}\;\left( {{c_{10}}{E_t}\left( {h_{t + s}^{5/2}} \right) + 2\omega \gamma {{\left( {\gamma  - \varphi } \right)}^{ - 1}}{c_9}{E_t}\left( {h_{t + s}^{3/2}} \right)} \right). \\
\end{array}$\\

We can now write: ${b_{l,s,n}} \le \sum\limits_{u = 1}^{n - s} {\theta _{su}^{\left( {3/2} \right)}}  \le {b_{u,s,n}}$, where, for $\gamma \neq 1$ \\
$\begin{array}{l}
{b_{l,s,n}} = \frac{3}{4}\left( {{c_9}{{\left[ {{q_{u,1,n - s}}\left( {\tilde \mu _{h,s + u}^{\left( 1 \right)}, - {c_9}} \right)} \right]}^{1/2}} + \omega \varphi {{\left( {\varphi  - \gamma } \right)}^{ - 1}}{c_9}{{\left[ {{q_{u,1,n - s}}\left( {\tilde \mu _{h,s + u}^{\left( 1 \right)},\left( {\varphi  - \gamma } \right){c_9}} \right)} \right]}^{ - 1/2}}} \right) \\
\qquad \quad \;{E_t}\left( {h_{t + s}^{3/2}} \right){\left( {1 - \varphi } \right)^{ - 1}}\left( {1 - {\varphi ^{n - s}}} \right) \\
\qquad \quad \; + \frac{3}{8}\left( \begin{array}{l}
{c_{10}}{\left[ {{q_{u,1,n - s}}\left( {\tilde \mu _{h,s + u}^{\left( 1 \right)},{c_{10}}} \right)} \right]^{ - 1/2}}{E_t}\left( {h_{t + s}^{5/2}} \right) \\
+ 2\omega \gamma {\left( {\gamma  - \varphi } \right)^{ - 1}}{c_9}{\left[ {{q_{u,1,n - s}}\left( {\tilde \mu _{h,s + u}^{\left( 1 \right)},\left( {\gamma  - \varphi } \right){c_9}} \right)} \right]^{ - 1/2}}{E_t}\left( {h_{t + s}^{3/2}} \right) \\
\end{array} \right){\left( {1 - \gamma } \right)^{ - 1}}\left( {1 - {\gamma ^{n - s}}} \right), \\
\end{array}$

$\begin{array}{l}
{b_{u,s,n}} = \frac{3}{4}\left( \begin{array}{l}
{c_9}{\left[ {{q_{u,1,n - s}}\left( {\tilde \mu _{h,s + u}^{\left( 1 \right)},{c_9}} \right)} \right]^{1/2}} + \omega \varphi {\left( {\varphi  - \gamma } \right)^{ - 1}} \\
{c_9}{\left[ {{q_{u,1,n - s}}\left( {\tilde \mu _{h,s + u}^{\left( 1 \right)}, - \left( {\varphi  - \gamma } \right){c_9}} \right)} \right]^{ - 1/2}} \\
\end{array} \right){E_t}\left( {h_{t + s}^{3/2}} \right){\left( {1 - \varphi } \right)^{ - 1}}\left( {1 - {\varphi ^{n - s}}} \right) \\
\qquad \quad  + \frac{3}{8}\left( \begin{array}{l}
{c_{10}}{\left[ {{q_{u,1,n - s}}\left( {\tilde \mu _{h,s + u}^{\left( 1 \right)}, - {c_{10}}} \right)} \right]^{ - 1/2}}{E_t}\left( {h_{t + s}^{5/2}} \right) + 2\omega \gamma {\left( {\gamma  - \varphi } \right)^{ - 1}} \\
{c_9}{\left[ {{q_{u,1,n - s}}\left( {\tilde \mu _{h,s + u}^{\left( 1 \right)}, - \left( {\gamma  - \varphi } \right){c_9}} \right)} \right]^{ - 1/2}}{E_t}\left( {h_{t + s}^{3/2}} \right) \\
\end{array} \right){\left( {1 - \gamma } \right)^{ - 1}}\left( {1 - {\gamma ^{n - s}}} \right), \\
\end{array}$ \\

and, for $\gamma = 1$ \\
$\begin{array}{l}
{b_{l,s,n}} = \frac{3}{4}\left( {{c_9}{{\left[ {{q_{u,1,n - s}}\left( {\tilde \mu _{h,s + u}^{\left( 1 \right)}, - {c_9}} \right)} \right]}^{1/2}} + \omega \varphi {{\left( {\varphi  - 1} \right)}^{ - 1}}{c_9}{{\left[ {{q_{u,1,n - s}}\left( {\tilde \mu _{h,s + u}^{\left( 1 \right)}, - {c_9}} \right)} \right]}^{ - 1/2}}} \right) \\
\qquad \quad {E_t}\left( {h_{t + s}^{3/2}} \right){\left( {1 - \varphi } \right)^{ - 1}}\left( {1 - {\varphi ^{n - s}}} \right) \\
\qquad \quad  + \frac{3}{8}\left( {n - s} \right)\left( \begin{array}{l}
{c_{10}}{\left[ {{q_{u,1,n - s}}\left( {\tilde \mu _{h,s + u}^{\left( 1 \right)},{c_{10}}} \right)} \right]^{ - 1/2}}{E_t}\left( {h_{t + s}^{5/2}} \right) \\
+ 2\omega {\left( {1 - \varphi } \right)^{ - 1}}{c_9}{\left[ {{q_{u,1,n - s}}\left( {\tilde \mu _{h,s + u}^{\left( 1 \right)},{c_9}} \right)} \right]^{ - 1/2}}{E_t}\left( {h_{t + s}^{3/2}} \right) \\
\end{array} \right), \\
\end{array}$

$\begin{array}{l}
{b_{u,s,n}} = \frac{3}{4}\left( {{c_9}{{\left[ {{q_{u,1,n - s}}\left( {\tilde \mu _{h,s + u}^{\left( 1 \right)},{c_9}} \right)} \right]}^{1/2}} + \omega \varphi {{\left( {\varphi  - 1} \right)}^{ - 1}}{c_9}{{\left[ {{q_{u,1,n - s}}\left( {\tilde \mu _{h,s + u}^{\left( 1 \right)},{c_9}} \right)} \right]}^{ - 1/2}}} \right) \\
\qquad \quad {E_t}\left( {h_{t + s}^{3/2}} \right){\left( {1 - \varphi } \right)^{ - 1}}\left( {1 - {\varphi ^{n - s}}} \right) \\
\qquad \quad  + \frac{3}{8}\left( {n - s} \right)\left( \begin{array}{l}
{c_{10}}{\left[ {{q_{u,1,n - s}}\left( {\tilde \mu _{h,s + u}^{\left( 1 \right)}, - {c_{10}}} \right)} \right]^{ - 1/2}}{E_t}\left( {h_{t + s}^{5/2}} \right) \\
+ 2\omega {\left( {1 - \varphi } \right)^{ - 1}}{c_9}{\left[ {{q_{u,1,n - s}}\left( {\tilde \mu _{h,s + u}^{\left( 1 \right)}, - {c_9}} \right)} \right]^{ - 1/2}}{E_t}\left( {h_{t + s}^{3/2}} \right) \\
\end{array} \right), \\
\end{array}$\\

with 	${q_{z,a,b}}\left( {{x_z},y} \right) = \left\{ \begin{array}{l}
\mathop {\max }\limits_{a \le z \le b} {x_z}\quad {\rm{if }}\; y \ge 0, \\
\mathop {\min }\limits_{a \le z \le b} {x_z}\quad {\rm{if }}\; y < 0, \\
\end{array} \right.$ where $z = u$ or $z = s$.\\

Also, ${b_{ul,n}} \le \sum\limits_{s = 1}^n {{b_{u,s,n}}}  \le
{b_{uu,n}}$ where for $\gamma \neq 1$ \\
$\begin{array}{l}
{b_{ul,n}} = \frac{3}{4}{\left( {1 - \varphi } \right)^{ - 1}}\sum\limits_{s = 1}^n {\left( {1 - {\varphi ^{n - s}}} \right)}  \\
\quad \;\;\;\;\left[ \begin{array}{l}
{c_9}{q_{s,1,n}}\left( {{E_t}\left( {h_{t + s}^{3/2}} \right), - {c_9}} \right){\left[ {{q_{s,1,n}}\left( {{q_{u,1,n - s}}\left( {\tilde \mu _{h,s + u}^{\left( 1 \right)},{c_9}} \right), - {c_9}} \right)} \right]^{1/2}} \\
+ \omega \varphi {\left( {\varphi  - \gamma } \right)^{ - 1}}{c_9}{q_{s,1,n}}\left( {{E_t}\left( {h_{t + s}^{3/2}} \right), - \left( {\varphi  - \gamma } \right){c_9}} \right) \\
{\left[ {{q_{s,1,n}}\left( {{q_{u,1,n - s}}\left( {\tilde \mu _{h,s + u}^{\left( 1 \right)}, - \left( {\varphi  - \gamma } \right){c_9}} \right),\left( {\varphi  - \gamma } \right){c_9}} \right)} \right]^{ - 1/2}} \\
\end{array} \right] \\
\quad \;\; + \frac{3}{8}{\left( {1 - \gamma } \right)^{ - 1}}\sum\limits_{s = 1}^n {\left( {1 - {\gamma ^{n - s}}} \right)}  \\
\quad \;\;\;\;\left[ \begin{array}{l}
{c_{10}}{q_{s,1,n}}\left( {{E_t}\left( {h_{t + s}^{5/2}} \right), - {c_{10}}} \right){\left[ {{q_{s,1,n}}\left( {{q_{u,1,n - s}}\left( {\tilde \mu _{h,s + u}^{\left( 1 \right)}, - {c_{10}}} \right),{c_{10}}} \right)} \right]^{ - 1/2}} \\
+ 2\omega \gamma {\left( {\gamma  - \varphi } \right)^{ - 1}}{c_9}{q_{s,1,n}}\left( {{E_t}\left( {h_{t + s}^{3/2}} \right), - \left( {\gamma  - \varphi } \right){c_9}} \right) \\
{\left[ {{q_{s,1,n}}\left( {{q_{u,1,n - s}}\left( {\tilde \mu _{h,s + u}^{\left( 1 \right)}, - \left( {\gamma  - \varphi } \right){c_9}} \right),\left( {\gamma  - \varphi } \right){c_9}} \right)} \right]^{ - 1/2}} \\
\end{array} \right] \\
\quad \quad  = \frac{3}{4}{l_{1,ul,n}}{\left( {1 - \varphi } \right)^{ - 1}}\left( {n - {{\left( {1 - \varphi } \right)}^{ - 1}}\left( {1 - {\varphi ^n}} \right)} \right) + \frac{3}{8}{l_{2,ul,n}}{\left( {1 - \gamma } \right)^{ - 1}}\left( {n - {{\left( {1 - \gamma } \right)}^{ - 1}}\left( {1 - {\gamma ^n}} \right)} \right) \\
\end{array}$ \\
with \\
$\begin{array}{l}
{l_{1,ul,n}} = {c_9}{q_{s,1,n}}\left( {{E_t}\left( {h_{t + s}^{3/2}} \right), - {c_9}} \right){\left[ {{q_{s,1,n}}\left( {{q_{u,1,n - s}}\left( {\tilde \mu _{h,s + u}^{\left( 1 \right)},{c_9}} \right), - {c_9}} \right)} \right]^{1/2}} \\
\quad \quad \;\; + \omega \varphi {\left( {\varphi  - \gamma } \right)^{ - 1}}{c_9}{q_{s,1,n}}\left( {{E_t}\left( {h_{t + s}^{3/2}} \right), - \left( {\varphi  - \gamma } \right){c_9}} \right) \\
\quad \quad \;\;{\left[ {{q_{s,1,n}}\left( {{q_{u,1,n - s}}\left( {\tilde \mu _{h,s + u}^{\left( 1 \right)}, - \left( {\varphi  - \gamma } \right){c_9}} \right),\left( {\varphi  - \gamma } \right){c_9}} \right)} \right]^{ - 1/2}} \\
\end{array}$ \\
and \\
$\begin{array}{l}
{l_{2,ul,n}} = {c_{10}}{q_{s,1,n}}\left( {{E_t}\left( {h_{t + s}^{5/2}} \right), - {c_{10}}} \right){\left[ {{q_{s,1,n}}\left( {{q_{u,1,n - s}}\left( {\tilde \mu _{h,s + u}^{\left( 1 \right)}, - {c_{10}}} \right),{c_{10}}} \right)} \right]^{ - 1/2}} \\
\quad \quad \;\; + 2\omega \gamma {\left( {\gamma  - \varphi } \right)^{ - 1}}{c_9}{q_{s,1,n}}\left( {{E_t}\left( {h_{t + s}^{3/2}} \right), - \left( {\gamma  - \varphi } \right){c_9}} \right) \\
\quad \quad \;\;{\left[ {{q_{s,1,n}}\left( {{q_{u,1,n - s}}\left( {\tilde \mu _{h,s + u}^{\left( 1 \right)}, - \left( {\gamma  - \varphi } \right){c_9}} \right),\left( {\gamma  - \varphi } \right){c_9}} \right)} \right]^{ - 1/2}}. \\
\end{array}$ \\
Also \\
$\begin{array}{l}
{b_{uu,n}} = \frac{3}{4}{\left( {1 - \varphi } \right)^{ - 1}}\sum\limits_{s = 1}^n {\left( {1 - {\varphi ^{n - s}}} \right)}  \\
\quad \quad \;\;\left[ \begin{array}{l}
{c_9}{q_{s,1,n}}\left( {{E_t}\left( {h_{t + s}^{3/2}} \right),{c_9}} \right){\left[ {{q_{s,1,n}}\left( {{q_{u,1,n - s}}\left( {\tilde \mu _{h,s + u}^{\left( 1 \right)},{c_9}} \right),{c_9}} \right)} \right]^{1/2}} + \omega \varphi {\left( {\varphi  - \gamma } \right)^{ - 1}}{c_9} \\
{q_{s,1,n}}\left( {{E_t}\left( {h_{t + s}^{3/2}} \right),\left( {\varphi  - \gamma } \right){c_9}} \right){\left[ {{q_{s,1,n}}\left( {{q_{u,1,n - s}}\left( {\tilde \mu _{h,s + u}^{\left( 1 \right)}, - \left( {\varphi  - \gamma } \right){c_9}} \right), - \left( {\varphi  - \gamma } \right){c_9}} \right)} \right]^{ - 1/2}} \\
\end{array} \right] \\
\quad \quad  + \frac{3}{8}{\left( {1 - \gamma } \right)^{ - 1}}\sum\limits_{s = 1}^n {\left( {1 - {\gamma ^{n - s}}} \right)}  \\
\quad \quad \;\;\left[ \begin{array}{l}
{c_{10}}{q_{s,1,n}}\left( {{E_t}\left( {h_{t + s}^{5/2}} \right),{c_{10}}} \right){\left[ {{q_{s,1,n}}\left( {{q_{u,1,n - s}}\left( {\tilde \mu _{h,s + u}^{\left( 1 \right)}, - {c_{10}}} \right), - {c_{10}}} \right)} \right]^{ - 1/2}} + 2\omega \gamma {\left( {\gamma  - \varphi } \right)^{ - 1}}{c_9} \\
{q_{s,1,n}}\left( {{E_t}\left( {h_{t + s}^{3/2}} \right),\left( {\gamma  - \varphi } \right){c_9}} \right){\left[ {{q_{s,1,n}}\left( {{q_{u,1,n - s}}\left( {\tilde \mu _{h,s + u}^{\left( 1 \right)}, - \left( {\gamma  - \varphi } \right){c_9}} \right), - \left( {\gamma  - \varphi } \right){c_9}} \right)} \right]^{ - 1/2}} \\
\end{array} \right] \\
\quad \quad {\rm{  }} = \frac{3}{4}{l_{1,uu,n}}{\left( {1 - \varphi } \right)^{ - 1}}\left( {n - {{\left( {1 - \varphi } \right)}^{ - 1}}\left( {1 - {\varphi ^n}} \right)} \right) + \frac{3}{8}{l_{2,uu,n}}{\left( {1 - \gamma } \right)^{ - 1}}\left( {n - {{\left( {1 - \gamma } \right)}^{ - 1}}\left( {1 - {\gamma ^n}} \right)} \right) \\
\end{array}$. \\
with \\
$\begin{array}{l}
{l_{1,uu,n}} = {c_9}{q_{s,1,n}}\left( {{E_t}\left( {h_{t + s}^{3/2}} \right),{c_9}} \right){\left[ {{q_{s,1,n}}\left( {{q_{u,1,n - s}}\left( {\tilde \mu _{h,s + u}^{\left( 1 \right)},{c_9}} \right),{c_9}} \right)} \right]^{1/2}} \\
\quad \quad \;\; + \omega \varphi {\left( {\varphi  - \gamma } \right)^{ - 1}}{c_9}{q_{s,1,n}}\left( {{E_t}\left( {h_{t + s}^{3/2}} \right),\left( {\varphi  - \gamma } \right){c_9}} \right) \\
\quad \quad \;\;{\left[ {{q_{s,1,n}}\left( {{q_{u,1,n - s}}\left( {\tilde \mu _{h,s + u}^{\left( 1 \right)}, - \left( {\varphi  - \gamma } \right){c_9}} \right), - \left( {\varphi  - \gamma } \right){c_9}} \right)} \right]^{ - 1/2}} \\
\end{array}$ \\
and \\
$\begin{array}{l}
{l_{2,uu,n}} = {c_{10}}{q_{s,1,n}}\left( {{E_t}\left( {h_{t + s}^{5/2}} \right),{c_{10}}} \right){\left[ {{q_{s,1,n}}\left( {{q_{u,1,n - s}}\left( {\tilde \mu _{h,s + u}^{\left( 1 \right)}, - {c_{10}}} \right), - {c_{10}}} \right)} \right]^{ - 1/2}} \\
\quad \quad \; + 2\omega \gamma {\left( {\gamma  - \varphi } \right)^{ - 1}}{c_9}{q_{s,1,n}}\left( {{E_t}\left( {h_{t + s}^{3/2}} \right),\left( {\gamma  - \varphi } \right){c_9}} \right) \\
\quad \quad \;{\left[ {{q_{s,1,n}}\left( {{q_{u,1,n - s}}\left( {\tilde \mu _{h,s + u}^{\left( 1 \right)}, - \left( {\gamma  - \varphi } \right){c_9}} \right), - \left( {\gamma  - \varphi } \right){c_9}} \right)} \right]^{ - 1/2}}\,. \\
\end{array}$.\\

For $\gamma = 1$: \\
$\begin{array}{l}
{b_{ul,n}} = \frac{3}{4}\left[ \begin{array}{l}
{c_9}{q_{s,1,n}}\left( {{E_t}\left( {h_{t + s}^{3/2}} \right), - {c_9}} \right){\left[ {{q_{s,1,n}}\left( {{q_{u,1,n - s}}\left( {\tilde \mu _{h,s + u}^{\left( 1 \right)},{c_9}} \right), - {c_9}} \right)} \right]^{1/2}} + \omega \varphi {\left( {\varphi  - 1} \right)^{ - 1}}{c_9} \\
{q_{s,1,n}}\left( {{E_t}\left( {h_{t + s}^{3/2}} \right),{c_9}} \right){\left[ {{q_{s,1,n}}\left( {{q_{u,1,n - s}}\left( {\tilde \mu _{h,s + u}^{\left( 1 \right)},{c_9}} \right), - {c_9}} \right)} \right]^{ - 1/2}} \\
\end{array} \right] \\
\quad \quad \;\; \; {\left( {1 - \varphi } \right)^{ - 1}}\sum\limits_{s = 1}^n {\left( {1 - {\varphi ^{n - s}}} \right) + \frac{3}{{16}}n\left( {n - 1} \right)}  \\
\quad \quad \; \; \left[ \begin{array}{l}
{c_{10}}{q_{s,1,n}}\left( {{E_t}\left( {h_{t + s}^{5/2}} \right), - {c_{10}}} \right){\left[ {{q_{s,1,n}}\left( {{q_{u,1,n - s}}\left( {\tilde \mu _{h,s + u}^{\left( 1 \right)}, - {c_{10}}} \right),{c_{10}}} \right)} \right]^{ - 1/2}} \\
+ 2\omega {\left( {1 - \varphi } \right)^{ - 1}}{c_9}{q_{s,1,n}}\left( {{E_t}\left( {h_{t + s}^{3/2}} \right), - {c_9}} \right){\left[ {{q_{s,1,n}}\left( {{q_{u,1,n - s}}\left( {\tilde \mu _{h,s + u}^{\left( 1 \right)}, - {c_9}} \right),{c_9}} \right)} \right]^{ - 1/2}} \\
\end{array} \right] \\
\quad \quad \; \; = \frac{3}{4}{{\tilde l}_{1,ul,n}}{\left( {1 - \varphi } \right)^{ - 1}}\left( {n - {{\left( {1 - \varphi } \right)}^{ - 1}}\left( {1 - {\varphi ^n}} \right)} \right) + \frac{3}{{16}}{{\tilde l}_{2,ul,n}}n\left( {n - 1} \right) \\
\end{array}$ \\
with \\
$\begin{array}{l}
{{\tilde l}_{1,ul,n}} = {c_9}{q_{s,1,n}}\left( {{E_t}\left( {h_{t + s}^{3/2}} \right), - {c_9}} \right){\left[ {{q_{s,1,n}}\left( {{q_{u,1,n - s}}\left( {\tilde \mu _{h,s + u}^{\left( 1 \right)},{c_9}} \right), - {c_9}} \right)} \right]^{1/2}} \\
\quad \quad \; \; + \omega \varphi {\left( {\varphi  - 1} \right)^{ - 1}}{c_9}{q_{s,1,n}}\left( {{E_t}\left( {h_{t + s}^{3/2}} \right),{c_9}} \right){\left[ {{q_{s,1,n}}\left( {{q_{u,1,n - s}}\left( {\tilde \mu _{h,s + u}^{\left( 1 \right)},{c_9}} \right), - {c_9}} \right)} \right]^{ - 1/2}} \\
\end{array}$ \\
and \\
$\begin{array}{l}
{{\tilde l}_{2,ul,n}} = {c_{10}}{q_{s,1,n}}\left( {{E_t}\left( {h_{t + s}^{5/2}} \right), - {c_{10}}} \right){\left[ {{q_{s,1,n}}\left( {{q_{u,1,n - s}}\left( {\tilde \mu _{h,s + u}^{\left( 1 \right)}, - {c_{10}}} \right),{c_{10}}} \right)} \right]^{ - 1/2}} \\
\quad \quad \; \; + 2\omega {\left( {1 - \varphi } \right)^{ - 1}}{c_9}{q_{s,1,n}}\left( {{E_t}\left( {h_{t + s}^{3/2}} \right), - {c_9}} \right){\left[ {{q_{s,1,n}}\left( {{q_{u,1,n - s}}\left( {\tilde \mu _{h,s + u}^{\left( 1 \right)}, - {c_9}} \right),{c_9}} \right)} \right]^{ - 1/2}}. \\
\end{array}$ \\
Also, \\
$\begin{array}{l}
{b_{uu,n}} = \frac{3}{4}{\left( {1 - \varphi } \right)^{ - 1}}\sum\limits_{s = 1}^n {\left( {1 - {\varphi ^{n - s}}} \right)}  \\
\quad \quad \left[ \begin{array}{l}
{c_9}{q_{s,1,n}}\left( {{E_t}\left( {h_{t + s}^{3/2}} \right),{c_9}} \right){\left[ {{q_{s,1,n}}\left( {{q_{u,1,n - s}}\left( {\tilde \mu _{h,s + u}^{\left( 1 \right)},{c_9}} \right),{c_9}} \right)} \right]^{1/2}} + \omega \varphi {\left( {\varphi  - 1} \right)^{ - 1}}{c_9} \\
{q_{s,1,n}}\left( {{E_t}\left( {h_{t + s}^{3/2}} \right), - {c_9}} \right){\left[ {{q_{s,1,n}}\left( {{q_{u,1,n - s}}\left( {\tilde \mu _{h,s + u}^{\left( 1 \right)},{c_9}} \right),{c_9}} \right)} \right]^{ - 1/2}} \\
\end{array} \right] \\
\quad \quad  + \frac{3}{{16}}n\left( {n - 1} \right) \\
\quad \quad \left[ \begin{array}{l}
{c_{10}}{q_{s,1,n}}\left( {{E_t}\left( {h_{t + s}^{5/2}} \right),{c_{10}}} \right){\left[ {{q_{s,1,n}}\left( {{q_{u,1,n - s}}\left( {\tilde \mu _{h,s + u}^{\left( 1 \right)}, - {c_{10}}} \right), - {c_{10}}} \right)} \right]^{ - 1/2}} \\
+ 2\omega {\left( {1 - \varphi } \right)^{ - 1}}{c_9}{q_{s,1,n}}\left( {{E_t}\left( {h_{t + s}^{3/2}} \right),{c_9}} \right){\left[ {{q_{s,1,n}}\left( {{q_{u,1,n - s}}\left( {\tilde \mu _{h,s + u}^{\left( 1 \right)}, - {c_9}} \right), - {c_9}} \right)} \right]^{ - 1/2}} \\
\end{array} \right] \\
\quad \quad  = \frac{3}{4}{{\tilde l}_{1,uu,n}}{\left( {1 - \varphi } \right)^{ - 1}}\left( {n - {{\left( {1 - \varphi } \right)}^{ - 1}}\left( {1 - {\varphi ^n}} \right)} \right) + \frac{3}{{16}}{{\tilde l}_{2,uu,n}}n\left( {n - 1} \right) \\
\end{array}$ \\
where \\
$\begin{array}{l}
{{\tilde l}_{1,uu,n}} = {c_9}{q_{s,1,n}}\left( {{E_t}\left( {h_{t + s}^{3/2}} \right),{c_9}} \right){\left[ {{q_{s,1,n}}\left( {{q_{u,1,n - s}}\left( {\tilde \mu _{h,s + u}^{\left( 1 \right)},{c_9}} \right),{c_9}} \right)} \right]^{1/2}} \\
\quad \quad \; \;\, + \omega \varphi {\left( {\varphi  - 1} \right)^{ - 1}}{c_9}{q_{s,1,n}}\left( {{E_t}\left( {h_{t + s}^{3/2}} \right), - {c_9}} \right){\left[ {{q_{s,1,n}}\left( {{q_{u,1,n - s}}\left( {\tilde \mu _{h,s + u}^{\left( 1 \right)},{c_9}} \right),{c_9}} \right)} \right]^{ - 1/2}} \\
\end{array}$ \\
and \\
$\begin{array}{l}
{{\tilde l}_{2,uu,n}} = {c_{10}}{q_{s,1,n}}\left( {{E_t}\left( {h_{t + s}^{5/2}} \right),{c_{10}}} \right){\left[ {{q_{s,1,n}}\left( {{q_{u,1,n - s}}\left( {\tilde \mu _{h,s + u}^{\left( 1 \right)}, - {c_{10}}} \right), - {c_{10}}} \right)} \right]^{ - 1/2}} \\
\quad \quad \;\; \; + 2\omega {\left( {1 - \varphi } \right)^{ - 1}}{c_9}{q_{s,1,n}}\left( {{E_t}\left( {h_{t + s}^{3/2}} \right),{c_9}} \right){\left[ {{q_{s,1,n}}\left( {{q_{u,1,n - s}}\left( {\tilde \mu _{h,s + u}^{\left( 1 \right)}, - {c_9}} \right), - {c_9}} \right)} \right]^{ - 1/2}}. \\
\end{array}$

We have previously shown that ${L_{\max }} = \mathop {\lim }\limits_{n \to \infty } \left( {\mathop {\max }\limits_{1 \le s \le n} \;\tilde \mu _{h,s}^{\left( 1 \right)}} \right) = \max \left( {\bar h,{h_{t + 1}}} \right)$ and \\${L_{\min }} = \mathop {\lim }\limits_{n \to \infty } \left( {\mathop {\min }\limits_{1 \le s \le n} \;\tilde \mu _{h,s}^{\left( 1 \right)}} \right) = \min \left( {\bar h,{h_{t + 1}}} \right)$. Also, we have:

$\begin{array}{l}
\mathop {\lim }\limits_{s \to \infty } {E_t}\left( {h_{t + s}^{3/2}} \right) = \frac{1}{8}\mathop {\lim }\limits_{s \to \infty } \left( {5{{\left( {\tilde \mu _{h,s}^{\left( 1 \right)}} \right)}^{3/2}} + 3\tilde \mu _{h,s}^{\left( 2 \right)}{{\left( {\tilde \mu _{h,s}^{\left( 1 \right)}} \right)}^{ - 1/2}}} \right) \\
\qquad \qquad \qquad \; \, = \left\{ \begin{array}{l}
\frac{1}{8}\left( {5{{\bar h}^{3/2}} + 3{{\bar h}^{ - 1/2}}\left( {{c_1} + \left( {h_{t + 1}^2 - {c_3}} \right)\mathop {\lim }\limits_{s \to \infty } {\gamma ^{s - 1}} + {c_2}\mathop {\lim }\limits_{s \to \infty } {\varphi ^{s - 1}}} \right)} \right)\qquad \,{\rm{if }}\; \gamma  \ne 1, \\
\frac{1}{8}\left( {5{{\bar h}^{3/2}} + 3{{\bar h}^{ - 1/2}}\mathop {\lim }\limits_{s \to \infty } \left( \begin{array}{l}
\left( {s - 1} \right)\left( {{\omega ^2} + 2\omega \varphi \bar h} \right) \\
+ 2\varphi \bar h\left( {1 - {\varphi ^{s - 1}}} \right)\left( {{h_{t + 1}} - \bar h} \right) + h_{t + 1}^2 \\
\end{array} \right)} \right)\quad {\rm{if }} \; \gamma  = 1, \\
\end{array} \right. \\
\end{array}$

$  \qquad \qquad \qquad \; \,
= \left\{ \begin{array}{l}
\frac{1}{8}\left( {5{{\bar h}^{3/2}} + 3{{\bar h}^{ - 1/2}}{c_1}} \right)\quad {\rm{if }} \; \gamma  \in \left( {0,1} \right), \\
\infty \qquad \qquad \quad \quad \quad \quad \quad {\rm{if }} \; \gamma  \in \left[ {1,\infty } \right). \\
\end{array} \right.$


$
\mathop {\lim }\limits_{s \to \infty } {E_t}\left( {h_{t + s}^{5/2}} \right) = \frac{1}{8}\mathop {\lim }\limits_{s \to \infty } \left( {\left( {15{{\left( {\tilde \mu _{h,s}^{\left( 1 \right)}} \right)}^{1/2}}\tilde \mu _{h,s}^{\left( 2 \right)} - 7{{\left( {\tilde \mu _{h,s}^{\left( 1 \right)}} \right)}^{5/2}}} \right)} \right)
= \left\{ \begin{array}{l}
\frac{1}{8}\left( { - 7{{\bar h}^{5/2}} + 15{{\bar h}^{1/2}}{c_1}} \right)\quad {\rm{if }} \; \gamma  \in \left( {0,1} \right), \\
\infty \qquad \qquad \quad \quad \quad \quad \quad \;\;\;{\rm{if }} \; \gamma  \in \left[ {1,\infty } \right)\,. \\
\end{array} \right.
$ \\
Hence, when $\gamma  \in \left( {0,1} \right)$,
$\mathop {\lim }\limits_{s \to \infty } {E_t}\left( {h_{t + s}^{3/2}} \right)$ and
$\mathop {\lim }\limits_{s \to \infty } {E_t}\left( {h_{t + s}^{5/2}} \right)$ exist and are finite. Furthermore, 
$\mathop {\max }\limits_{1 \le s \le n} {E_t}\left( {h_{t + s}^{i/2}} \right)$ and $\mathop {\min }\limits_{1 \le s \le n} {E_t}\left( {h_{t + s}^{i/2}} \right)$, $i = 3$ and $5$ are bounded. Thus,
$\mathop {\lim }\limits_{n \to \infty } \left( {{l_{j,uu,n}}} \right)$
and $\mathop {\lim }\limits_{n \to \infty } \left( {{l_{j,ul,n}}} \right)$, $j = 1$ and $2$ exist and are finite. We get: \\
$
\mathop {\lim }\limits_{n \to \infty } {n^{ - 2}}{b_{ul,n}} = \mathop {\lim }\limits_{n \to \infty } {n^{ - 2}}\left( \begin{array}{l}
\frac{3}{4}{l_{1,ul,n}}{\left( {1 - \varphi } \right)^{ - 1}}\left( {n - {{\left( {1 - \varphi } \right)}^{ - 1}}\left( {1 - {\varphi ^n}} \right)} \right) \\
+ \frac{3}{8}{l_{2,ul,n}}{\left( {1 - \gamma } \right)^{ - 1}}\left( {n - {{\left( {1 - \gamma } \right)}^{ - 1}}\left( {1 - {\gamma ^n}} \right)} \right) \\
\end{array} \right) =0$ \\
$
\mathop {\lim }\limits_{n \to \infty } {n^{ - 2}}{b_{uu,n}} = \mathop {\lim }\limits_{n \to \infty } {n^{ - 2}}\left[ \begin{array}{l}
\frac{3}{4}{l_{1,uu,n}}{\left( {1 - \varphi } \right)^{ - 1}}\left( {n - {{\left( {1 - \varphi } \right)}^{ - 1}}\left( {1 - {\varphi ^n}} \right)} \right) \\
+ \frac{3}{8}{l_{2,uu,n}}{\left( {1 - \gamma } \right)^{ - 1}}\left( {n - {{\left( {1 - \gamma } \right)}^{ - 1}}\left( {1 - {\gamma ^n}} \right)} \right) \\
\end{array} \right] =0.$ \\
Thus $\mathop {\lim }\limits_{n \to \infty } {n^{ - 2}}\sum\limits_{s = 1}^n {{b_{u,s,n}}}  = 0$. This translates into: $$\mathop {\lim }\limits_{n \to \infty } {n^{ - 2}}\sum\limits_{s = 1}^n {\sum\limits_{u = 1}^{n - s} {\theta _{su}^{\left( {3/2} \right)}} }  \le \mathop {\lim }\limits_{n \to \infty } {n^{ - 2}}\sum\limits_{s = 1}^n {{b_{u,s,n}}}  = 0.$$ Similarly, it can be shown that: $0 = \mathop {\lim }\limits_{n \to \infty } {n^{ - 2}}\sum\limits_{s = 1}^n {{b_{l,s,n}}}  \le \mathop {\lim }\limits_{n \to \infty } {n^{ - 2}}\sum\limits_{s = 1}^n {\sum\limits_{u = 1}^{n - s} {\theta _{su}^{\left( {3/2} \right)}} }$. Finally, we obtain that $\mathop {\lim }\limits_{n \to \infty } {n^{ - 2}}\sum\limits_{s = 1}^n {\sum\limits_{u = 1}^{n - s} {\theta _{su}^{\left( {3/2} \right)}} }  = 0$ for $\gamma  \in \left( {0,1} \right)$. Similarly it can be shown that $\mathop {\lim }\limits_{n \to \infty } {n^{ - 2}}\sum\limits_{s = 1}^n {\sum\limits_{u = 1}^{n - s} {{\varphi ^{n - s - u}}\theta _{su}^{\left( {3/2} \right)}} }  = 0$ for $\gamma  \in \left( {0,1} \right)$.\\

 Consider now: ${B_{l,s,n}} \le \sum\limits_{u = 1}^{n - s} {{\varphi ^{n - s - u}}\theta _{su}^{3/2}}  \le {B_{u,s,n}}$ where, for any $\gamma$: \\
$\begin{array}{l}
{B_{l,s,n}} = \frac{3}{4}\left( {{c_9}{{\left[ {{q_{u,1,n - s}}\left( {\tilde \mu _{h,s + u}^{\left( 1 \right)}, - {c_9}} \right)} \right]}^{1/2}} + \omega \varphi {{\left( {\varphi  - \gamma } \right)}^{ - 1}}{c_9}{{\left[ {{q_{u,1,n - s}}\left( {\tilde \mu _{h,s + u}^{\left( 1 \right)},\left( {\varphi  - \gamma } \right){c_9}} \right)} \right]}^{ - 1/2}}} \right) \\
\qquad \quad \;\;{E_t}\left( {h_{t + s}^{3/2}} \right)\left( {n - s} \right){\varphi ^{n - s - 1}} \\
\qquad \quad + \frac{3}{8}\left( \begin{array}{l}
{c_{10}}{\left[ {{q_{u,1,n - s}}\left( {\tilde \mu _{h,s + u}^{\left( 1 \right)},{c_{10}}} \right)} \right]^{ - 1/2}}{E_t}\left( {h_{t + s}^{5/2}} \right) + 2\omega \gamma {\left( {\gamma  - \varphi } \right)^{ - 1}}{c_9} \\
{\left[ {{q_{u,1,n - s}}\left( {\tilde \mu _{h,s + u}^{\left( 1 \right)},\left( {\gamma  - \varphi } \right){c_9}} \right)} \right]^{ - 1/2}}{E_t}\left( {h_{t + s}^{3/2}} \right) \\
\end{array} \right){\left( {\varphi  - \gamma } \right)^{ - 1}}\left( {{\varphi ^{n - s}} - {\gamma ^{n - s}}} \right), \\
\end{array}$

$\begin{array}{l}
{B_{u,s,n}} = \frac{3}{4}\left( {{c_9}{{\left[ {{q_{u,1,n - s}}\left( {\tilde \mu _{h,s + u}^{\left( 1 \right)},{c_9}} \right)} \right]}^{1/2}} + \omega \varphi {{\left( {\varphi  - \gamma } \right)}^{ - 1}}{c_9}{{\left[ {{q_{u,1,n - s}}\left( {\tilde \mu _{h,s + u}^{\left( 1 \right)}, - \left( {\varphi  - \gamma } \right){c_9}} \right)} \right]}^{ - 1/2}}} \right) \\
\qquad \quad \; \,{E_t}\left( {h_{t + s}^{3/2}} \right)\left( {n - s} \right){\varphi ^{n - s - 1}} \\
\qquad \quad + \frac{3}{8}\left( \begin{array}{l}
{c_{10}}{\left[ {{q_{u,1,n - s}}\left( {\tilde \mu _{h,s + u}^{\left( 1 \right)}, - {c_{10}}} \right)} \right]^{ - 1/2}}{E_t}\left( {h_{t + s}^{5/2}} \right) + 2\omega \gamma {\left( {\gamma  - \varphi } \right)^{ - 1}}{c_9} \\
{\left[ {{q_{u,1,n - s}}\left( {\tilde \mu _{h,s + u}^{\left( 1 \right)}, - \left( {\gamma  - \varphi } \right){c_9}} \right)} \right]^{ - 1/2}}{E_t}\left( {h_{t + s}^{3/2}} \right) \\
\end{array} \right){\left( {\varphi  - \gamma } \right)^{ - 1}}\left( {{\varphi ^{n - s}} - {\gamma ^{n - s}}} \right). \\
\end{array}$

Also, ${B_{ul,n}} \le \sum\limits_{s = 1}^n {{B_{u,s,n}}}  \le {B_{uu,n}}$ where for $\gamma \neq 1$:

$\begin{array}{l}
{B_{ul,n}} = \frac{3}{4}{l_{1,ul,n}}\left[ {{{\left( {1 - \varphi } \right)}^{ - 2}}\left( {1 - {\varphi ^n}} \right) - {{\left( {1 - \varphi } \right)}^{ - 1}}n{\varphi ^{n - 1}}} \right] \\
\qquad \;\, + \frac{3}{8}{l_{2,ul,n}}\left[ {{{\left( {\varphi  - \gamma } \right)}^{ - 1}}\left( {{{\left( {1 - \varphi } \right)}^{ - 1}}\left( {1 - {\varphi ^n}} \right) - {{\left( {1 - \gamma } \right)}^{ - 1}}\left( {1 - {\gamma ^n}} \right)} \right)} \right] \\
\end{array}$ \\
and \\
$\begin{array}{l}
{B_{uu,n}} = \frac{3}{4}{l_{1,uu,n}}\left[ {{{\left( {1 - \varphi } \right)}^{ - 2}}\left( {1 - {\varphi ^n}} \right) - {{\left( {1 - \varphi } \right)}^{ - 1}}n{\varphi ^{n - 1}}} \right] \\
\qquad \;\,\,  + \frac{3}{8}{l_{2,uu,n}}\left[ {{{\left( {\varphi  - \gamma } \right)}^{ - 1}}\left( {{{\left( {1 - \varphi } \right)}^{ - 1}}\left( {1 - {\varphi ^n}} \right) - {{\left( {1 - \gamma } \right)}^{ - 1}}\left( {1 - {\gamma ^n}} \right)} \right)} \right]. \\
\end{array}$\\

For $\gamma = 1$: \\
$\begin{array}{l}
{B_{ul,n}} = \frac{3}{4}{{\tilde l}_{1,ul,n}}\left[ {{{\left( {1 - \varphi } \right)}^{ - 2}}\left( {1 - {\varphi ^n}} \right) - {{\left( {1 - \varphi } \right)}^{ - 1}}n{\varphi ^{n - 1}}} \right] \\
\qquad \;\, + \frac{3}{8}{{\tilde l}_{2,ul,n}}{\left( {1 - \varphi } \right)^{ - 1}}\left( {n - {{\left( {1 - \varphi } \right)}^{ - 1}}\left( {1 - {\varphi ^n}} \right)} \right) \\
\end{array}$ \\
and \\
$\begin{array}{l}
{B_{uu,n}} = \frac{3}{4}{{\tilde l}_{1,uu,n}}\left[ {{{\left( {1 - \varphi } \right)}^{ - 2}}\left( {1 - {\varphi ^n}} \right) - {{\left( {1 - \varphi } \right)}^{ - 1}}n{\varphi ^{n - 1}}} \right] \\
\qquad \;\, + \frac{3}{8}{{\tilde l}_{2,uu,n}}{\left( {1 - \varphi } \right)^{ - 1}}\left( {n - {{\left( {1 - \varphi } \right)}^{ - 1}}\left( {1 - {\varphi ^n}} \right)} \right). \\
\end{array}$

For $\gamma  \in \left( {0,1} \right)$ we obtained that $\mathop {\lim }\limits_{n \to \infty } \left( {{l_{j,uu,n}}} \right)$, $\mathop {\lim }\limits_{n \to \infty } \left( {{l_{j,ul,n}}} \right)$, $j = 1$, $2$ exist and are finite. Thus:

$
\mathop {\lim }\limits_{n \to \infty } {n^{ - 2}}{B_{ul,n}} = \mathop {\lim }\limits_{n \to \infty } {n^{ - 2}}\left[ \begin{array}{l}
\frac{3}{4}{l_{1,ul,n}}\left[ {{{\left( {1 - \varphi } \right)}^{ - 2}}\left( {1 - {\varphi ^n}} \right) - {{\left( {1 - \varphi } \right)}^{ - 1}}n{\varphi ^{n - 1}}} \right] + \frac{3}{8}{l_{2,ul,n}}\\
\left[ {{{\left( {\varphi  - \gamma } \right)}^{ - 1}}\left( {{{\left( {1 - \varphi } \right)}^{ - 1}}\left( {1 - {\varphi ^n}} \right) - {{\left( {1 - \gamma } \right)}^{ - 1}}\left( {1 - {\gamma ^n}} \right)} \right)} \right] \\
\end{array} \right] =0$ \\
and \\
$
\mathop {\lim }\limits_{n \to \infty } {n^{ - 2}}{B_{uu,n}} = \mathop {\lim }\limits_{n \to \infty } {n^{ - 2}}\left[ \begin{array}{l}
\frac{3}{4}{l_{1,uu,n}}\left[ {{{\left( {1 - \varphi } \right)}^{ - 2}}\left( {1 - {\varphi ^n}} \right) - {{\left( {1 - \varphi } \right)}^{ - 1}}n{\varphi ^{n - 1}}} \right] + \frac{3}{8}{l_{2,uu,n}} \\
\left[ {{{\left( {\varphi  - \gamma } \right)}^{ - 1}}\left( {{{\left( {1 - \varphi } \right)}^{ - 1}}\left( {1 - {\varphi ^n}} \right) - {{\left( {1 - \gamma } \right)}^{ - 1}}\left( {1 - {\gamma ^n}} \right)} \right)} \right] \\
\end{array} \right] =0.$ \\

Thus $\mathop {\lim }\limits_{n \to \infty } {n^{ - 2}}\sum\limits_{s = 1}^n {{B_{u,s,n}}}  = 0$, yielding: $\mathop {\lim }\limits_{n \to \infty } {n^{ - 2}}\sum\limits_{s = 1}^n {\sum\limits_{u = 1}^{n - s} {{\varphi ^{n - s - u}}\theta _{su}^{\left( {3/2} \right)}} }  \le \mathop {\lim }\limits_{n \to \infty } {n^{ - 2}}\sum\limits_{s = 1}^n {{B_{u,s,n}}}  = 0$. Similarly, it can be shown that: $0 = \mathop {\lim }\limits_{n \to \infty } {n^{ - 2}}\sum\limits_{s = 1}^n {{B_{l,s,n}}}  \le \mathop {\lim }\limits_{n \to \infty } {n^{ - 2}}\sum\limits_{s = 1}^n {\sum\limits_{u = 1}^{n - s} {{\varphi ^{n - s - u}}\theta _{su}^{\left( {3/2} \right)}} }$. Finally, we get: $\mathop {\lim }\limits_{n \to \infty } {n^{ - 2}}\sum\limits_{s = 1}^n {\sum\limits_{u = 1}^{n - s} {{\varphi ^{n - s - u}}\theta _{su}^{\left( {3/2} \right)}} }  = 0$; we also showed above that $\mathop {\lim }\limits_{n \to \infty } {n^{ - 2}}\sum\limits_{s = 1}^n {\sum\limits_{u = 1}^{n - s} {\theta _{su}^{\left( {3/2} \right)}} }  = 0$, therefore, $A_3$ = 0 for $\gamma  \in \left( {0,1} \right)$.\\

For $\gamma  \in \left( {1,\infty } \right)$, we showed that ${A_1} = {A_2} =\infty$ . Thus if we can show that $A_3$ is bounded below, then the kurtosis of aggregated returns will diverge to plus infinity in this case.
Setting ${c_{17}} = {\tau _z} + 3{\left( {1 - \varphi } \right)^{ - 1}}{c_9}$, we can then write:

${A_3} = \mathop {\lim }\limits_{n \to \infty } {n^{ - 2}}\sum\limits_{s = 1}^n {\sum\limits_{u = 1}^{n - s} {\left[ \begin{array}{l}
	\frac{3}{4}{\left( {\tilde \mu _{h,s + u}^{\left( 1 \right)}} \right)^{ - 1/2}}{c_9}\left( {{c_{17}} + \left( {{\tau _z} - {c_{17}}} \right){\varphi ^{n - s - u}}} \right) \\
	\left[ {\tilde \mu _{h,s + u}^{\left( 1 \right)}{\varphi ^{u - 1}} + \omega {{\left( {\varphi  - \gamma } \right)}^{ - 1}}\left( {{\varphi ^u} - {\gamma ^u}} \right)} \right]{E_t}\left( {h_{t + s}^{3/2}} \right) \\
	+ \frac{3}{8}{\left( {\tilde \mu _{h,s + u}^{\left( 1 \right)}} \right)^{ - 1/2}}{\gamma ^{u - 1}}{c_{10}}\left( {{c_{17}} + \left( {{\tau _z} - {c_{17}}} \right){\varphi ^{n - s - u}}} \right){E_t}\left( {h_{t + s}^{5/2}} \right) \\
	\end{array} \right]} }. $\\


It is reasonable to assume that ${\mathop{\rm sgn}} \left( {{\tau _z}} \right) =  - {\mathop{\rm sgn}} \left( \lambda  \right)$
when $\tau_z \neq 0$ and $\lambda \neq 0$, since a positive $\lambda$ means that volatility is more responsive to negative shocks rather than positive shocks of the same magnitude and translates into a negative skew of the aggregated returns. \\
If ${\mathop{\rm sgn}} \left( {{\tau _z}} \right) =  - {\mathop{\rm sgn}} \left( \lambda  \right) \neq 0$, then
${\rm{sgn}}\left( {{\tau _{\rm{z}}}} \right){\rm{ = sgn}}\left( {{{\rm{c}}_9}} \right) = {\rm{sgn}}\left( {{{\rm{c}}_{{\rm{17}}}}} \right)$, and consequently \\ ${\mathop{\rm sgn}} \left( {{c_9}\left( {{c_{17}} + \left( {{\tau _z} - {c_{17}}} \right){\varphi ^{n - s - u}}} \right)} \right) =
{\mathop{\rm sgn}} \left( {{c_9}\left( {{\tau _z} + 3{{\left( {1 - \varphi } \right)}^{ - 1}}{c_9}\left( {1 - {\varphi ^{n - s - u}}} \right)} \right)} \right) = 1$. \\For
${\mathop{\rm sgn}} \left( {{\tau _z}} \right) = {\mathop{\rm sgn}} \left( {\mu _z^{\left( 5 \right)}} \right)$,\footnote{This is a sufficient but not necessary condition.} ${\rm{sgn}}\left( {{{\rm{c}}_9}} \right) = {\rm{sgn}}\left( {{{\rm{c}}_{{\rm{10}}}}} \right)$
and \\ ${\mathop{\rm sgn}} \left( {{c_{10}}\left[ {{c_{17}} + {\varphi ^{n - s - u}}\left( {{\tau _z} - {c_{17}}} \right)} \right]} \right) = {\mathop{\rm sgn}} \left( {{c_{10}}\left( {{\tau _z} + 3{{\left( {1 - \varphi } \right)}^{ - 1}}{c_9}\left( {1 - {\varphi ^{n - s - u}}} \right)} \right)} \right) = 1$. \\
Furthermore, for either $\{\tau_z=0 \;{\rm {and}}\; \lambda \ne 0 \}$ or $\{\tau_z \ne 0 \;{\rm {and}}\; \lambda = 0 \}$, we still get that \\ ${\mathop{\rm sgn}} \left( {{c_9}\left( {{c_{17}} + \left( {{\tau _z} - {c_{17}}} \right){\varphi ^{n - s - u}}} \right)} \right) = {\mathop{\rm sgn}} \left( {{c_{10}}\left( {{c_{17}} + \left( {{\tau _z} - {c_{17}}} \right){\varphi ^{n - s - u}}} \right)} \right) = 1$, if ${\mathop{\rm sgn}} \left( {{\tau _z}} \right) = {\mathop{\rm sgn}} \left( {\mu _z^{\left( 5 \right)}} \right)$; hence all terms in $A_3$ are positive if ${\rm{ sgn}}\left( {\left| \lambda  \right| + \left| {{\tau _z}} \right|} \right) \ne 0$. Finally, for $\tau_z=\lambda=0$, $A_3 = 0$.
We have thus shown that $A_3$ is bounded below (by 0), and hence the limit $\mathop {\lim }\limits_{n \to \infty } {{\rm K}_{r,n}} = \infty $ for $\gamma  \in \left( {1,\infty } \right)$.\\

For $\gamma = 1$,
we can write:

$\begin{array}{l}
\theta _{su}^{\left( {3/2} \right)} = \frac{3}{4}{c_9}\left[ {{{\left( {\tilde \mu _{h,s + u}^{\left( 1 \right)}} \right)}^{1/2}} + \omega \varphi {{\left( {\varphi  - 1} \right)}^{ - 1}}{{\left( {\tilde \mu _{h,s + u}^{\left( 1 \right)}} \right)}^{ - 1/2}}} \right]{\varphi ^{u - 1}}{E_t}\left( {h_{t + s}^{3/2}} \right) \\
\qquad \; + \frac{3}{8}{\left( {\tilde \mu _{h,s + u}^{\left( 1 \right)}} \right)^{ - 1/2}}\left( {{c_{10}}{E_t}\left( {h_{t + s}^{5/2}} \right) + 2\omega {{\left( {1 - \varphi } \right)}^{ - 1}}{c_9}{E_t}\left( {h_{t + s}^{3/2}} \right)} \right) \\
\end{array}$\\
and \\
$\begin{array}{l}
{\varphi ^{n - s - u}}\theta _{su}^{\left( {3/2} \right)} = \frac{3}{4}{c_9}\left[ {{{\left( {\tilde \mu _{h,s + u}^{\left( 1 \right)}} \right)}^{1/2}} + \omega \varphi {{\left( {\varphi  - 1} \right)}^{ - 1}}{{\left( {\tilde \mu _{h,s + u}^{\left( 1 \right)}} \right)}^{ - 1/2}}} \right]{\varphi ^{n - s - 1}}{E_t}\left( {h_{t + s}^{3/2}} \right) \\
\qquad \qquad \quad \;\; + \frac{3}{8}{\left( {\tilde \mu _{h,s + u}^{\left( 1 \right)}} \right)^{ - 1/2}}{\varphi ^{n - s - u}}\left( {{c_{10}}{E_t}\left( {h_{t + s}^{5/2}} \right) + 2\omega {{\left( {1 - \varphi } \right)}^{ - 1}}{c_9}{E_t}\left( {h_{t + s}^{3/2}} \right)} \right). \\
\end{array}$\\

Thus, for $\gamma = 1$, the expression for $A_3$ is:

${A_3} = \mathop {\lim }\limits_{n \to \infty } {n^{ - 2}}\sum\limits_{s = 1}^n {\sum\limits_{u = 1}^{n - s} {\left( \begin{array}{l}
	\frac{3}{4}{c_9}\left( {{c_{17}} + \left( {{\tau _z} - {c_{17}}} \right){\varphi ^{n - s - u}}} \right){\left( {\tilde \mu _{h,s + u}^{\left( 1 \right)}} \right)^{ - 1/2}} \\
	\left[ {\tilde \mu _{h,s + u}^{\left( 1 \right)}{\varphi ^{u - 1}} + \omega {{\left( {1 - \varphi } \right)}^{ - 1}}\left( {1 - {\varphi ^u}} \right)} \right]{E_t}\left( {h_{t + s}^{3/2}} \right) \\
	+ \frac{3}{8}{c_{10}}\left( {{c_{17}} + \left( {{\tau _z} - {c_{17}}} \right){\varphi ^{n - s - u}}} \right){\left( {\tilde \mu _{h,s + u}^{\left( 1 \right)}} \right)^{ - 1/2}}{E_t}\left( {h_{t + s}^{5/2}} \right) \\
	\end{array} \right)} }.$

Since ${c_9}$, ${c_{10}}$, and ${c_{17}}$ do not depend on  $\gamma$, we still have that
${\mathop{\rm sgn}} \left( {{c_9}} \right) = {\mathop{\rm sgn}} \left( {{c_{10}}} \right) = {\mathop{\rm sgn}} \left( {{c_{17}}} \right)$
and that:
${\mathop{\rm sgn}} \left( {{c_9}\left( {{c_{17}} + \left( {{\tau _z} - {c_{17}}} \right){\varphi ^{n - s - u}}} \right)} \right) = {\mathop{\rm sgn}} \left( {{c_{10}}\left( {{c_{17}} + \left( {{\tau _z} - {c_{17}}} \right){\varphi ^{n - s - u}}} \right)} \right) = 1$. Now we write:

$\begin{array}{l}
{A_3} \ge \mathop {\lim }\limits_{n \to \infty } {n^{ - 2}}\sum\limits_{s = 1}^n {\sum\limits_{u = 1}^{n - s} {\frac{3}{8}{c_{10}}\left( {{c_{17}} + \left( {{\tau _z} - {c_{17}}} \right){\varphi ^{n - s - u}}} \right){{\left( {\tilde \mu _{h,s + u}^{\left( 1 \right)}} \right)}^{ - 1/2}}{E_t}\left( {h_{t + s}^{5/2}} \right) \ge } }  \\
\qquad\frac{3}{8}\mathop {\lim }\limits_{n \to \infty } {\left[ {\mathop {\max }\limits_{1 \le s \le n} \left( {\tilde \mu _{h,s + u}^{\left( 1 \right)}} \right)} \right]^{ - 1/2}}\mathop {\lim }\limits_{n \to \infty } {n^{ - 2}}\sum\limits_{s = 1}^n {\sum\limits_{u = 1}^{n - s} {{c_{10}}\left( {{c_{17}} + \left( {{\tau _z} - {c_{17}}} \right){\varphi ^{n - s - u}}} \right)} } {E_t}\left( {h_{t + s}^{5/2}} \right) \ge  \\
\qquad \frac{3}{{64}}\mathop {\lim }\limits_{n \to \infty } {\left[ {\mathop {\max }\limits_{1 \le s \le n} \left( {\tilde \mu _{h,s + u}^{\left( 1 \right)}} \right)} \right]^{ - 1/2}}\mathop {\lim }\limits_{n \to \infty } \left[ {\mathop {\min }\limits_{1 \le s \le n} \left[ {\mathop {\min }\limits_{1 \le u \le n - s} \left[ {{c_{10}}\left( {{c_{17}} + \left( {{\tau _z} - {c_{17}}} \right){\varphi ^{n - s - u}}} \right)} \right]} \right]} \right] \\
\qquad \mathop {\lim }\limits_{n \to \infty } {n^{ - 2}}\sum\limits_{s = 1}^n {\sum\limits_{u = 1}^{n - s} {\left( {\left( {15{{\left( {\tilde \mu _{h,s}^{\left( 1 \right)}} \right)}^{1/2}}\tilde \mu _{h,s}^{\left( 2 \right)} - 7{{\left( {\tilde \mu _{h,s}^{\left( 1 \right)}} \right)}^{5/2}}} \right)} \right)} }  \\
\end{array}$\\

Since $\varphi  \in \left( {0,1} \right)$, both
$\mathop {\lim }\limits_{n \to \infty } \left[ {\mathop {\min }\limits_{1 \le s \le n} \left[ {\mathop {\min }\limits_{1 \le u \le n - s} \left[ {{c_{10}}\left( {{c_{17}} + \left( {{\tau _z} - {c_{17}}} \right){\varphi ^{n - s - u}}} \right)} \right]} \right]} \right]$
and
$\mathop {\lim }\limits_{n \to \infty } \left[ {\mathop {\max }\limits_{1 \le s \le n} \left( {\tilde \mu _{h,s + u}^{\left( 1 \right)}} \right)} \right]$
are finite and we have shown above that they are positive, if $\tau_z \neq 0$  or $\lambda \neq 0$. In this case,

$\begin{array}{l}
\;\mathop {\lim }\limits_{n \to \infty } {n^{ - 2}}\sum\limits_{s = 1}^n {\sum\limits_{u = 1}^{n - s} {\left( {\left( {15{{\left( {\tilde \mu _{h,s}^{\left( 1 \right)}} \right)}^{1/2}}\tilde \mu _{h,s}^{\left( 2 \right)} - 7{{\left( {\tilde \mu _{h,s}^{\left( 1 \right)}} \right)}^{5/2}}} \right)} \right)} }  \\
= \;\mathop {\lim }\limits_{n \to \infty } {n^{ - 2}}\sum\limits_{s = 1}^n {\sum\limits_{u = 1}^{n - s} {\left( {\left( {15{{\left( {\tilde \mu _{h,s}^{\left( 1 \right)}} \right)}^{1/2}}\left( \begin{array}{l}
			\left( {s - 1} \right)\left( {{\omega ^2} + 2\omega \varphi \bar h} \right) \\
			+ 2\varphi \bar h\left( {1 - {\varphi ^{s - 1}}} \right)\left( {{h_{t + 1}} - \bar h} \right) + h_{t + 1}^2 \\
			\end{array} \right) - 7{{\left( {\tilde \mu _{h,s}^{\left( 1 \right)}} \right)}^{5/2}}} \right)} \right)} }  \\
= 15\left( {{\omega ^2} + 2\omega \varphi \bar h} \right)\;\mathop {\lim }\limits_{n \to \infty } {n^{ - 2}}\sum\limits_{s = 1}^n {{{\left( {\tilde \mu _{h,s}^{\left( 1 \right)}} \right)}^{1/2}}\left( {n - s} \right)\left( {s - 1} \right)}  \\
\ge 15\left( {{\omega ^2} + 2\omega \varphi \bar h} \right)\;\mathop {\lim }\limits_{n \to \infty } \left[ {\mathop {\min }\limits_{1 \le s \le n} \left( {\tilde \mu _{h,s + u}^{\left( 1 \right)}} \right)} \right]\mathop {\lim }\limits_{n \to \infty } {n^{ - 2}}\left ( \frac{{{n^2}\left( {n - 5} \right)}}{6} \right), \\
\end{array}$ \\
where $15\left( {{\omega ^2} + 2\omega \varphi \bar h} \right)\mathop {\lim }\limits_{n \to \infty } \left[ {\mathop {\min }\limits_{1 \le s \le n} \left( {\tilde \mu _{h,s + u}^{\left( 1 \right)}} \right)} \right]$
is positive and finite. Thus $A_3 = \infty$ for $\gamma = 1$ and $\tau_z \neq 0$ and\textbackslash or $\lambda \neq 0$, while for $\gamma = 1$ and $\tau_z=\lambda=0$, $c_9=c_{10}=0$ and $A_3=0$. This completes our proof for the final expression for the limit of aggregated kurtosis given in Theorem 3, equation (10).\\


For the normal GJR, $\tau_z=0$ (but $\lambda \ne 0$), thus the expression for the limit of the aggregated kurtosis in his case simplifies to:
\begin{center}
$\mathop {\lim }\limits_{n \to \infty } {{\rm K}_{r,n}} = \left\{ \begin{array}{l}
3\qquad {\rm{if}}\;\gamma  \in \left( {0,1} \right), \\
\infty \quad \; \;  {\rm{if}}\;\gamma  \in \left[ {1,\infty } \right). \\
\end{array} \right.$
\end{center}
For the normal GARCH(1,1), we have $\tau_z=\lambda = 0$ and the limit of the aggregated kurtosis is now finite for $\gamma = 1$, although different from the normal value of 3. Thus, for the normal GARCH(1,1), we get:
\begin{center}
$\mathop {\lim }\limits_{n \to \infty } {{\rm K}_{r,n}} = \left\{ \begin{array}{l}
3\quad \qquad \qquad \qquad \quad \quad \quad \quad \quad \quad \quad \quad \; {\rm{if}}\,\gamma  \in \left( {0,1} \right), \\
3\left[ {1 + \frac{1}{2}\left( {1 + \alpha  + \beta } \right)\left( {1 + 5\alpha  + \beta } \right)} \right]\quad \;{\rm{if}}\,\gamma  = 1, \\
\infty \qquad \qquad \qquad \quad \quad \quad \quad \quad \quad \quad \quad \;\;\,{\rm{if}}\,\gamma  \in \left( {1,\infty } \right)\,. \\
\end{array} \right.$
\end{center}

\subsection*{T.A.4: Limits of the Moments of Variances}

This appendix derives the limits of the conditional moments of forward and aggregated variances of the generic GJR model as the time horizon increases. In what follows we use the notation and assumptions defined at the start of the Appendix; additionally, we assume  $c_4\neq \varphi$, $c_4 \neq \gamma$. \\

(a) \textit{Limit of the Variance of Forward Variance} \\
$ \begin{array}{l}
\mathop {\lim }\limits_{s \to \infty } \mu _{h,s}^{\left( 2 \right)} = \mathop {\lim }\limits_{s \to \infty } \left( {\tilde \mu _{h,s}^{\left( 2 \right)} - {{\left( {\tilde \mu _{h,s}^{\left( 1 \right)}} \right)}^2}} \right) \\
\qquad \qquad = \left\{ \begin{array}{l}
\mathop {\lim }\limits_{s \to \infty } \left[ \begin{array}{l}
\left( {{c_1} - {{\bar h}^2}} \right) + \left( { - {c_3} + h_{t + 1}^2} \right){\gamma ^{s - 1}} \\
+ \left[ {{c_2} - 2\bar h\left( {{h_{t + 1}} - \bar h} \right)} \right]{\varphi ^{s - 1}} - {\varphi ^{2\left( {s - 1} \right)}}{\left( {{h_{t + 1}} - \bar h} \right)^2} \\
\end{array} \right]\quad \quad \quad {\rm{if }} \; \gamma  \ne 1, \\
\mathop {\lim }\limits_{s \to \infty } \left[ \begin{array}{l}
\left( {2\omega \varphi \bar h\left( {{h_{t + 1}} - \bar h} \right) + h_{t + 1}^2 - {{\bar h}^2}} \right) + \left( {s - 1} \right)\left( {{\omega ^2} + 2\omega \varphi \bar h} \right) \\
- 2\bar h\left( {\omega \varphi  - 1} \right)\left( {{h_{t + 1}} - \bar h} \right){\varphi ^{s - 1}} - {\varphi ^{2\left( {s - 1} \right)}}{\left( {{h_{t + 1}} - \bar h} \right)^2} \\
\end{array} \right]\quad {\rm{if }}\; \gamma  = 1, \\
\end{array} \right. \\
\qquad \qquad = \left\{ \begin{array}{l}
\left( {{c_1} - {{\bar h}^2}} \right)\quad  {\rm{if }}\; \gamma  \in \left( {0,1} \right), \\
\infty \qquad \qquad {\rm{if }}\; \gamma  \in \left[ {1,\infty } \right), \\
\end{array} \right.
\end{array}$

where ${c_1} - {{\bar h}^2}>0$ and ${h_{t + 1}^2 - {c_3}}>0$. \\

(b)\textit{Limit of the Variance of Aggregated Variance}

%

For $\gamma \neq 1$, using (\ref{eq:3.41}) and (\ref{eq:3.40}) - (\ref{eq:3.43}), the expression for the variance of the aggregated variance can be written as:\footnote{The $n^2$ terms cancel out.}
$M_{h,n}^{\left( 2 \right)} = An + B{\gamma ^n} + {C_n}$
where	
$A = \left( {{c_1} - {{\bar h}^2}} \right)\left( {1 + 2\varphi {{\left( {1 - \varphi } \right)}^{ - 1}}} \right)$, \\
$B =  - \left( {\tilde \mu _{h,1}^{\left( 2 \right)} - {c_3}} \right){\left( {1 - \gamma } \right)^{ - 1}} - 2\varphi {\left( {1 - \varphi } \right)^{ - 1}}\left[ {\left( {\tilde \mu _{h,1}^{\left( 2 \right)} - {c_3}} \right){{\left( {1 - \gamma } \right)}^{ - 1}} - {{\left( {\varphi  - \gamma } \right)}^{ - 1}}} \right]$,
${C_n} = f\left( {n{\varphi ^n},{\varphi ^n}} \right)$.
Since $\varphi  \in \left( {0,1} \right)$, $\mathop {\lim }\limits_{n \to \infty } {\varphi ^n} = \mathop {\lim }\limits_{n \to \infty } n{\varphi ^n} = 0$ and $\exists \;C$ finite such that\footnote{Showing that $\mathop {\lim }\limits_{n \to \infty } n{\varphi ^n} = 0$ for $\varphi  \in \left( {0,1} \right)$ is rather immediate. If we define $y = {1 \mathord{\left/ {\vphantom {1 \varphi }} \right. \kern-\nulldelimiterspace} \varphi }$, then $y > 1$. We now have$\mathop {\lim }\limits_{n \to \infty } n{\varphi ^n} = \mathop {\lim }\limits_{n \to \infty } \frac{n}{{{y^n}}} = 0.$ }:
$\mathop {\lim }\limits_{n \to \infty } {C_n} = {\rm{C}}$.
Thus, the limit of the conditional variance of the aggregated conditional variance becomes:
\begin{equation}
\mathop {\lim }\limits_{n \to \infty } M_{h,n}^{\left( 2 \right)} = \left\{ \begin{array}{l}
{\mathop{\rm sgn}} \left( A \right)\infty ,\quad {\rm{if }}\;\gamma  \in \left( {0,1} \right), \\
{\mathop{\rm sgn}} \left( B \right)\infty \quad \; \, {\rm{if }}\; \gamma  \in \left( {1,\infty } \right), \\
\end{array} \right.
\label{eq:3.49}
\end{equation}
where it can be easily seen that ${\mathop{\rm sgn}} \left( A \right) = 1$ and ${\mathop{\rm sgn}} \left( B \right) = 1$.\\

For $\gamma = 1$, using (\ref{eq:3.55}) - (\ref{eq:3.56}), (\ref{eq:3.54}), and  (\ref{eq:3.42}) - (\ref{eq:3.53}), the expression for the variance of the aggregated variance becomes:
$M_{h,n}^{\left( 2 \right)} = A'{n^2} + B'n + C'$
where\footnote{As in the $\gamma \neq 1$ case above, $C' = f\left( {n{\varphi ^n},{\varphi ^n}} \right)$ and $\mathop {\lim }\limits_{n \to \infty } C' = {\rm{constant}}\,{\rm{.}}$ $B'$ is a constant, but its value and sign are irrelevant for the limit above.}
$A' = \frac{1}{2}\left[ {2\varphi {{\left( {1 - \varphi } \right)}^{ - 1}} + 1} \right]\left( {{\omega ^2} + 2\omega \varphi \bar h} \right) > 0$.
Hence:
$\mathop {\lim }\limits_{n \to \infty } M_{h,n}^{\left( 2 \right)} = \infty$
for any $\gamma$. \\

As with the variance of aggregated returns, the conditional variance of the aggregated conditional variance diverges to infinity when we increase the time horizon infinitely.
It is meaningful to compute the limit of the daily variance, i.e. the variance divided by time. However, unlike the daily (one-period) variance of aggregated returns which converges to the level of (daily) unconditional variance, the daily conditional variance of aggregated conditional variance diverges to infinity under certain parameter conditions:\footnote{Unlike the conditional variance of the aggregated returns, which only depends on (powers of) the $\varphi$ parameter which takes values only between 0 and 1, the conditional variance of the aggregated variance also depends on (powers of) the $\gamma$ parameter, which can take any positive value.}
\begin{center}
$\begin{array}{l}
\mathop {\lim }\limits_{n \to \infty } \frac{{M_{h,n}^{\left( 2 \right)}}}{n} = \left\{ \begin{array}{l}
\mathop {\lim }\limits_{n \to \infty } \left( {A + B\frac{{{\gamma ^n}}}{n} + \frac{{{C_n}}}{n}} \right)\quad \quad \,{\rm{if}}\;\gamma  \ne 1, \\
\mathop {\lim }\limits_{n \to \infty } \left( {A'n + B' + \frac{{C'}}{n}} \right)\quad \quad \,{\rm{if}}\,\gamma  = 1, \\
\end{array} \right.
= \left\{ \begin{array}{l}
A\quad {\rm{if}}\,\gamma  \in \left( {0,1} \right), \\
\infty \quad {\rm{if}}\,\gamma  \in \left[ {1,\infty } \right). \\
\end{array} \right. \\
\end{array}$
\end{center}
\vspace{0.5cm}
(c) \textit{Limit of the Skewness of Forward Variance}
\begin{center}
$\mathop {\lim }\limits_{s \to \infty } {\tau _{h,s}} =\mathop {\lim }\limits_{s \to \infty } {\left( {\mu _{h,s}^{\left( 3 \right)}}{{{{\left( {\mu _{h,s}^{\left( 2 \right)}} \right)}^{-3/2}}}}\right) }$
\end{center}
For $\gamma  \in \left( {0,1} \right)$, we can write: $\mathop {\lim }\limits_{s \to \infty } {\tau _{h,s}} = \frac{{\mathop {\lim }\limits_{s \to \infty } \left( {\mu _{h,s}^{\left( 3 \right)}} \right)}}{{\mathop {\lim }\limits_{s \to \infty } {{\left( {\mu _{h,s}^{\left( 2 \right)}} \right)}^{3/2}}}}$, where $\mathop {\lim }\limits_{s \to \infty } \mu _{h,s}^{\left( 2 \right)} = {c_1} - {\bar h^2}$. Using (\ref{third_centr_mom_fwd_var}), we get:
\vspace{-0.25cm}
\begin{equation}
\mathop {\lim }\limits_{s \to \infty } {\tau _{h,s}} = \left\{ \begin{array}{l}
M_1\quad {\rm{if}}\;\gamma  \in \left( {0,1} \right)\;{\rm{and}}\;{c_4} \in \left( {0,1} \right),\\
\infty \; \quad {\rm{if}}\;\gamma  \in \left( {0,1} \right)\;{\rm{and}}\;{c_4} \in \left[ {1,\infty } \right), \\
\end{array} \right.
\label{eq:3.62}
\end{equation}
where $M_1 = \frac{{\omega \left( {{\omega ^2} + 3\omega \varphi \bar h + 3\gamma {c_1}} \right){{\left( {1 - {c_4}} \right)}^{ - 1}} - 3\bar h{c_1} + 2{{\bar h}^3}}}{{{{\left( {{c_1} - {{\bar h}^2}} \right)}^{3/2}}}}$. \\
For $\gamma  \in \left( {1,\infty } \right)$, we can write:
$\mathop {\lim }\limits_{s \to \infty } {\tau _{h,s}} = \frac{{\mathop {\lim }\limits_{s \to \infty } \left( {{\gamma ^{ - (3/2)s}}\mu _{h,s}^{\left( 3 \right)}} \right)}}{{{{\left[ {\mathop {\lim }\limits_{s \to \infty } \left( {{\gamma ^{ - s}}\mu _{h,s}^{\left( 2 \right)}} \right)} \right]}^{3/2}}}}$
where, $\mathop {\lim }\limits_{s \to \infty } \left( {{\gamma ^{ - s}}\mu _{h,s}^{\left( 2 \right)}} \right) = {\gamma ^{ - 1}}\left( { - {c_3} + h_{t + 1}^2} \right)$. Using (\ref{third_centr_mom_fwd_var}), we get:
\begin{equation}
\mathop {\lim }\limits_{s \to \infty } {\tau _{h,s}} = \left\{ \begin{array}{l}
\infty \quad \;\;\,{\rm{if}}\;\gamma  \in \left( {1,\infty} \right)\;{\rm{and}}\;{c_4} > {\gamma ^{3/2}}, \\
0\quad \quad {\rm{if}}\;\gamma  \in \left( {1,\infty} \right)\;{\rm{and}}\;{c_4} \in \left( {0,{\gamma ^{3/2}}} \right), \\
M_2\quad \;\,{\rm{if}}\;\gamma  \in \left( {1,\infty} \right)\;{\rm{and}}\;{c_4} = {\gamma ^{3/2}}\,. \\
\end{array} \right.
\label{eq:3.64}
\end{equation}
where
$M_2 = \frac{{{c_{18}}}}{{{{\left( { - {c_3} + h_{t + 1}^2} \right)}^{3/2}}}} =
\frac{{h_{t + 1}^3 - \omega \left( {{\omega ^2} + 3\omega \varphi \bar h + 3\gamma {c_1}} \right){{\left( {1 - {c_4}} \right)}^{ - 1}} - \left( {3{\omega ^2}\varphi \left( {{h_{t + 1}} - \bar h} \right) + 3\omega \gamma \left( { - {c_1} + h_{t + 1}^2} \right)} \right){{\left( {\varphi  - {c_4}} \right)}^{ - 1}}}}{{{{\left( { - {c_3} + h_{t + 1}^2} \right)}^{3/2}}}}\;.
$ \\
For $\gamma = 1$, we can write: $\mathop {\lim }\limits_{s \to \infty } {\tau _{h,s}} = \frac{{\mathop {\lim }\limits_{s \to \infty } \left( {{s^{ - 3/2}}\mu _{h,s}^{\left( 3 \right)}} \right)}}{{{{\left[ {\mathop {\lim }\limits_{s \to \infty } \left( {{s^{ - 1}}\mu _{h,s}^{\left( 2 \right)}} \right)} \right]}^{3/2}}}}$, where,	$\mathop {\lim }\limits_{s \to \infty } \left( {{s^{ - 1}}\mu _{h,s}^{\left( 2 \right)}} \right) = \left( {{\omega ^2} + 2\omega \varphi \bar h} \right)$.
In this case, the limit becomes:
\begin{equation}
\mathop {\lim }\limits_{s \to \infty } {\tau _{h,s}} = \mathop {\lim }\limits_{s \to \infty } \left( {{s^{ - 3/2}}\tilde \mu _{h,s}^{\left( 3 \right)}} \right) = \left\{ \begin{array}{l}
0\quad {\rm{if}}\;{c_4} \in \left( {0,1} \right), \\
\infty \quad {\rm{if}}\;{c_4} \in \left( {1,\infty } \right). \\
\end{array} \right.
\label{eq:3.x}
\end{equation}
Now, (\ref{eq:3.62}) - (\ref{eq:3.x}) give (13) in Theorem 4. \\
%
(d) \textit{Limit of the Skewness of Aggregated Variance}
\begin{center}
$\mathop {\lim }\limits_{n \to \infty } {{\rm T}_{h,n}} = \mathop {\lim }\limits_{n \to \infty } \left[ {M_{h,n}^{\left( 3 \right)}{{\left( {M_{h,n}^{\left( 2 \right)}} \right)}^{ - 3/2}}} \right]$
\end{center}

For $\gamma \in \left( {0,1} \right)$, we can write: $\mathop {\lim }\limits_{n \to \infty } {{\rm T}_{h,n}} = \frac{{\mathop {\lim }\limits_{n \to \infty } \left( {{n^{ - 3/2}}M_{h,n}^{\left( 3 \right)}} \right)}}{{{{\left[ {\mathop {\lim }\limits_{n \to \infty } \left( {{n^{ - 1}}M_{h,n}^{\left( 2 \right)}} \right)} \right]}^{3/2}}}}$, \\
where $\mathop {\lim }\limits_{n \to \infty } \left( {{n^{ - 1}}M_{h,n}^{\left( 2 \right)}} \right) = \left( {{c_1} - {{\bar h}^2}} \right)\left( {1 + 2\varphi {{\left( {1 - \varphi } \right)}^{ - 1}}} \right)$. Also, we write: \\
$\begin{array}{l}
{{n^{ - 3/2}}M_{h,n}^{\left( 3 \right)}}= {L_1} - 3{L_2} + 2{L_3} + 3\left( {{L_4} + {L_5} + 2\left( {{L_6} + {L_7} - {L_8} - {L_9}} \right) - {L_{10}} - {L_{11}}} \right) \\
\quad \quad \quad \quad \quad \;\; + 6\left( {{L_{12}} - {L_{13}} - {L_{14}} - {L_{15}} + 2{L_{16}}} \right)\,, \\
\end{array}$ \\
where \\
${L_1} =  {\frac{{\sum\limits_{s = 1}^n {\tilde \mu _{h,s}^{\left( 3 \right)}} }}{{{n^{3/2}}}}} \,,$
${L_2} = {\frac{{\sum\limits_{s = 1}^n {\tilde \mu _{h,s}^{\left( 2 \right)}\tilde \mu _{h,s}^{\left( 1 \right)}} }}{{{n^{3/2}}}}} \,,$
${L_3} = {\frac{{\sum\limits_{s = 1}^n {{{\left( {\tilde \mu _{h,s}^{\left( 1 \right)}} \right)}^3}} }}{{{n^{3/2}}}}}$
${L_4} = {\frac{{\sum\limits_{s = 1}^n {\sum\limits_{u = 1}^{n - s} {\tilde \mu _{h,su}^{\left( {2,1} \right)}} } }}{{{n^{3/2}}}}}\,,$
${L_5} = {\frac{{\sum\limits_{s = 1}^n {\sum\limits_{u = 1}^{n - s} {\tilde \mu _{h,su}^{\left( {1,2} \right)}} } }}{{{n^{3/2}}}}}\,,$ \\
${L_6} = {\frac{{\sum\limits_{s = 1}^n {\sum\limits_{u = 1}^{n - s} {{{\left( {\tilde \mu _{h,s}^{\left( 1 \right)}} \right)}^2}\tilde \mu _{h,s + u}^{\left( 1 \right)}} } }}{{{n^{3/2}}}}} \,,$
${L_7} = {\frac{{\sum\limits_{s = 1}^n {\sum\limits_{u = 1}^{n - s} {{{\left( {\tilde \mu _{h,s + u}^{\left( 1 \right)}} \right)}^2}\tilde \mu _{h,s}^{\left( 1 \right)}} } }}{{{n^{3/2}}}}}\,,$
${L_8} = {\frac{{\sum\limits_{s = 1}^n {\sum\limits_{u = 1}^{n - s} {\tilde \mu _{h,s}^{\left( 1 \right)}\tilde \mu _{h,su}^{\left( {1,1} \right)}} } }}{{{n^{3/2}}}}} \,,$
${L_9} =  {\frac{{\sum\limits_{s = 1}^n {\sum\limits_{u = 1}^{n - s} {\tilde \mu _{h,s + u}^{\left( 1 \right)}\tilde \mu _{h,su}^{\left( {1,1} \right)}} } }}{{{n^{3/2}}}}}\,,$
${L_{10}} = {\frac{{\sum\limits_{s = 1}^n {\sum\limits_{u = 1}^{n - s} {\tilde \mu _{h,s}^{\left( 1 \right)}\tilde \mu _{h,s + u}^{\left( 2 \right)}} } }}{{{n^{3/2}}}}}\,,$
${L_{11}} =  {\frac{{\sum\limits_{s = 1}^n {\sum\limits_{u = 1}^{n - s} {\tilde \mu _{h,s + u}^{\left( 1 \right)}\tilde \mu _{h,s}^{\left( 2 \right)}} } }}{{{n^{3/2}}}}}\,,$
${L_{12}} = {\frac{{\sum\limits_{s = 1}^n {\sum\limits_{u = 1}^{n - s} {\sum\limits_{v = 1}^{n - s - u} {\tilde \mu _{h,suv}^{\left( {1,1,1} \right)}} } } }}{{{n^{3/2}}}}} \,,$
${L_{13}} =  {\frac{{\sum\limits_{s = 1}^n {\sum\limits_{u = 1}^{n - s} {\sum\limits_{v = 1}^{n - s - u} {\tilde \mu _{h,s}^{\left( 1 \right)}\tilde \mu _{h,\left( {s + u} \right)v}^{\left( {1,1} \right)}} } } }}{{{n^{3/2}}}}} \,,$
${L_{14}} = {\frac{{\sum\limits_{s = 1}^n {\sum\limits_{u = 1}^{n - s} {\sum\limits_{v = 1}^{n - s - u} {\tilde \mu _{h,\left( {s + u} \right)}^{\left( 1 \right)}\tilde \mu _{h,s\left( {u + v} \right)}^{\left( {1,1} \right)}} } } }}{{{n^{3/2}}}}}\,,$
${L_{15}} = {\frac{{\sum\limits_{s = 1}^n {\sum\limits_{u = 1}^{n - s} {\sum\limits_{v = 1}^{n - s - u} {\tilde \mu _{h,\left( {s + u + v} \right)}^{\left( 1 \right)}} } } \tilde \mu _{h,su}^{\left( {1,1} \right)}}}{{{n^{3/2}}}}} \,,$
${L_{16}} = {\frac{{\sum\limits_{s = 1}^n {\sum\limits_{u = 1}^{n - s} {\sum\limits_{v = 1}^{n - s - u} {\tilde \mu _{h,s}^{\left( 1 \right)}\tilde \mu _{h,\left( {s + u} \right)}^{\left( 1 \right)}\tilde \mu _{h,\left( {s + u + v} \right)}^{\left( 1 \right)}} } } }}{{{n^{3/2}}}}}\,.$

Performing the necessary (tedious but straightforward) calculations and using the notation ${R_i}$, $\tilde{R_i}$, $i \in \left \{1,2 , 3....16 \right\}$ with $\mathop {\lim}\limits_{n\to \infty}{R_i}=\mathop{\lim}\limits_{n \to \infty}{\tilde{R_i}}=0$, we get:\\

${L_1} = \left\{ \begin{array}{l}
{c_{18}}{\left( {{c_4} - 1} \right)^{ - 1}}\frac{{c_4^n}}{{{n^{3/2}}}}+{R_1}\, \quad \quad {\rm{if}}\;{c_4} \ne 1, \\
\frac{{\omega \left( {{\omega ^2} + 3\omega \varphi \bar h + 3\gamma {c_1}} \right)}}{2} \frac{{{n^2}}}{{{n^{3/2}}}}+\tilde{R_1}\quad \,{\rm{if}}\;{c_4} = 1, \\
\end{array} \right.$
${L_2} = {R_2}$,
${L_3} ={R_3}$, \\
${L_4} = \left\{ \begin{array}{l}
\frac{{\varphi {c_{18}}}}{{\left( {1 - {c_4}} \right)\left( {\varphi  - {c_4}} \right)}}\frac{{c_4^n}}{{{n^{3/2}}}} + \frac{{\bar h}}{2}{c_1}\frac{{{n^2}}}{{{n^{3/2}}}}+{R_4}\qquad \qquad \qquad {\rm{if}}\;{c_4} \ne 1, \\
\frac{{\bar h}}{2}\left( {{c_1} + \varphi \left( {{\omega ^2} + 3\bar h\varphi \omega  + 3\gamma {c_1}} \right)} \right) \frac{{{n^2}}}{{{n^{3/2}}}}+\tilde{R_4} \quad \;\,\,{\rm{if}}\;{c_4} = 1, \\
\end{array} \right.$

${L_5} = \left\{ \begin{array}{l}
\frac{{\gamma {c_{18}}}}{{\left( {1 - {c_4}} \right)\left( {\gamma  - {c_4}} \right)}}\frac{{c_4^n}}{{{n^{3/2}}}} + {c_1}\frac{{\bar h}}{2}\frac{{{n^2}}}{{{n^{3/2}}}}+{R_5}\quad \;\;\;\;\qquad \qquad {\rm{if}}\;{c_4} \ne 1, \\
\left[ \begin{array}{l}
\gamma {\left( {1 - \gamma } \right)^{ - 1}}\frac{{\omega \left( {{\omega ^2} + 3\bar h\varphi \omega  + 3\gamma {c_1}} \right)}}{2} \\
{\omega ^2}{\left( {1 - \gamma } \right)^{ - 1}}\frac{{\bar h}}{2} + \omega \varphi {{\bar h}^2} \\
\end{array} \right]\frac{{{n^2}}}{{{n^{3/2}}}}+\tilde{R_5}\;\;\;\quad {\rm{if}}\;{c_4} = 1, \\
\end{array} \right.$\\

${L_6} = \frac{{{{\bar h}^3}}}{2}\frac{{{n^2}}}{{{n^{3/2}}}}+{R_6}$,
${L_7} = \frac{{{{\bar h}^3}}}{2}\frac{{{n^2}}}{{{n^{3/2}}}}+{R_7}$,
${L_8} = \frac{{{{\bar h}^3}}}{2}\frac{{{n^2}}}{{{n^{3/2}}}} +{R_8}$,
${L_9} = \frac{{{{\bar h}^3}}}{2} \frac{{{n^2}}}{{{n^{3/2}}}}+{R_9}$,
${L_{10}} = \frac{{{c_1}\bar h}}{2} \frac{{{n^2}}}{{{n^{3/2}}}} +{R_{10}}$, \\
${L_{11}} = \frac{{{c_1}\bar h}}{2} \frac{{{n^2}}}{{{n^{3/2}}}}+R_{11}$,
${L_{12}} = \left\{ \begin{array}{l}
\begin{array}{l}
\varphi \gamma {c_{18}}{\left( {{c_4} - 1} \right)^{ - 1}}{\left( {{c_4} - \varphi } \right)^{ - 1}}{\left( {{c_4} - \gamma } \right)^{ - 1}}\frac{{c_4^n}}{{{n^{3/2}}}} + \frac{{{{\bar h}^3}}}{6}\frac{{{n^3}}}{{{n^{3/2}}}} \\
+ \frac{{\bar h}}{2}{\left( {1 - \varphi } \right)^{ - 1}}\left(  2\varphi{\left( {c_1} - {{\bar h}^2}\right)  + \bar h\left( {\left( {{h_{t + 1}} - \bar h} \right) - \omega } \right)} \right)\frac{{{n^2}}}{{{n^{3/2}}}}+{R_{12}} \\
\end{array}\quad \;{\rm{if}}\,{c_4} \ne 1,\quad  \\
\frac{{{{\bar h}^3}}}{6}\frac{{{n^3}}}{{{n^{3/2}}}} + \frac{{\bar h}}{2}\left[ \begin{array}{l}
{\left( {1 - \varphi } \right)^{ - 1}} \\
\left( {{c_1}\left( {1 + \varphi } \right) - 2\varphi {{\bar h}^2} + \bar h\left( {\left( {{h_{t + 1}} - \bar h} \right) - \omega } \right)} \right) \\
+ \varphi \gamma {\left( {1 - \gamma } \right)^{ - 1}}\left( {{\omega ^2} + 3\omega \varphi \bar h + 3\gamma {c_1}} \right) \\
\end{array} \right]\frac{{{n^2}}}{{{n^{3/2}}}}+\tilde{R_{12}} \;\;{\rm{if}}\,{c_4} = 1, \\
\end{array} \right.$
${L_{13}} = \frac{{{{\bar h}^3}}}{6} \frac{{{n^3}}}{{{n^{3/2}}}} + \frac{{\bar h}}{2}\left( { - {{\bar h}^2} + {{\left( {1 - \varphi } \right)}^{ - 1}}\left( {\bar h\left( {{h_{t + 1}} - \bar h} \right) + \varphi \left( {{c_1} - {{\bar h}^2}} \right)} \right)} \right)\frac{{{n^2}}}{{{n^{3/2}}}} +{R_{13}}$, \\
${L_{14}} = \frac{{{{\bar h}^3}}}{6}\frac{{{n^3}}}{{{n^{3/2}}}} + \frac{{{{\bar h}^2}}}{2}\left( { - \bar h + \left( {{h_{t + 1}} - \bar h} \right){{\left( {1 - \varphi } \right)}^{ - 1}}} \right)\frac{{{n^2}}}{{{n^{3/2}}}}+{R_{14}}$, \\
${L_{15}} = \frac{{{{\bar h}^3}}}{6} \frac{{{n^3}}}{{{n^{3/2}}}} + \frac{{\bar h}}{2}{\left( {1 - \varphi } \right)^{ - 1}}\left( { - 2{{\bar h}^2} + \bar h{h_{t + 1}} + {c_1}\varphi } \right) \frac{{{n^2}}}{{{n^{3/2}}}}+{R_{15}}$, \\
${L_{16}} = \frac{{{{\bar h}^3}}}{6} \frac{{{n^3}}}{{{n^{3/2}}}} + \frac{{{{\bar h}^2}}}{2}\left( { - \bar h + \left( {{h_{t + 1}} - \bar h} \right){{\left( {1 - \varphi } \right)}^{ - 1}}} \right)\frac{{{n^2}}}{{{n^{3/2}}}} +{R_{16}}$.\\

Performing the necessary calculations, we obtain the expression in Theorem 4, equation (14), where: \\
$\begin{array}{l}
N =  \frac{{\omega \left( {{\omega ^2} + 3\omega \varphi \bar h + 3\gamma {c_1}} \right)}}{2} + 3\frac{{\bar h}}{2}\left( {{c_1} + \varphi \left( {{\omega ^2} + 3\omega \varphi \bar h + 3\gamma {c_1}} \right)+ {\omega ^2}{{\left( {1 - \gamma } \right)}^{ - 1}}} + 2\omega \varphi {\bar h}\right) \\
\quad + 3 {\gamma {{\left( {1 - \gamma } \right)}^{ - 1}}\frac{{\left( {{\omega ^2} + 3\omega \varphi \bar h + 3\gamma {c_1}} \right)}}{2}}\left(\omega +2\varphi\bar h\right)  \\
\quad + 3{\left( {1 - \varphi } \right)^{ - 1}}\bar h\left[ {{c_1}\left( {1 + \varphi } \right) - 2\varphi {{\bar h}^2} + \bar h\left( {\left( {{h_{t + 1}} - \bar h} \right) - \omega } \right)} - \varphi \left( {2{c_1} - {{\bar h}^2}} \right)\right]. \\
\end{array}$

For $\gamma  \in \left( {1,\infty } \right)$, we can write: $\mathop {\lim }\limits_{n \to \infty } {{\rm T}_{h,n}} = \frac{{\mathop {\lim }\limits_{n \to \infty } \left( {{\gamma ^{ - 3/2n}}M_{h,n}^{\left( 3 \right)}} \right)}}{{{{\left[ {\mathop {\lim }\limits_{n \to \infty } \left( {{\gamma ^{ - n}}M_{h,n}^{\left( 2 \right)}} \right)} \right]}^{3/2}}}}$, \\
where
$\mathop {\lim }\limits_{n \to \infty } \left( {{\gamma ^{ - n}}M_{h,n}^{\left( 2 \right)}} \right) =  - \left( {\tilde \mu _{h,1}^{\left( 2 \right)} - {c_3}} \right){\left( {1 - \gamma } \right)^{ - 1}} - 2\varphi {\left( {1 - \varphi } \right)^{ - 1}}\left[ {\left( {\tilde \mu _{h,1}^{\left( 2 \right)} - {c_3}} \right){{\left( {1 - \gamma } \right)}^{ - 1}} - {{\left( {\varphi  - \gamma } \right)}^{ - 1}}} \right]$.
For $\gamma = 1$, we can write: $\mathop {\lim }\limits_{n \to \infty } {{\rm T}_{h,n}} = \frac{{\mathop {\lim }\limits_{n \to \infty } \left( {{n^{ - 3}}M_{h,n}^{\left( 3 \right)}} \right)}}{{{{\left[ {\mathop {\lim }\limits_{n \to \infty } \left( {{n^{ - 2}}M_{h,n}^{\left( 2 \right)}} \right)} \right]}^{3/2}}}}$,\\
where
$\mathop {\lim }\limits_{n \to \infty } \left( {{n^{ - 2}}M_{h,n}^{\left( 2 \right)}} \right) = \frac{1}{2}\left[ {2\varphi {{\left( {1 - \varphi } \right)}^{ - 1}} + 1} \right]\left( {{\omega ^2} + 2\omega \varphi \bar h} \right)$.

\end{document}